\documentclass[12pt, reqno, a4paper]{report}

\usepackage{amsfonts}

\usepackage{amssymb}

\usepackage{amsmath}

\usepackage{amsthm}

\usepackage{makeidx}

\numberwithin{equation}{section}

\newcommand{\Space}[2]{ \mathbb{#1}^{#2} }

\newcommand{\ub}[3][]{\left\{\!#1\left[#2,#3\right]\!#1\right\}}

\newcommand{\halg}[1]{ \mathfrak{h}_{#1} }

\newcommand{\horb}{\mathcal{O}_{h}}

\newtheorem{Thm}{Theorem}[section]

\newtheorem{Defn}[Thm]{Definition}

\newtheorem{Lemma}[Thm]{Lemma}

\newtheorem{Example}[Thm]{Example}

\newtheorem{Proposition}[Thm]{Proposition}

\newtheorem{Corollary}[Thm]{Corollary}

\newtheorem{Remark}[Thm]{Remark}

\newcommand{\Partial}[1]{ \frac{\partial}{\partial #1} }

\newcommand{\Fracpartial}[2]{\frac{\partial #1}{\partial #2} }

\newcommand{\Partialtwo}[1]{ \frac{\partial^2}{\partial #1^2} }

\newcommand{\Diffl}[1]{\frac{d}{d #1}}

\newcommand{\Fracdiffl}[2]{\frac{d #1}{d #2}}

\newcommand{\Lalg}[2]{ \mathfrak{#1}^{#2} }

\newcommand{\LieA}{\mathfrak{g}}

\newcommand{\Symp}[1]{ Sp(#1, \Space{R}{} ) }

\newcommand{\Heisn}{\Space{H}{n}}

\newcommand{\heisn}{\mathfrak{h}_{n}}

\newcommand{\heisnstar}{\Lalg{h}{*}_{n}}

\newcommand{\uorb}{\mathcal{O}_{h}}

\newcommand{\hilbh}{\mathcal{H}_{h}^2}

\newcommand{\hilbho}{\mathcal{H}_{h}}

\newcommand{\lkerh}{\mathcal{L}_{h}}

\newcommand{\lkero}{\mathcal{L}_{0}}

\newcommand{\loneh}{L^1(\Heisn)}

\newcommand{\ltwoh}{L^2(\Heisn)}

\newcommand{\ltwohnstar}{L^2 (\heisnstar) }

\newcommand{\ltworn}{L^2(\Space{R}{n})}

\newcommand{\ltworthree}{L^2(\Space{R}{3})}

\newcommand{\cinf}{C^{\infty}}

\newcommand{\czeroinf}{C^{\infty}_{0}(\Space{R}{n})}

\newcommand{\czeroinfg}{C^{\infty}_{0}(G)}

\newcommand{\fock}{F^2(\uorb)}

\newcommand{\focko}{F(\uorb)}

\newcommand{\antid}{\mathcal{A}}

\newcommand{\zerodel}{\delta(s) \delta(x) \delta(y) }

\newcommand{\zerodelsone}{\delta^{(1)}(s) \delta(x) \delta(y) }

\newcommand{\zerodelxone}{\delta(s) \delta^{(1)}(x) \delta(y) }

\newcommand{\zerodelyone}{\delta(s) \delta(x) \delta^{(1)}(y) }

\newcommand{\zerodelxtwo}{\delta(s) \delta^{(2)}(x) \delta(y) }

\newcommand{\zerodelytwo}{\delta(s) \delta(x) \delta^{(2)}(y) }

\newcommand{\zerodelky}{\delta (s) \delta (x) \delta_{(k)}^{(1)} (y) }

\newcommand{\zerodelix}{\delta (s) \delta_{(i)}^{(1)} (x) \delta (y) }

\newcommand{\zerodeljx}{\delta (s) \delta_{(j)}^{(1)} (x) \delta (y) }

\newcommand{\zerodeljy}{\delta (s) \delta (x) \delta_{(j)}^{(1)} (y) }

\newcommand{\fort}{\mathcal{F}}

\newcommand{\vab}{v_{(h,a,b)}}

\newcommand{\vabd}{v_{(h,a',b')}}

\newcommand{\vabdd}{v_{(h,a'',b'')}}

\newcommand{\voo}{v_{(h,0,0)}}

\newcommand{\lab}{l_{(h,a,b)}}

\newcommand{\labo}{l_{(0,a,b)}}

\newcommand{\qporb}{\mathcal{O}_{(q,p)}}

\newcommand{\statem}{\mathcal{S}_h}

\newcommand{\statemo}{\mathcal{S}_0}

\newcommand{\proj}{\mathcal{P}}

\newcommand{\sokh}{\mathcal{L}_h}

\newcommand{\Nat}{\Space{N}{}}

\newcommand{\schw}{\mathcal{S} (\Space{R}{n})}

\newcommand{\schwp}{\mathcal{S}'(\Space{R}{n})}

\newcommand{\schwg}{\mathcal{S} (G)}

\newcommand{\schwpg}{\mathcal{S}'(G)}

\newcommand{\kephamwkp}{\delta (s) \frac{1}{\pi \|x\|^2 } \delta (y)}

\newcommand{\fsp}{\mathcal{F} (\Space{SP}{3})}

\newcommand{\msp}{\mathcal{M}_{S} }

\newcommand{\mspi}{\mathcal{M}_{S}^{-1} }

\newcommand{\mgc}{\mathcal{M}}

\newcommand{\mgci}{\mathcal{M}^{-1}}

\newcommand{\fgc}{\mathcal{F}}

\newcommand{\pzang}{\delta(s) \delta_{(1)}^{(1)}(x) \delta_{(2)}^{(1)}(y) - \delta(s) \delta_{(2)}^{(1)}(x) \delta_{(1)}^{(1)}(y)}

\newcommand{\kcs}{\mathbb{KC}}

\newcommand{\kcmap}{\mathcal{K}_1}

\newcommand{\bssp}{\mathbb{BS}}

\newcommand{\sgob}{\sigma,\gamma,\overline{\Omega}}

\newcommand{\rtph}{r,\theta,\phi}

\newcommand{\vlam}{v_{\lambda}}

\newcommand{\wavtrans}{\mathcal{W}_h}

\newcommand{\invwavtrans}{\mathcal{M}_h}

\newcommand{\qmap}{\mathcal{Q}}

\newcommand{\rhm}{\rho_h^{\mathcal{M}}}

\newcommand{\rn}{\Space{R}{n}}

\makeindex

\begin{document}

\begin{center}

\textbf{{\LARGE Relationships Between Quantum and Classical Mechanics using the Representation Theory of the Heisenberg Group.}}

\vspace{1cm}

{\Huge Alastair Robert Brodlie}

\vspace{1cm}

{\LARGE Submitted in accordance with the requirements for the degree of Doctor of Philosophy.}

\vspace{1cm}

{\LARGE School of Mathematics, \\  The University of Leeds.}

\vspace{1cm}

{\LARGE September 2004}

\vspace{1cm}

{\large The candidate confirms that the work submitted is his own and that appropriate credit has been given where reference has been made to the work of others.}

\vspace{1cm}

{\large This copy has been supplied on the understanding that it is copyright material and that no quotation from the thesis may be published without proper acknowledgment.}

\end{center}

\newpage
\begin{center}
\textbf{{\LARGE Acknowledgments}}
\end{center}

\indent

First of all I would like to thank my supervisor, Vladimir Kisil, for all his support and encouragement throughout my PhD studies. Without his guidance and useful discussions this thesis would never have appeared. I would also like to thank all the other members of the Functional Analysis group at Leeds, especially my co-supervisor Jonathan Partington.

I would also like to thank my parents and the rest of my family for all their support throughout the course of my PhD studies. Many thanks go to those postgraduates who have studied in the Leeds maths department at the same time as me. The last three years would have been much less enjoyable without the numerous football matches and pub trips. Further thanks go to all my friends outside the maths department who have given me many welcome distractions throughout my PhD studies. I would also like to thank EPSRC for funding my research.

\newpage

\begin{center}
\textbf{{\LARGE Abstract}}
\end{center}

\indent

This thesis is concerned with the representation theory of the Heisenberg group and its applications to both classical and quantum mechanics. We continue the development of $p$-mechanics which is a consistent physical theory capable of describing both classical and quantum mechanics simultaneously. $p$-Mechanics starts from the observation that the one dimensional representations of the Heisenberg group play the same role in classical mechanics which the infinite dimensional representations play in quantum mechanics.

In this thesis we introduce the idea of states to $p$-mechanics. $p$-Mechanical states come in two forms: elements of a Hilbert space and integration kernels. In developing $p$-mechanical states we show that quantum probability amplitudes can be obtained using solely functions/distributions on the Heisenberg group. This theory is applied to the examples of the forced, harmonic and coupled oscillators. In doing so we show that both the quantum and classical dynamics of these systems can be derived from the same source. Also using $p$-mechanics we simplify some of the current quantum mechanical calculations.

We also analyse the role of both linear and non-linear canonical transformations in $p$-mechanics. We enhance a method derived by Moshinsky for studying the passage of canonical transformations from classical to quantum mechanics. The Kepler/Coulomb problem is also examined in the $p$-mechanical context. In analysing this problem we show some limitations of the current $p$-mechanical approach. We then use Klauder's coherent states to generate a Hilbert space which is particularly useful for  the Kepler/Coulomb problem.

\newpage

\tableofcontents

\chapter{Introduction}

Since the time of von-Neumann the infinite dimensional Schr\"odinger representation of the Heisenberg group on $\ltworn$ has been used in quantum mechanics. By the Stone-von Neumann theorem all unitary irreducible infinite dimensional representations of the Heisenberg group are unitarily equivalent to the Schr\"odinger representation. This means that up to unitary equivalence all unitary irreducible infinite dimensional representations of the Heisenberg group are ``the same''. In Bargmann's 1961 paper \cite{Bargmann61} a unitary irreducible representation of the Heisenberg group was defined on the Fock--Segal--Bargmann space of entire analytic functions on $\Space{C}{n}$. This representation despite being unitarily equivalent to the Schr\"odinger representation was shown to be especially useful when considering particular systems. For example the dynamics of a state evolving in the harmonic oscillator system is given by a rotation of the function's coordinates.

In the Stone-von Neumann theorem it is also stated that there exists a family of one dimensional representations of the Heisenberg group. These representations are largely ignored; however in \cite{Kisil96.1,Kisil02,Kisil02.1,Kisil97} it is shown that the one dimensional representations can play the same role in classical mechanics which the infinite dimensional representations play in quantum mechanics. This led to the development of $p$-mechanics and the theory that both classical and quantum mechanics are derived from the same source being separated by the one and infinite dimensional representations respectively.

In this thesis we continue the development of expanding the representation theory of the Heisenberg group beyond the infinite dimensional Schr\"odinger representation. We show how coherent states and canonical transformations can be made clearer using different representations of the Heisenberg group. We consider the examples of the forced, coupled and harmonic oscillators along with the Kepler/Coulomb problem. In doing so we show how both the quantum and classical behaviour of these systems can be modelled using $p$-mechanics and the representation theory of the Heisenberg group. By analysing these problems we show that $p$-mechanics can be applied to actual physical systems and is not a purely theoretical concept.

We now give a summary and overview of the thesis. In Chapter \ref{chap:classandquant} we start the thesis by presenting some background material on the mathematical foundations of both classical and quantum mechanics. In this chapter we include all the formulae and results from classical and quantum mechanics which are needed in the thesis. We also give a little history on the development of quantum theory and discuss the limitations of the current mathematical framework.

Chapter \ref{chap:pmechandhn} is mainly preliminary material and is split into two sections. Section \ref{sect:hnanditsreps} contains definitions and results on the representation theory of the Heisenberg group. The majority of the results in this section are known but everything is presented in a way which is accessible for the rest of the thesis. In Section \ref{sect:pmechintro} we present a summary of $p$-mechanics. The majority of this summary is known material, but we present a new definition of $p$-mechanical observables.

Chapter \ref{chap:staesandpic} is the first chapter of entirely new material. In this chapter we introduce the concept of states to $p$-mechanics. In doing so we show that quantum mechanical probability amplitudes can be calculated using solely representations/distributions on the Heisenberg group. We also introduce a system of coherent states; in doing so we give a simple proof of the classical limit of coherent states. Chapter \ref{chap:staesandpic} also contains a description of the interaction picture in $p$-mechanics using both the kernel and Hilbert space states. Also contained within this chapter are relationships between our new Hilbert space and the usual $\ltworn$ model of quantum mechanics. The chapter concludes with a discussion about the rigged Hilbert spaces associated with $p$-mechanics. The majority of the work in this chapter was published in the papers \cite{Brodlie02,KisilBrodlie02}.

In Chapter \ref{chap:forcedosc} we divert from deriving the general theory of $p$-mechanics  to consider a few examples of physical systems. This illuminates the theory and shows how $p$-mechanics is applicable to actual physical systems. The problems we consider are the harmonic oscillator and the forced oscillator. By considering these systems in $p$-mechanics we are able to obtain some new and interesting relations between quantum and classical mechanics. It is shown that both the quantum and classical dynamics of these systems are generated from the same source. Also this chapter demonstrates that by using the machinery of $p$-mechanics we can simplify some of the calculations which are given in the standard quantum mechanical literature. The work in Chapter \ref{chap:forcedosc} is entirely new and was published in \cite{Brodlie02,KisilBrodlie02}.

Chapter \ref{chap:cantransf} is another chapter of entirely new material. In this chapter we investigate how canonical transformations should be modelled in $p$-mechanics. In doing so we obtain relations between classical and quantum canonical transformations. We also shed some light on the long standing problem of how classical canonical transformations should be passed into quantum mechanics. We look at both linear and non-linear transformations separately. Linear canonical transformations are shown to be closely linked to the metaplectic representation of the symplectic group. For dealing with non-linear transformations we enhance a method derived by Moshinsky through using the coherent states which were introduced in Chapter \ref{chap:staesandpic}. We apply our theory to several examples. In particular we show how two coupled oscillators can be decoupled in $p$-mechanics. Some of the work in this chapter was published in \cite{Brodlie04}.

In Chapter \ref{chap:keplercoulomb} we consider the problem of  the Kepler/Coulomb problem using $p$-mechanics. We initially describe the non-trivial nature of  the problem and show that the machinery in this thesis so far is insufficient for dealing with this problem. Next we introduce spherical polar coordinates to $p$-mechanics and show that this helps to simplify the problem. In doing so we show that the spherical polar coordinates pictures of both classical and quantum mechanics can be derived from the same source using the representation theory of the Heisenberg group. We also construct a new Hilbert space which plays a similar role for the Kepler/Coulomb system which the Fock--Segal--Bargmann space plays for the harmonic oscillator system. This space is only suitable for modelling a subset of the quantum mechanical states/observables and does not possess a representation of the Heisenberg group. Some of the work in this chapter has been presented in \cite{Brodlie04.1}.

\chapter{Classical and Quantum Physics} \label{chap:classandquant}

\section{Classical Mechanics} \label{sect:classmech}

\indent

 Since the time of Newton till the start of the twentieth century the majority of physics was assumed to be governed by the laws of classical mechanics \cite{Arnold90,Jose98}.
In classical mechanics the state of a system with $n$ independent particles is given by $3n$ position coordinates and $3n$ velocity coordinates.

One formulation of classical mechanics is Hamiltonian mechanics \cite[Part 3]{Arnold90} which was originated by Hamilton in the early nineteenth century. At the centre of Hamiltonian Mechanics\index{Hamiltonian mechanics} is a phase space. For a system with $n$ degrees of freedom the phase space\index{phase space} is a $2n$ dimensional manifold consisting of all the possible position and velocity coordinates. Throughout this thesis we use the simplest case of $\Space{R}{2n}$ for phase space.

It is common practice  to notate the coordinates of phase space as \newline
$(q_1, \cdots , q_n)$ for the position coordinates and $(p_1 , \cdots , p_n)$ for the velocity co-ordinates --- throughout this thesis we use this notation. Observables in Hamiltonian mechanics are real functions defined on phase space\footnote{Certain conditions are needed on these functions such as differentiability and continuity. For the purposes of this thesis the only conditions we require for our classical observables are that they are differentiable everywhere and can be realised as elements of $\mathcal{S}'(\Space{R}{2n})$, this is discussed in Section \ref{subsect:obsvinpmech}. }, $\Space{R}{2n}$. Some examples of observables are position, momentum, energy, angular momentum.
\begin{Defn}[Poisson bracket] \cite[Sect. 9.5]{Goldstein80}
The Poisson bracket \index{Poisson bracket} of two observables $f,g$ is defined as
\begin{equation} \label{eq:poissonbrackets}
\{ f , g \} = \sum_{i=1}^{n} \left( \Fracpartial{f}{q_i} \Fracpartial{g}{p_i} - \Fracpartial{g}{q_i} \Fracpartial{f}{p_i} \right).
\end{equation}
\end{Defn}
In a system with energy $H$ the time evolution of an arbitrary observable $f$ is defined by Hamilton's equation \cite[Sect. 9.6]{Goldstein80}
\begin{equation} \label{eq:dynforclassobsv}
\Fracdiffl{f}{t} = \{ f , H \}.
\end{equation}
We now continue our discussion using the theory of vector fields and differential forms on $\Space{R}{2n}$ as described in Appendix \ref{app:vfsanddfs}. On $\Space{R}{2n}$ there exists a two-form, $\omega$, defined as
\begin{equation} \label{eq:sympformonrn}
\omega ( (q,p), (q',p')) = qp'- pq'
\end{equation}
where $(q,p)$ and $(q',p')$ are elements of $\Space{R}{2n}$. $\omega$ is referred to as the symplectic form on $\Space{R}{2n}$.
 For a classical observable $f$ the Hamiltonian vector field $X_f$ is the vector field which satisfies
\begin{equation} \nonumber
df(Y) = \omega (Y,X_f)
\end{equation}
for any vector field $Y$. By a simple calculation it can be seen that the Hamiltonian vector field $X_f$ will be of the form
\begin{equation} \nonumber
X_f = \sum_{i=1}^n \left( \Fracpartial{f}{p_i} \Partial{q_i} - \Fracpartial{f}{q_i} \Partial{p_i} \right).
\end{equation}
Another straightforward calculation will verify that
\begin{equation} \nonumber
\{ f, g \} = \omega (X_f,X_g)
\end{equation}
for any two observables $f,g$. Hamiltonian mechanics is often extended by using a symplectic manifold other than $\Space{R}{2n}$ as phase space. We do not describe this here but it may be found in \cite{Arnold90,Jose98,MarsdenRatiu99}.

Empirically an important point in classical mechanics is if you know all the forces acting on a particle, and its initial position and velocity, then you will know exactly its position and velocity at any time after this. This is called determinism \cite[Sect. 2.7]{Liboff80}.

\section{The Birth and Development of Quantum Mechanics}

\indent

In this section we give a brief overview of the origins and developments of quantum mechanics. This will give some motivation for why we are developing the mathematical model which is discussed in this thesis.

At the start of the twentieth century a number of experiments took place which showed the classical theory (see Section \ref{sect:classmech}) was insufficient for describing nature on the macroscopic level. This led to the gradual development of a new theory which would come to be known as quantum theory. We now give a very brief outline of these developments --- for a complete description see \cite{BransdenJoachin00,Liboff80,Messiah61}.

Quantum theory is widely regarded to have originated at the turn of the century when scientists started investigating blackbody radiation\index{blackbody radiation} \cite[Sect. 2.2]{Liboff80} \cite[Sect. 1.1]{BransdenJoachin00}. A blackbody is a hypothetical body which absorbs all the radiation which falls on to it. It was shown that classical mechanics was incapable of explaining the spectrum of the radiation emitted by a blackbody when heated. The problem was in finding a spectral distribution function $\rho ( \lambda , T)$ which gave the energy density  at temperature $T$ of the radiation with wavelength $\lambda$. Lord Rayleigh and J. Jeans derived a spectral distribution using thermodynamic reasoning but this failed to model the situation for small $\lambda$. These inconsistencies at small wavelengths were called the ``ultraviolet catastrophe". A solution to this problem was proposed by Max Planck in December 1900. Planck proposed that the energy of an oscillator cannot take arbitrary values between zero and infinity, instead it can only take a discrete set of values. As a consequence of this he derived a spectral distribution function which satisfied all the requirements. Also this led to the introduction of a new constant, $h=6.62618 \times 10^{-34} J s$, called Planck's constant\footnote{Planck's constant is also known as the fundamental quantum of action.}. The physical dimensions of $h$ are those of action\footnote{For a discussion of dimensionality in $p$-mechanics see \cite{Kisil02.1}.}. Planck proposed that the energy of radiation with frequency $\nu$ could only exist in multiples of $h\nu$ ($h\nu$ is called a quantum of radiation or a photon).

The next step in the development of quantum theory was Einstein's work on the photoelectric effect. For conciseness we do not go into a description of the experimental observations concerning the photoelectric effect, but we refer the reader to \cite[Sect. 1.4]{Messiah61}, \cite[Sect. 1.2]{BransdenJoachin00} for these. Instead we state the hypotheses which Einstein used to explain the photoelectric effect. Einstein proposed that light of frequency $\nu$ must come in discrete corpuscles of energy $h\nu$.  Further evidence of the corpuscular nature of electromagnetic waves was demonstrated by A.H. Compton's observations about the scattering of x-rays. This is known as the Compton effect and a description can be found in \cite{Messiah61}.

These experiments contradicted the classical assumption that light acted like waves. However interference and diffraction phenomena showed that in some circumstances light must act like waves. The work of De Broglie, Davisson and Germer showed that electrons must act like both waves and particles. This led to the necessity of a wave-particle duality theory.

Further experimental evidence which supported quantum theory was found when Niels Bohr studied the structure of atoms. It had been observed that when light is emitted from hydrogen only a certain discrete set of frequencies occur. Bohr explained this by claiming hydrogen could only exist in a discrete set of energy states, $\{ E_n : n \in \Nat \}$. Along with this he claimed that to move from an energy state $E_n$ to another energy state $E_m$ a photon with frequency $\nu$ would be emitted, where $h \nu = E_n - E_m$. This was yet more evidence that certain physical constants took discrete values. The discrete spectrum of the hydrogen atom is described in the context of $p$-mechanics in Section \ref{sect:klaudrep}.

All these new phenomena were only noticeable on the microscopic level. On a scale in which Planck's constant, $h$, was negligible, all these new phenomena did not arise --- that is physics obeyed the laws of classical mechanics. This meant that quantum mechanics must agree with classical mechanics as $h$ tends towards $0$. This led to the development of what is now known as old quantum theory \cite[Sect. 1.15]{Messiah61}. The mathematical framework of old quantum theory was successful in deriving the energy levels of the hydrogen atom. However for more complex problems it proved insufficient. It took the work of Dirac, Schr\"odinger, Heisenberg and many others to develop a mathematical and physical theory which resolved many --- but not all --- of these new problems.

In 1925 Heisenberg explained these phenomena using his uncertainty principle. The uncertainty principle\index{uncertainty principle} stated that
if you knew exactly the position of a particle then its momentum is completely unknown and vice versa. More generally the principle states that the more we know about the position of a particle the less we know about its momentum and vice versa. The exact formulation of the uncertainty principle is
\begin{equation} \label{eq:uncenprin}
\triangle x \triangle p \geq h
\end{equation}
where $\triangle x$ and $\triangle p$ are the uncertainty in the position and momentum respectively of a particle. Below in equation (\ref{eq:uncertainty}) we give a mathematical definition of the uncertainty of an observable. This coincides with the wave-particle duality since if, for example an electron is acting like a particle, then we have a good idea of its position, but not a good idea of its momentum. If an electron is acting like a wave, then we have a good idea of its momentum, but not its position.

The mathematical framework that was developed contained both states and observables as in classical mechanics (see Section \ref{sect:classmech}). In a lot of the literature quantum mechanics is described as starting from several axioms or postulates. We describe the postulates which are given in \cite{Liboff80}. The first postulate of quantum mechanics is that the state of a quantum mechanical system is given by a wave function $\psi$ which is the element of a Hilbert space\footnote{For full mathematical rigour wave functions must be elements of a rigged Hilbert space see Section \ref{sect:rhs}. Another approach which can be taken is to use densely defined unbounded operators as described in \cite[Chap. 8]{ReedSimon80}.}. In Section \ref{subsect:whatarestates} we show that in some cases it is easier to define states directly as functionals on the set of observables rather than as elements of a Hilbert space. The second postulate is that observables are represented by operators on this set of wave functions (observables in $p$-mechanics are discussed in Section \ref{subsect:obsvinpmech}). The expectation value of an observable $A$ in state $\psi$ is given by
\begin{equation} \nonumber
\langle A \rangle = \langle A \psi , \psi \rangle.
\end{equation}
The uncertainty of an observable $A$ in a state $\psi$ is defined as
\begin{equation} \label{eq:uncertainty}
\triangle A = \sqrt{\langle A^2 \rangle - \langle A \rangle^2}.
\end{equation}
The third postulate of quantum mechanics is that if an observable, $A$, in a system is measured to be $a$ then the system will be in state $\psi_a$, where $\psi_a$ is the eigenfunction of $A$ with eigenvalue $a$. Eigenvalues and eigenfunctions in $p$-mechanics are discussed in Section \ref{sect:eigenvalueeigenfunction}. The fourth postulate of quantum mechanics is on the time evolution of states and observables. If a system is governed by a Hamiltonian $H$, which is an operator on the Hilbert space, then the time evolution of a state, $\psi$, is governed by the Schr\"odinger equation
\begin{equation} \label{eq:schrodingerequation}
\Fracdiffl{\psi}{t} = \frac{2\pi}{ih} H \psi.
\end{equation}
The time evolution of an observable, $A$, is governed by the Heisenberg equation
\begin{equation} \label{eq:heisenbergequation}
\Fracdiffl{A}{t} = \frac{2\pi}{ih} [ A, H] = \frac{2\pi}{ih} (AH-HA)
\end{equation}
the right hand side of this equation is called the quantum commutator. Time evolution in $p$-mechanics is defined in Sections \ref{sect:timeevolofobsv} and \ref{subsect:timeevolofstates}.

\section{Quantisation}

\indent

Quantisation\index{quantisation} \cite[Sect. 1.1]{Folland89} is the problem of deriving the mathematical framework of a quantum mechanical system from the mathematical framework of the corresponding classical mechanical system. A method of quantisation must contain a map $\mathcal{Q}$ from the set of classical observables to the set of quantum observables with the following properties:
\begin{itemize}
\item $\mathcal{Q} (f+g) = \mathcal{Q}(f) + \qmap (g)$
\item $\qmap (\lambda f) = \lambda \qmap (f)$
\item $\qmap ( \{ f,g \} ) = \frac{2\pi}{ih} [ \qmap (f) , \qmap (g) ]$
\item $\qmap (1) = I_d$
\item $\qmap (q_i)$ and $\qmap (p_i)$ are represented irreducibly on the Hilbert space in question.
\end{itemize}
The Groenwold-von Hove ``no-go'' theorem \cite{Gotay80} \cite[Thm. 4.59]{Folland89} proves that it is impossible to do this if we want to quantise every single classical observable. Instead the best we can hope for is to quantise a subset of the set of classical mechanical observables. Various methods with varying levels of success have been established since the start of quantum mechanics to obtain a clear method of quantisation. Geometric quantisation \cite{Woodhouse92,Sniatycki80}, deformation quantisation \cite{Fedosov02,Zachos02a}, Berezin quantisation \cite{Berezin72,Berezin75}, Weyl quantisation \cite[Chap. 2]{Folland89} are some of the more famous methods of quantisation. In \cite{KisilBrodlie02,Kisil02.1} relations between $p$-mechanics and these various methods of quantisation are realised.

\chapter{$p$-Mechanics and the Heisenberg Group} \label{chap:pmechandhn}

\indent

This chapter has two purposes. The first purpose is to introduce the Heisenberg group and its representation theory; this is the content of Section \ref{sect:hnanditsreps}. The second purpose is to introduce the ideas behind the theory of $p$-mechanics; this is contained in Section \ref{sect:pmechintro}.

\section{The Heisenberg Group and its Representations} \label{sect:hnanditsreps}

\indent

In this section we give some preliminary results on the Heisenberg group and its representation theory. Many results in this section are similar to those readily available in the literature. However throughout this section we present the results in a form which will make them accessible for the rest of this thesis. The main purpose of this section is to set up the machinery which will help us prove the main results of the thesis.

In Subsection \ref{subsect:hnanditsliealg} the Heisenberg group and its Lie algebra are introduced along with concepts such as Haar measure and convolution. Kirrilov's method of orbits is applied to the Heisenberg group in Subsection \ref{subsect:methodoforbits}. This allows us to obtain irreducible representations of the Heisenberg group using the theory of induced representations --- this is explained in \ref{sect:indreponhn}. In doing so we define a new Hilbert space $\fock$ and a unitary irreducible representation of the Heisenberg group on this space. In Subsection \ref{sect:relbetweenfockandotherhilb} we exhibit relations between $\ltworn$ and our new Hilbert space, $\fock$. We show that they can be mapped into each other using an integration kernel which intertwines the Schr\"odinger representation with our new representation. The sole purpose of Subsection \ref{subsect:svn} is to describe the Stone-von Neumann theorem. The Stone-von Neumann theorem about the unitary irreducible representations of the Heisenberg group motivates the whole of $p$-mechanics and the majority of the work in this thesis. In Subsection \ref{sect:csinfock} we introduce a system of square integrable coherent states for $\fock$ -- this allows us to calculate a reproducing kernel for $\fock$.

\subsection{The Heisenberg Group and its Lie Algebra} \label{subsect:hnanditsliealg}

\indent

At the heart of this thesis is the Heisenberg group (\cite{Folland89}, \cite{Taylor86}).
\begin{Defn}
The Heisenberg group\index{Heisenberg group} (denoted $\Space{H}{n}$\index{$\Space{H}{n}$}) is the set of all triples in $\Space{R}{} \times \Space{R}{n} \times \Space{R}{n}$ under the law of multiplication
\begin{equation} \label{heismult}
(s,x,y) \cdot (s',x',y') = \left( s+s' + \frac{1}{2} (x \cdot y' -x' \cdot y),x+x',y+y' \right) .
\end{equation}
\end{Defn}
The non-commutative convolution\index{convolution} of two functions $B_1,B_2 \in \loneh$ is defined as
\begin{equation} \nonumber
(B_1 * B_2)(g) = \int_{\Heisn} B_1 (h) B_2 (h^{-1} g) dh = \int_{\Heisn} B_1 (g h^{-1}) B_2 (h) dh,
\end{equation}
where $dh$ is Haar Measure on $\Heisn$, which is just Lebesgue measure on $\Space{R}{2n+1}$, $ds \, dx \, dy$. Using the left regular representation\index{left regular representation} $\lambda_l$ of $\Heisn$
\begin{equation} \label{eq:leftregrep}
\lambda_l (g') f(g) = f(g'^{-1} g )
\end{equation}
we can write the convolution of two functions in $\loneh$ as
\begin{equation} \label{eq:altconvol}
B_1*B_2 (g) = \int_{\Heisn} B_1 (h) \lambda_l (h) \, dh \, B_2 (g).
\end{equation}
The convolution of two distributions is defined in equation (\ref{eq:convoftwodistsforgeng}) of Appendix \ref{app:distributions}.
The Lie Algebra $\heisn$\index{$\heisn$} can be realised by the left invariant vector fields\index{left invariant vector fields!Heisenberg group}
\begin{equation} \label{eq:leftinvvecfields}
\mathfrak{S}^l = \Partial{s},  \hspace{1cm} \mathfrak{X}^l_j = \Partial{x_j} - \frac{y_j}{2} \Partial{s},\hspace{1cm}  \mathfrak{Y}^l_j = \Partial{y_j} + \frac{x_j}{2} \Partial{s},
\end{equation}
with the Heisenberg commutator relations
\begin{equation} \label{eq:commrelforlinvvf}
[ \mathfrak{X}^l_i , \mathfrak{Y}^l_j ] = \delta_{ij} \mathfrak{S}^l \hspace{0.5cm} ; \hspace{1cm} [\mathfrak{S}^l ,\mathfrak{X}^l_i] = [ \mathfrak{S}^l , \mathfrak{Y}^l_i] = 0.
\end{equation}
The Lie algebra $\heisn$ can be realised as $\Space{R}{2n+1}$ (the vector $(r,a,b)$ corresponds to the vector field  $r\mathfrak{S} + \sum_{j=1}^n a_j \mathfrak{X}_j + \sum_{j=1}^n b_j \mathfrak{Y}_j$). In this realisation the exponential map from $\heisn$ to $\Heisn$ is just the identity map on $\Space{R}{2n+1}$. The dual space to the Lie Algebra $\heisn^*$\index{$\heisn^*$} is spanned by the left invariant first order differential forms $dS,dX,dY$. $\heisn^*$ can also be realised as $\Space{R}{2n+1}$ (the vector $(h,q,p)$ corresponds to the differential form $hdS + \sum_{j=1}^n [ q_j dX_j + p_j dY_j ]$). In this thesis the right invariant vector fields for the Heisenberg group will also be of use; they are
\begin{equation} \label{eq:rightinvvecfields}
\mathfrak{S}^r = - \Partial{s},  \hspace{1cm} \mathfrak{X}_j^r = \Partial{x_j} + \frac{y_j}{2} \Partial{s},\hspace{1cm}  \mathfrak{Y}_j^r = \Partial{y_j} - \frac{x_j}{2} \Partial{s},
\end{equation}
with the commutator relations
\begin{equation} \label{eq:commrelforrinvvf}
[ \mathfrak{X}^r_i , \mathfrak{Y}^r_j ] = \delta_{i,j} \mathfrak{S}^r.
\end{equation}
One of the principal ways of transferring between $\ltwoh$ and $\ltwohnstar$ is by the Fourier transform on $\Heisn$ \cite[Eq. 2.3.4]{Kirillov99}
\begin{equation} \label{eq:liegpfourier}
\hat{\phi}(F) = \int_{\heisn} \phi (\exp X) e^{-2 \pi i \langle X,F \rangle } \, dX
\end{equation}
where $F \in \heisn^*$, $X \in \heisn$, $\phi \in \ltwoh$ and $\hat{\phi} \in \ltwohnstar$. This has the simple form
\begin{equation} \nonumber
\hat{\phi}(h,q,p) = \int_{\Space{R}{2n+1}} \phi (s,x,y) e^{-2\pi i (hs+q.x+p.y)} \, ds \, dx \, dy
\end{equation}
which is just the usual Fourier transform on $\Space{R}{2n+1}$.
The most common representation of the Heisenberg group is the Schr\"odinger representation\index{Schr\"odinger representation}. The Schr\"odinger representation \cite[Sect. 1.3]{Folland89} for $h>0$ is defined on $\ltworn$ as
\begin{equation} \label{eq:schrorep}
\left( \rho_h^S (s,x,y) \psi \right) (\xi) = e^{-2\pi ihs + 2\pi ix\xi + \pi ihxy} \psi(\xi+hy).
\end{equation}
It has been shown that this representation is unitary \cite[Sect. 1.3]{Folland89} and irreducible \cite[Prop. 1.43]{Folland89}. In this thesis we only briefly look at this infinite dimensional representation; instead we concentrate on other forms of the infinite dimensional representation and also the often neglected family of one dimensional representations.

\subsection{The Method of Orbits Applied to the Heisenberg group} \label{subsect:methodoforbits}

\indent

We now derive another infinite dimensional representation of the Heisenberg group which is unitarily equivalent to the Schr\"odinger representation. Before we can derive this representation we need to describe how Kirillov's method of orbits\index{method of orbits} can be applied to the Heisenberg group. For a discussion of the method of orbits, see \cite{Kirillov99} or \cite[Chap 15]{Kirillov76}; its relation to $p$-mechanics is described in \cite{Kisil02.1}. The method of orbits is at the centre of geometric quantisation \cite{Sniatycki80,Woodhouse92} and plays an important role in the representation theory of Lie groups \cite[Chap. 7]{Kirillov94}.

A Lie group can act on itself by conjugation (that is $g \in G$ acts on $x \in G$ by $x \mapsto g^{-1} x g$ ). For the Heisenberg group the action of $(s,x,y)$ by conjugation on $(s',x',y')$ is
\begin{equation} \nonumber
(s',x',y') \mapsto (s'+x'y-xy',x',y').
\end{equation}
This action clearly preserves the identity and therefore we can take the derivative of this at the identity (see equation (\ref{eq:diffofmapintermsofdiff})). This gives us a representation of $\Space{H}{n}$ on $\halg{n}$.
\begin{equation} \nonumber
Ad_{(s,x,y)} (r,a,b) = (r+ay-bx,a,b).
\end{equation}
To get the coadjoint representation, $Ad^*$, we take the map $Ad$ over to the dual space $\halg{n^*}$ in the natural way:
\begin{equation} \label{adstar}
Ad^{*}_{(s,x,y)}(h,q,p) = (h,q+hy,p-hx).
\end{equation}
From this it can be seen that the orbits of $Ad^*$ are all of the form
\begin{itemize}
\item $\horb = \{ (h,q,p):q,p \in \Space{R}{n} \}$ for a particular $h \in \Space{R}{} \setminus \{ 0 \}$ or
\item the singleton sets $\mathcal{O}_{(q,p)} = \{ (0,q,p) \}$ where $q,p \in \Space{R}{n}$.
\end{itemize}
It is clear that $\horb$ is isomorphic to $\Space{R}{2n}$ which is the phase space of a system with $n$ degrees of freedom. A natural symplectic form can be found on these orbits \cite[Chap. 15]{Kirillov76} -- this is the starting point of geometric quantisation. On the contrary $p$-mechanics \cite[Eq. 2.17]{Kisil02.1} utilises the fact that the union of all the $\mathcal{O}_{(q,p)}$ orbits is the classical phase space $\Space{R}{2n}$. Note here that $(s,x,y) \in \Heisn$ and $(h,q,p) \in \heisn^* $ --- this choice of letters will be used throughout this thesis.

\subsection{Induced Representations of the Heisenberg \newline Group} \label{sect:indreponhn}

\indent

To get a new form of an infinite dimensional representation for the Heisenberg group we use the theory of induced representations\index{induced representation} (see Appendix \ref{app:indrep}). Before we can embark on generating this representation we need to give the definition of a subordinate subalgebra\index{subordinate subalgebra}.
\begin{Defn}[Subordinate Subalgebra]
If $\mathfrak{h}$ is the subalgebra of a Lie algebra $\mathfrak{g}$ then $\mathfrak{h}$ is subordinate to a functional $f \in \mathfrak{g}^*$ if and only if
\begin{equation} \nonumber
\langle f , [x,y] \rangle = 0 \hspace{2cm} \forall x,y \in \mathfrak{h}.
\end{equation}
\end{Defn}
In the case of the Heisenberg group for any $f \in \horb$ \begin{equation} \nonumber
\{(r,0,0):r\in \Space{R}{} \} \subset \heisn
\end{equation}
is the only nontrivial subordinate subalgebra. The exponential of this subalgebra is $Z=\{(h,0,0):h\in \Space{R}{} \} \subset \Heisn$ which is the centre of $\Heisn$.

A one dimensional representation of $Z$ on $\Space{C}{}$ is given by $\rho^Z_h ((s,0,0)) = e^{2\pi ihs}$. It is shown in \cite[Thm. 7.2]{Kirillov94} that irreducible representations are given by the induced representations of $\rho^Z_h$. We now construct the two equivalent forms of this representation using the method outlined in Appendix \ref{app:indrep}. The space $L( \Heisn, Z, \rho_h^Z )$ is the set of measurable functions on $\Heisn$ such that $F(s+s',x,y)=e^{2\pi ihs'} F(s,x,y)$. The representation by left shifts $\lambda_l$ is a representation on this space. The space $L^2 ( \Heisn, Z, \rho_h^Z )$ is the subset of $L( \Heisn, Z, \rho_h^Z )$ containing the functions which are square integrable with respect to the inner product
\begin{equation} \label{eq:iponindrepspace}
\langle F_1 , F_2 \rangle_{L^2 \left( \Heisn, Z, \rho_h^Z \right) } = \int_{\Heisn} F_1  \overline{F_2} \, dx \, dy.
\end{equation}
The measure in question here is of the form $\alpha(g) dg$ where $dg$ is Haar measure and $\alpha(g)$ is the function $\frac{e^{-s^2}}{\sqrt{\pi}}$.

For the second realisation of this representation (that is the one described in Theorem \ref{thm:secondindrep}) we let $X$ denote the Homogeneous space $G/Z$. Any coset $x \in X$ will be of the form $\{ (s,x,y):s\in \Space{R}{} \}$ for a particular $(x,y)\in \Space{R}{2n}$, so $X$ can be associated with $\Space{R}{2n}$. Under the association $(x,y) \leftrightarrow \{ (s,x,y):s\in \Space{R}{} \}$ the measure on $X$ is Lebesgue measure $dx\, dy$ on $\Space{R}{2n}$. $L^2 (X)$ is the space of square integrable functions with respect to this measure. The projection $\sigma:X \rightarrow \Heisn$ is given by
\begin{equation} \label{eq:projtohomsp}
\sigma ((x,y))=(0,x,y).
\end{equation}
For this choice of projection Lemma \ref{eq:decompforindrep} takes the simple form that every $(s,x,y) \in \Heisn$ can be written in the form $(0,x,y)(s,0,0)= \sigma(x,y) (s,0,0)$. By the construction in Appendix \ref{app:indrep} the representation $\phi$ on $L^2(X)$ for $F \in L^2 (X)$ is
\begin{eqnarray} \nonumber
\phi ((s',x',y')) F (x,y) &=& \lambda_l ((s',x',y')) f(0,x,y) \\ \nonumber
&=& f((s',x',y')^{-1} (0,x,y)) \\ \nonumber
&=& f \left( -s'+\frac{1}{2} (xy'-x'y),x-x',y-y' \right) \\ \nonumber
&=& f \left( \left( 0 ,x-x',y-y' \right) \left( -s'+\frac{1}{2} (xy'-x'y),0,0 \right) \right) \\ \nonumber
&=& e^{2\pi ih \left( -s'+\frac{1}{2} (xy'-x'y) \right) } f(0,x-x',y-y') \\ \label{eq:lastofindreponcoset}
&=& e^{2\pi ih \left( -s'+\frac{1}{2} (xy'-x'y) \right) } F(x-x',y-y')
\end{eqnarray}

If we intertwine the Fourier transform with this representation we get a representation on the orbit $L^2 (\uorb )$. The Fourier transform of (\ref{eq:lastofindreponcoset}) is
\begin{equation} \label{eq:startofindreponoh}
\int_{\Space{R}{2n}} e^{2\pi ih \left( -s'+\frac{1}{2} (xy'-x'y) \right) } F(x-x',y-y') e^{-2\pi i (qx+py)} \, dx \, dy.
\end{equation}
By the change of variable $a=x-x'$, $b=y-y'$ (\ref{eq:startofindreponoh}) becomes
\begin{eqnarray} \nonumber
\lefteqn{e^{-2\pi ihs'} \int_{\Space{R}{2n}} e^{\pi ih [ (a+x')y' - (b+y')x']} F(a,b) e^{-2\pi i [ q(a+x') +p(b+y')]} \, da \, db }\\ \nonumber
&& = e^{-2\pi ihs'} e^{-2\pi i (qx' + py')} \int_{\Space{R}{2n}} F(a,b) e^{-2\pi i \left[ a \left( q-\frac{h}{2} y' \right) +b \left(  p + \frac{h}{2} x' \right) \right] } \, da \, db \\ \label{eq:leftshiftsandtherhohrep}
&& = e^{-2\pi i (hs' +qx'+py') } \hat{F} \left( q-\frac{h}{2} y' , p + \frac{h}{2} x' \right).
\end{eqnarray}
Throughout this thesis we denote this representation by $\rho_h$\index{$\rho_h$}. $\rho_h$ can be written neatly as
\begin{equation} \label{eq:infdimrep}
\rho_h (s,x,y): f_h (q,p) \mapsto e^{-2 \pi i (hs + qx +py)} f_h \left( q-\frac{h}{2} y, p+\frac{h}{2} x \right).
\end{equation}
In \cite{Kisil02.1} it is shown that this representation is reducible on $L^2 (\horb)$. To get an irreducible representation we need to reduce the size of the Hilbert space it acts upon. To do this we use the idea of a polarisation from geometric quantisation. We define the operator $D_h^j$ on $L^2 (\uorb )$ by
\begin{eqnarray} \label{eq:ftwoohpolarz}
D_h^j &=& -X^r_{\rho_h} + i Y^r_{\rho_h} \\ \nonumber
&=& \frac{h}{2} \left( \Partial{p_j} +  i \Partial{q_j} \right) + 2 \pi (p_j +i q_j)I.
\end{eqnarray}
$\fock$\index{$\fock$} is the subspace of $L^2 (\uorb )$ defined by
\begin{eqnarray} \nonumber
\fock &=& \{ f_h (q,p) \in L^2 (\uorb)  : \textrm{$f$ is differentiable and} \\ \nonumber
&& \hspace{4cm} D^j_h f_h =0 ,\textrm{ for } 1 \leq j \leq n \}.
\end{eqnarray}
In Section \ref{sect:relbetweenfockandotherhilb} we show that $\rho_h$ is unitary and irreducible on $\fock$. The inner product on $\fock$ is given by
\begin{equation} \label{eq:fockip}
\langle v_1, v_2 \rangle_{\fock} = \left( \frac{4}{ h } \right)^{n} \int_{\Space{R}{2n}} v_1 (q,p) \overline{v_2 (q,p)} \, dq \, dp.
\end{equation}
$\fock$ is a reproducing kernel Hilbert space; we postpone proving this until Subsection \ref{sect:csinfock}.

One advantage of using the $\rho_h$ representation (\ref{eq:infdimrep}) over the Schr\"odinger representation (\ref{eq:schrorep}) is made apparent when taking the limit as $h \rightarrow 0$. If we take the direct integral \cite[Chap. 6 Sect. 1.5]{Kirillov94} of all the one dimensional representations $\rho_{(q,p)}$ we clearly get the representation $\rho_0$. To prove a similar result for the Schr\"odinger representation requires a lengthy argument \cite[Example 7.11]{Kirillov94}. More advantages of the $\rho_h$ representation are described in \cite{Kisil02.1}. Furthermore throughout this thesis there will be many situations in which the $\rho_h$ representation is shown to be more convenient than the Schr\"odinger representation.

\subsection{Relationships Between $\fock$ and Other Hilbert Spaces} \label{sect:relbetweenfockandotherhilb}

\indent

$\fock$ is closely related to the Fock-Segal-Bargmann space\index{Fock-Segal-Bargmann space} \cite{Bargmann61,Folland89,Taylor86}.
\begin{Defn}
\cite{Folland89,Taylor86,Howe80.1} The Fock-Segal-Bargmann space, $SB_h^2 ( \Space{C}{n} )$, with parameter $h>0$ and dimension $n \in \Nat$ consists of all functions on $\Space{C}{n}$ which are analytic everywhere and square integrable with respect to the measure $e^{-|z|^2 / h}$. The inner product on $SB_h^2 (\Space{C}{n})$ is
\begin{equation} \nonumber
\langle f,g \rangle_{SB_h^2} = \int_{\Space{C}{n}} f \bar{g} e^{-|z|^2 / h}dz.
\end{equation}
\end{Defn}
It is shown in \cite{Kisil02.1} that a function $f_h (q,p)$ is in $\fock$ if and only if $z \mapsto f_h (Im(z),Re(z)) e^{|z|^2/h}$, $z=p+iq$ is in $SB_h^2 (\Space{C}{n})$.

We now show how we can map $\ltworn$ into $\fock$ by the integration kernel
\begin{equation} \nonumber
K_I(q,p,\xi) = e^{\frac{4\pi i}{h} (p\xi + qp)} e^{-\frac{\pi}{h}(\xi + 2q)^2} .
\end{equation}

\begin{Lemma} \label{lem:kernelsatisfiesthepolarz}
$K_I$ satisfies the polarization from equation (\ref{eq:ftwoohpolarz}) for any $\xi$.
\end{Lemma}
\begin{proof}
By direct calculations
\begin{eqnarray} \nonumber
\Fracpartial{K_I}{q_j} &=& \left( \frac{4\pi i}{h} p_j - \frac{4\pi}{h} (\xi_j + 2q_j) \right) K_I \\ \nonumber
\Fracpartial{K_I}{p_j} &=& \left( \frac{4\pi i}{h} \xi_j + \frac{4\pi i}{h}q_j \right) K_I.
\end{eqnarray}
So
\begin{eqnarray} \nonumber
D_j^h K_I &=& \left( \frac{h}{2}  \left(\Partial{p_j} + i \Partial{q_j} \right) + 2\pi (p_j+iq_j) \right) K_I \\ \nonumber
&=& \left( \frac{h}{2} \left[ \frac{4\pi i}{h} \xi_j + \frac{4\pi i}{h} q_j - \frac{4\pi }{h}p_j - \frac{4\pi i}{h} \xi_j - \frac{8\pi i}{h}q_j \right] +2\pi (p_j+iq_j) \right) K_I \\ \nonumber
&=& (-2\pi p_j - 2\pi i q_j + 2\pi p_j + 2\pi i q_j) K_I \\ \nonumber
&=& 0.
\end{eqnarray}
\end{proof}
The map $\mathcal{T}$ from $\ltworn$ to a subset of the set of functions on $\Space{R}{2n}$  is\footnote{We will see later that this subset is precisely $\fock$.} defined by
\begin{equation} \label{eq:defoft}
(\mathcal{T} \psi) (q,p) = \left( \frac{2}{h} \right)^{n/4} \int K_I (q,p,\xi) \psi(\xi) \, d\xi
\end{equation}
where $\psi$ is any element of $\ltworn$.
\begin{Lemma} \label{lem:fisalsobyrepandgs}
An equivalent form of $\mathcal{T}$ is given by the wavelet transform
\begin{equation} \nonumber
\mathcal{T} (\psi) (q,p) \mapsto \left( \frac{2}{h} \right)^{n/4} \left\langle \rho_h^S \left( 0, \frac{2}{h} p , -\frac{2}{h} q \right) \psi , \phi_0 \right\rangle_{\ltworn} ,
\end{equation}
where $\rho_h^S$ is the Schr\"odinger representation of the Heisenberg group as defined in equation (\ref{eq:schrorep}) while $\phi_0= e^{-\frac{\pi}{h} \xi^2}$ is the ground state of the harmonic oscillator (see Section \ref{sect:harmosc}) in $\ltworn$.
\end{Lemma}
\begin{proof}
This follows by a direct calculation
\begin{eqnarray} \nonumber
\lefteqn{\left( \frac{2}{h} \right)^{n/4} \left\langle \rho_h^S \left( 0, \frac{2}{h} p , -\frac{2}{h} q \right) \psi , \phi_0 \right\rangle_{\ltworn} } \\ \nonumber
&=& \left( \frac{2}{h} \right)^{n/4} \int_{\Space{R}{n}} e^{-\frac{4\pi i}{h} qp} e^{\frac{4\pi i}{h} p \xi} \psi (\xi -2q) e^{-\frac{\pi}{h} \xi^2} \, d\xi \\ \nonumber
&=& \left( \frac{2}{h} \right)^{n/4} \int_{\Space{R}{n}} e^{\frac{4\pi i}{h} [p(\xi+2q) -qp]} \psi (\xi) e^{-\frac{\pi}{h} (\xi + 2q)^2} \, d\xi \\ \nonumber
&=& \left( \frac{2}{h} \right)^{n/4} \int_{\Space{R}{n}} K(q,p,\xi) \psi(\xi) \, d\xi.
\end{eqnarray}
\end{proof}
\begin{Thm} \label{thm:fockandltwointertwine}
The map $\mathcal{T}$ intertwines the representations $\rho_h^S$ and $\rho_h$, that is
\begin{equation} \nonumber
\mathcal{T} \rho_h^S = \rho_h \mathcal{T}.
\end{equation}
\end{Thm}
\begin{proof}
By Lemma \ref{lem:fisalsobyrepandgs} for any $\psi \in \ltworn$
\begin{eqnarray} \nonumber
\lefteqn{(\mathcal{T} \rho_h^S (s,x,y) \psi)} \\ \nonumber
&=& \left( \frac{2}{h} \right)^{n/4} \left\langle \rho_h^S \left( 0 , \frac{2}{h} p , -\frac{2}{h}q \right) \rho_h^S (s,x,y) \psi , \phi_0 \right\rangle_{\ltworn} \\ \nonumber
&=& \left( \frac{2}{h} \right)^{n/4} \left\langle \rho_h^S \left( s+ \frac{1}{h}(yp+xq), x + \frac{2}{h}p , y - \frac{2}{h}q \right) \psi , \phi_0 \right\rangle_{\ltworn} \\ \nonumber
&=& \left( \frac{2}{h} \right)^{n/4} \left\langle \rho_h^S \left( s+\frac{1}{h} (yp + xq),0,0 \right) \rho_h^S \left( 0,x+ \frac{2}{h} p , y - \frac{2}{h} q \right) \psi , \phi_0 \right\rangle_{\ltworn} \\ \nonumber
&=& \left( \frac{2}{h} \right)^{n/4} e^{-2\pi ihs} e^{-2\pi i (qx+py)} \\ \nonumber
&& \hspace{2cm} \times \left\langle \rho_h^S \left( 0, \frac{2}{h} \left( p+\frac{2}{h} x \right) , -\frac{2}{h} \left( q - \frac{h}{2}y \right) \right) \psi , \phi_0 \right\rangle_{\ltworn} \\ \nonumber
&=& \rho_h \mathcal{T} \psi.
\end{eqnarray}
\end{proof}
\begin{Thm} \label{thm:tisunitary}
$\mathcal{T}$ is a unitary operator from $\ltworn$ to $\fock$.
\end{Thm}
\begin{proof}
We need to show
\begin{equation} \label{eq:whatweneedtosatisfyforunitary}
\langle \mathcal{T} \psi_1 , \mathcal{T} \psi_2 \rangle_{\fock} = \langle \psi_1 , \psi_2 \rangle_{\ltworn}
\end{equation}
for any $\psi_1, \psi_2 \in \ltworn$. First we state a preliminary result. If we do a change of variable $\xi \mapsto \xi-q$ we get
\begin{eqnarray} \nonumber
\int_{\Space{R}{n}} K_I (q,p,\xi) \psi (\xi) \, d\xi &=&
\int_{\Space{R}{n}} e^{\frac{4\pi i}{h} (p\xi + qp)} e^{-\frac{\pi}{h}(\xi + 2q)^2} \psi (\xi) \, d\xi \\ \nonumber
&=& \int_{\Space{R}{n}} e^{\frac{4\pi i}{h} p \xi} \phi_0 (\xi +q) \psi (\xi - q) \, d\xi.
\end{eqnarray}
By another change of variable $\xi \mapsto \frac{h\xi}{2}$ we get
\begin{equation}
\int_{\Space{R}{n}} K_I (q,p,\xi) \psi (\xi) \, d\xi = \left(\frac{h}{2}\right)^n \int_{\Space{R}{n}} e^{2\pi i p\xi} \phi_0 \left( \frac{h}{2}\xi + q \right) \psi \left(\frac{h}{2}\xi - q \right) \, d\xi.
\end{equation}
This is just the inverse Fourier transform, $\mathcal{F}^{-1}$, of the function \newline $\phi \left( \frac{h}{2} \xi + q \right) \psi \left( \frac{h}{2} \xi - q \right)$, with respect to the $\xi$ variable.  Since the inverse Fourier transform is a unitary operator, the left hand side of (\ref{eq:whatweneedtosatisfyforunitary}) takes the form
\begin{eqnarray} \nonumber
\lefteqn{\left(\frac{2}{h} \right)^{n/2} \left( \frac{h}{2} \right)^{2n} \left( \frac{4}{h} \right)^n } \\ \nonumber
&& \hspace{0.5cm} \times \int_{\Space{R}{2n}} \phi_0 \left( \frac{h}{2}\xi + q \right) \psi_1 \left(\frac{h}{2}\xi - q \right) \overline{\phi_0 \left( \frac{h}{2}\xi + q \right) \psi_2 \left(\frac{h}{2}\xi - q \right)} \, d\xi \, dq \\ \nonumber
&=& (2h)^{n/2} \int_{\Space{R}{2n}} \phi_0 \left( \frac{h}{2}\xi + q \right) \psi_1 \left(\frac{h}{2}\xi - q \right) \overline{\phi_0 \left( \frac{h}{2}\xi + q \right) \psi_2 \left(\frac{h}{2}\xi - q \right)} \, d\xi \, dq.
\end{eqnarray}
By another change of variable $u= \frac{h}{2} \xi -q$ and $v= \frac{h}{2} \xi + q$ this becomes
\begin{eqnarray} \nonumber
&& (2h)^{n/2} \left( \frac{1}{h} \right)^n \int_{\Space{R}{2n}} \phi_0 (v) \psi_1 (u) \overline{\phi_0 (v)} \overline{\psi_2 (u)} \, du \, dv \\ \nonumber
&& \hspace{1.5cm} = \left( \frac{2}{h} \right)^{n/2} \int_{\Space{R}{n}} \psi_1 (u) \overline{\psi_2 (u) } \, du \int_{\Space{R}{n}} \phi_0 (v) \overline{ \phi_0 (v)} \, dv \\ \nonumber
&& \hspace{1.5cm} = \left( \frac{2}{h} \right)^{n/2} \langle \psi_1 , \psi_2 \rangle_{\ltworn} \langle \phi_0 , \phi_0 \rangle_{\ltworn} \\ \nonumber
&& \hspace{1.5cm}= \left( \frac{2}{h} \right)^{n/2} \langle \psi_1 , \psi_2 \rangle_{\ltworn} \left(\frac{h}{2} \right)^{n/2} \\ \nonumber
&& \hspace{1.5cm}= \langle \psi_1 , \psi_2 \rangle_{\ltworn}.
\end{eqnarray}
\end{proof}
The next theorem proves that $\mathcal{T}$ maps functions in $\ltworn$ into functions in $L^2 (\Space{R}{2n})$.
\begin{Thm} \label{thm:tissquareint}
If $\psi \in \ltworn$ then $(\mathcal{T} \psi)(q,p) \in L^2 (\Space{R}{2n})$.
\end{Thm}
\begin{proof}
By a direct calculation
\begin{eqnarray} \nonumber
(\mathcal{T} \psi)(q,p) &=& \left( \frac{2}{h} \right)^{n/4} \left\langle \rho_h^S \left( 0, \frac{2}{h} p , - \frac{2}{h} q \right) \psi , \phi_0 \right\rangle \\ \nonumber
&=& \left( \frac{2}{h} \right)^{n/4} \int_{\Space{R}{n}} e^{-\frac{4\pi i}{h} qp} e^{\frac{4\pi i}{h} p \xi} \psi (\xi -2q) \phi_o (\xi) \, d\xi \\ \label{eq:firstneweqn}
&=& \left( \frac{2}{h} \right)^{n/4} \int_{\Space{R}{n}} e^{\frac{4\pi i}{h} p \xi} \psi (\xi -q) \phi_o (\xi + q) \, d\xi.
\end{eqnarray}
Since $\phi$ and $\psi$ are both elements of $\ltworn$ the function $\phi(\xi) \phi_0(q)$ is in $L^2 (\Space{R}{2n})$. Clearly the function $\psi (\xi -q) \phi_0 (\xi +q)$ is also in $L^2 (\Space{R}{2n})$. Since (\ref{eq:firstneweqn}) is just the Fourier transform of $\psi (\xi -q) \phi_0 (\xi +q)$, $(\mathcal{T} \psi)(q,p)$ must be square integrable.
\end{proof}
By Lemma \ref{lem:kernelsatisfiesthepolarz}, Theorem \ref{thm:tisunitary} and Theorem \ref{thm:tissquareint} we see that $\mathcal{T}$ maps $\ltworn$ into $\fock$ unitarily. We now present the inverse of $\mathcal{T}$.
\begin{Thm} \label{thm:inverseoft}
The map $\mathcal{T}^{-1}$ from $\fock$ to $\ltworn$ given by
\begin{equation} \nonumber
(\mathcal{T}^{-1} f) = \int_{\Space{R}{2n}} f(q',p')e^{-\frac{4\pi i}{h} (p'\xi + q'p')} e^{-\frac{\pi}{h} (\xi + 2q')^2} \, dq' \, dp'.
\end{equation}
for any $f \in \fock$ is the inverse of $\mathcal{T}$.
\end{Thm}
\begin{proof}
Since $\mathcal{T}$ is a unitary operator for any $f \in \fock$ and $\psi \in \ltworn$
\begin{eqnarray} \nonumber
\langle \mathcal{T}^{-1} f, \psi \rangle_{\ltworn} &=& \langle f, \mathcal{T} \psi \rangle_{\fock} \\ \nonumber
&=& \left( \frac{2}{h} \right)^{n/4} \int_{\Space{R}{2n}} f(q', p') \overline{ e^{\frac{4\pi i}{h} (p'\xi + q'p')} e^{-\frac{\pi}{h}(\xi + 2q')^2} \psi (\xi)} \, d\xi\, dq' \, dp' \\ \nonumber
&=& \left( \frac{2}{h} \right)^{n/4} \int_{\Space{R}{2n}} f(q', p') e^{-\frac{4\pi i}{h} (p'\xi + q'p')} e^{-\frac{\pi}{h}(\xi + 2q')^2}  \, dq' \, dp' \, \overline{\psi (\xi)} \, d\xi.
\end{eqnarray}
We can use Fubini's theorem in the above calculation since we are taking the inner product of two square integrable functions.
\end{proof}

\subsection{The Stone-von Neumann Theorem} \label{subsect:svn}

\indent

We now present the crucial theorem which motivates the whole of $p$-mechanics.
\begin{Thm}[The Stone-von Neumann Theorem]\index{Stone-von Neumann Theorem} All unitary irreducible representations of the Heisenberg group, $\Space{H}{n}$, up to unitary equivalence, are either:

(i) of the form $\rho_h$, for $h \neq 0$
\begin{equation} \nonumber
\rho_h (s,x,y): f_h (q,p) \mapsto e^{-2 \pi i (hs + qx +py)} f_h \left( q-\frac{h}{2} y, p+\frac{h}{2} x \right)
\end{equation}
on $\fock$ or
\newline
(ii) for $(q,p) \in \Space{R}{2n}$ the commutative  one-dimensional representations\index{$\rho_{(q,p)}$} on $\Space{C}{} = L^2 (\qporb )$
\begin{equation} \label{eq:1drep}
\rho_{(q,p)} (s,x,y)u = e^{- 2 \pi i(q.x + p.y)} u.
\end{equation}
\end{Thm}
\begin{proof}
We know by Theorems \ref{thm:fockandltwointertwine} and \ref{thm:tisunitary} that $\rho_h$ is unitarily equivalent to $\rho_h^S$ and so the result follows by \cite[Thm. 1.50]{Folland89}.
\end{proof}
The representations $\rho_h$ and $\rho_{(q,p)}$ can be used to represent functions and distributions as outlined in equations
(\ref{eq:repofafctnonaliegp}) and (\ref{eq:repofadistnonaliegp}) respectively.

\subsection{Square Integrable Covariant Coherent States in $\fock$} \label{sect:csinfock}

\indent

In \cite{Kisil02.1} a set of coherent states\index{coherent states!$\fock$} in $\fock$ was introduced. We first give the definition of an overcomplete system of coherent states which suits our purposes..
\begin{Defn} \label{def:overcompcs}
Let $H$ be a Hilbert space and $G$ a group with Haar measure $dg$. A system of vectors $\{ v_g \in H : g \in G \}$ are an overcomplete system of coherent states if they span $H$ and for any $v \in H$
\begin{equation} \label{eq:thefirstdefnofovercomp}
\int_G \langle v, v_g \rangle v_g \, dg = v.
\end{equation}
\end{Defn}
The relation given by (\ref{eq:thefirstdefnofovercomp}) is called the resolution of the identity. In the literature various other constraints are used to define a system of coherent states -- see \cite{AliAntGaz00,GazeauMonceau02,Perelomov86} for some examples of this.

We now show that the set states introduced in \cite{Kisil02.1} are square integrable covariant coherent states. This set of coherent states can be generated using the homogeneous space $X= \Heisn /Z$ (defined in Subsection \ref{sect:indreponhn}) and the projection $\sigma$ (from equation (\ref{eq:projtohomsp})). We begin with the ground state of the harmonic oscillator in $\fock$ (see Chapter \ref{chap:forcedosc})
\begin{equation} \label{eq:gstateforharmoscinfock}
f_{(0,0)} (q,p) = \exp \left( -\frac{2\pi}{h} \left( q^2 + p^2 \right) \right).
\end{equation}
The set of coherent states $f_{(x,y)}(q,p) \in \fock$ generated by applying the $\rho_h$ representation to $f_{(0,0)}(q,p)$ are
\begin{eqnarray} \label{eq:fockcohstates}
f_{(x,y)} (q,p) &=& \left( \rho_h ( \sigma (x,y) ) f_{(0,0)} \right) (q,p) \\ \nonumber
&=& \exp \left( -2\pi i (qx+py) - \frac{2\pi}{h} \left( \left( q - \frac{h}{2} y \right)^2 + \left( p+ \frac{h}{2} x \right)^2 \right) \right).
\end{eqnarray}
From this set of coherent states we get a wavelet transform $\wavtrans : f(q,p) \mapsto \breve{f} (x,y)$
\begin{equation} \nonumber
\breve{f} (x,y) = \langle f , f_{(x,y)} \rangle.
\end{equation}
Let $\wavtrans \left( \fock \right)$ denote the image of $\fock$ under this wavelet transform and define $\phi_{(q,p)}(x,y)$ to be the element of $\wavtrans \left( \fock \right)$ which is equal to $f_{(x,y)} (q,p)$.  We get an inverse wavelet transform $\invwavtrans:\wavtrans(\fock) \rightarrow \fock$ by
\begin{equation} \nonumber
\invwavtrans (\phi) (q,p) = \langle \phi , \phi_{(q,p)} \rangle.
\end{equation}
The inner product in the above equation is given by the invariant measure on $X$ described in Subsection \ref{sect:indreponhn} equation (\ref{eq:iponindrepspace}).

It is shown in \cite{Kisil02.1} that the maps $\wavtrans$ and $\invwavtrans$ are inverses of each other. This implies that the operator
\begin{equation} \label{eq:resofidforfock}
f \mapsto \int_{\Space{R}{2n}} \rho_h (\sigma (x,y) ) f_{(0,0)} \langle f, \rho_h (\sigma (x,y) ) f_{(0,0)} \rangle \, dx \, dy
\end{equation}
is the identity operator on $\fock$. So using the terminology of \cite[Chap. 7]{AliAntGaz00} the representation $\rho_h$ is square integrable $mod(Z,\sigma)$. In the language of Klauder \cite{GazeauMonceau02,GazeauKlauder99,Klauder96} equation (\ref{eq:resofidforfock}) implies that the $f_{(x,y)}$ coherent states satisfy a resolution of unity. Using \cite[Thm. 7.3.1]{AliAntGaz00} we can conclude that
\begin{eqnarray} \nonumber
\lefteqn{ K(q,p,q',p') } \\ \nonumber
&=& \int_{\Space{R}{2n}} f_{(q,p)} (x,y) \overline{f_{(q',p')} (x,y) } \, dx \, dy \\ \nonumber
&=& \int_{\Space{R}{2n}} \exp \left(-2\pi i \left[ x(q-q') +y(p-p') \right] \right) \\ \nonumber
&& \hspace{1cm} \times \exp \left( -\frac{2\pi}{h} \left[ \left( q -\frac{h}{2} y \right)^2 + \left( p + \frac{h}{2} x \right)^2 \right. \right.  \\ \nonumber
&& \left. \left. \hspace{4cm} + \left( q' -\frac{h}{2} y \right)^2 + \left( p' + \frac{h}{2} x \right)^2 \right] \right) \, dx \, dy
\end{eqnarray}
\begin{eqnarray} \nonumber
&=& \exp \left( -\frac{2\pi}{h} (q^2 + p^2 + q'^2 + p'^2) \right) \\ \nonumber
&& \hspace{0.5cm} \times \int_{\Space{R}{2n}} \exp \left( x \left[ -2\pi i (q-q') - 2\pi (p+p') \right] \right) \\ \nonumber
&&  \hspace{2cm} \times \exp \left( y \left[ -2\pi i (p-p') +2\pi (q+q') \right] \right) \\ \nonumber
&& \hspace{2.8cm} \times \exp \left( -\pi h \left[x^2 + y^2 \right] \right) \, dx \, dy \\ \label{eq:usedwavtransfinrepker}
&=& \exp \left( -\frac{2\pi}{h} (q^2 + p^2 + q'^2 + p'^2) \right) \\ \nonumber
&& \left( \frac{1}{h} \right)^n \exp \left( \frac{\pi}{h} \left( [i(q'-q)+(p+p')]^2 + [ i(p'-p) - (q+q') ]^2 \right) \right) \\ \nonumber
&=& \left( \frac{1}{h} \right)^n \exp \left( -\frac{2\pi}{h} \left( q^2 +p^2 +q'^2 + p'^2 - 2qq' -2pp' -2iq'p +2iqp' \right) \right)
\end{eqnarray}
is a reproducing kernel\index{reproducing kernel!$\fock$} for $\fock$. At (\ref{eq:usedwavtransfinrepker}) we have used identity (\ref{eq:waveletwithx}).

These coherent states do not have the correct classical limit \cite{Kisil02.1}. In Section \ref{sect:cstates} we introduce another set of coherent states which are better in this sense. Before we can do this we need a better understanding of what states are in $p$-mechanics --- this is the main content of Chapter \ref{chap:staesandpic}.

\section{$p$-Mechanics} \label{sect:pmechintro}

\indent

In this thesis we continue the development of $p$-mechanics\index{p-mechanics} \cite{Kisil02,Kisil02.1}. $p$-Mechanics is  a consistent physical theory which simultaneously describes both quantum and classical mechanics. It uses the representation theory of the Heisenberg group to show that both quantum and classical mechanics can be derived from the same source.

In this section we give a brief summary of the foundations of $p$-mechanics. In Subsection \ref{subsect:obsvinpmech} we describe the role of observables in $p$-mechanics. In this subsection we also show how to choose a $p$-mechanical observable corresponding to a classical mechanical observable. In Subsection \ref{sect:timeevolofobsv} we define the universal brackets and in doing so the time evolution of $p$-mechanical observables. We show that the time evolution of both quantum and classical observables can be derived from the time evolution of $p$-mechanical observables.

\subsection{Observables in $p$-Mechanics} \label{subsect:obsvinpmech}

\indent

The basic idea of $p$-mechanics is to choose particular functions or distributions on $\Heisn$ which under the infinite dimensional representation will give quantum mechanical observables while under the one dimensional representation will give classical mechanical observables.

The observables\index{observables} can be realised as operators on subsets of $L^2 (\Heisn)$ generated by convolutions of the chosen functions or distributions. Before we can rigorously define $p$-mechanical observables we need to introduce a map from the set of classical mechanical observables to the set of $p$-mechanical observables. We call this the map of $p$-mechanisation\index{p-mechanisation}.
\begin{Defn}[$p$-Mechanisation]
In \cite{KisilBrodlie02,Kisil02.1} the $p$-mechanisation map, $\proj$, is defined as
\begin{equation} \label{eq:pmechmap}
(\proj f) (s,x,y) = \delta (s) \check{f} (x,y)
\end{equation}
where $f$ is any classical observable and $\check{f}$ is the inverse Fourier transform of $f$ (that is $\check{f} (x,y) = \int_{\Space{R}{2n}} f(q,p) e^{2\pi i (qx+py)} \, dq \, dp$).
\end{Defn}
\begin{Example} \label{exam:classicalposition}
The $p$-mechanisation of the $j$-th classical position coordinate is
\begin{equation} \label{eq:pmechposobsv}
\proj(q_j) = X_j = \frac{1}{2\pi i} \Partial{x_j} \zerodel
\end{equation}
while the $p$-mechanisation of the $j$th classical momentum coordinate is
\begin{equation} \label{eq:pmechmomobsv}
\proj (p_j) = Y_j = \frac{1}{2\pi i} \Partial{y_j} \zerodel
\end{equation}
which are both elements of $\mathcal{S}'(\Heisn)$ (see Appendix \ref{app:distributions}).
\end{Example}
Another map of $p$-mechanisation where $\delta(s)$ is replaced by a more general function function $c(s)$ is also discussed in \cite{KisilBrodlie02,Kisil02.1}. $\proj_c$, the map of $p$-mechanisation with function $c$, is defined as
\begin{equation} \label{eq:pmechmapwc}
(\proj_c f) (s,x,y) = c(s) \breve{f} (x,y)
\end{equation}
where $c$ is a real function of a single real variable, $s$, which vanishes as $s \rightarrow \pm \infty$. We are now in a position to define the set of $p$-mechanical observables.
\begin{Defn}[$p$-Mechanical Observables]
The set of $p$-mechanical observables is the image of the set of classical observables under the map $\proj$ from equation (\ref{eq:pmechmap}).
\end{Defn}
Clearly this definition depends on how the set of classical observables is defined. Any physically reasonable classical mechanical observable can be realised as an element of $\mathcal{S}' (\Space{R}{2n})$ (see Appendix \ref{app:distributions} for a definition of this space). Since the Fourier transform maps $\mathcal{S}'(\Space{R}{2n})$ into itself, $\mathcal{S}'(\Heisn)$ is a natural choice for the set of $p$-mechanical observables. It includes the image of all classical observables which are polynomials or exponentials of the variables $q$ and $p$.

The majority of $p$-mechanical observables will generate unbounded operators when realised as convolution operators on $L^2 (\Heisn)$. For example the $p$-mechanical position and momentum observables $X_j$ and $Y_j$ generate right and left invariant vector fields\index{left invariant vector field}\index{right invariant vector field} (\ref{eq:leftinvvecfields}, \ref{eq:rightinvvecfields}) under left and right convolution respectively. That is, if $B$ is an element of $L^2 (\Heisn)$
\begin{eqnarray} \label{eq:deltageninvvfxl}
X_j*B &=& \mathfrak{X}^r_j B = \frac{1}{2 \pi i}\left(\Partial{x_j} + \frac{y_j}{2} \Partial{s}\right)B, \\ \label{eq:deltageninvvfxr}
B*X_j &=& \mathfrak{X}^r_j B = \frac{1}{2 \pi i}\left(\Partial{x_j} - \frac{y_j}{2} \Partial{s}\right)B, \\ \label{eq:deltageninvvfyl}
Y_j*B &=& \mathfrak{Y}_j^r B = \frac{1}{2 \pi i}\left(\Partial{y_j} - \frac{x_j}{2} \Partial{s}\right)B,  \\ \label{eq:deltageninvvfyr}
B*Y_j &=& \mathfrak{Y}_j^l B = \frac{1}{2 \pi i}\left(\Partial{y_j} + \frac{x_j}{2} \Partial{s}\right)B.
\end{eqnarray}
These are clearly unbounded operators which are not defined on the whole of\footnote{If $B$ is a distribution then the convolution of $B$ and an element $v$ of $L^2 (\Heisn)$ is only defined if $v$ is in the test space of the distribution $B$.} $L^2 (\Heisn)$. This technical problem can be solved by the usual method of rigged Hilbert spaces\index{rigged Hilbert spaces} (also known as Gelfand triples) \cite{Roberts66,Ruelle66} which uses the theory of distributions. In \cite{Bohm02} the use of symmetry groups in rigged Hilbert spaces is explored, while \cite{IguriCastagnino99} extends operator algebras into this approach. In the literature on the representation theory of Lie groups this method of dealing with unbounded operators is described using the G\aa rding space --- this is explained in \cite[Chap. 0]{Taylor86}. Furthermore if we take the $\rho_h$ representation (\ref{eq:infdimrep}) of many of the distributions described above we get unbounded operators on $\fock$. For example the distribution $X_j$ under the $\rho_h$ representation will generate the unbounded operator $\frac{h}{4\pi i} \Partial{p_j} - q_j I$ which is not defined on the whole of $\fock$. Again this technicality can be solved using either rigged Hilbert spaces or the G\aa rding space. The use of rigged Hilbert spaces in $p$-mechanics is discussed in Section \ref{sect:rhs}.

It is shown in \cite[Sect. 3.3]{Kisil02.1} that we can also obtain a $p$-mechanical observable from a quantum observable (that is an operator on $\fock$). The map $\proj$ followed by the $\rho_h$ representation is the Weyl quantisation \cite[Chap.2]{Folland89}.

\subsection{$p$-Mechanical Brackets and the Time Evolution of Observables} \label{sect:timeevolofobsv}

\indent

One of the main developments in $p$-mechanics came in the paper \cite{Kisil02} when the universal brackets\index{universal brackets} (also known as $p$-mechanical brackets\index{p-mechanical brackets}) were introduced to describe the dynamics of a $p$-mechanical observable. Before we can define the universal brackets we need to define the operator $\antid$\index{$\antid$}. $\antid$ is defined on exponents by (recall that $\mathfrak{S}=\Partial{s}$ -- see equation (\ref{eq:leftinvvecfields}))
\begin{equation}
  \label{eq:defnofantid}
 \mathfrak{S} \antid=4\pi^2 I, \qquad \textrm{ where }\quad
  \antid e^{ 2\pi i h s}=\left\{
    \begin{array}{ll}
      \displaystyle
      \frac{2\pi}{i h \strut} e^{2\pi i h s}, & \textrm{if }
h \neq 0,\\
      4\pi^2 s\strut, & \textrm{if } h=0.
    \end{array}
    \right.
\end{equation}
This can be realised as an operator on a subset of $L^2 (\Heisn)$ -- this will be described in Section \ref{subsect:timeevolofstates}. $\antid$ is called the antiderivative operator since it is a right inverse to $\Partial{s}$. If we realise our $p$-mechanical observables as convolution operators on $L^2 (\Heisn)$ then we can define a universal bracket on this set of operators.
\begin{Defn}[Universal Brackets] \label{def:pmechbrackets}
The $p$-mechanical brackets of two $p$-mechanical observables, $B_1,B_2$ are defined by the equation
\begin{equation} \label{eq:pmechbrackets}
\ub{B_1}{B_2}{} = ( B_1 * B_2 - B_2 * B_1 ) \antid.
\end{equation}
\end{Defn}
We now state the main result of \cite{Kisil02}.
\begin{Thm}
The image of the $p$-mechanical brackets (see (\ref{eq:pmechbrackets})) under the representations $\rho_h$ and $\rho_{(q,p)}$ give the quantum commutator (see (\ref{eq:heisenbergequation})) and the Poisson brackets (see (\ref{eq:poissonbrackets})) respectively.
\end{Thm}
It is also proved in \cite{Kisil02} that the universal brackets satisfy both the Liebniz and Jacobi identities along with being anticommutative. Note that the $p$-mechanical bracket of two observables realised as elements of $\mathcal{S}'(\Heisn)$ will be an operator on a subset of $L^2 (\Heisn)$ which is not necessarily a convolution operator. For a $p$-mechanical system with energy $B_H$, the $p$-mechanical brackets give us a $p$-dynamic equation for an observable $B$:
\begin{equation} \label{eq:pdyneqn}
\frac{d B}{dt} = \ub{B}{B_H}{}.
\end{equation}
More discussion of the universal brackets and dynamics in $p$-mechanics is given in \cite{KisilBrodlie02,Kisil02.1}. Equation (\ref{eq:pdyneqn}) is extremely useful since when it is solved it will give immediately both the quantum and classical dynamics through the infinite and one dimensional representations respectively. All the machinery and working is in $p$-mechanics, but the results are in classical and quantum mechanics. The applicability of the universal brackets is demonstrated in Chapter \ref{chap:forcedosc} when applied to some examples.

\chapter{States and the Pictures of p-Mechanics} \label{chap:staesandpic}

\indent

In this chapter we introduce the concept of states\index{states} to $p$-mechanics. These are defined in Section \ref{subsect:whatarestates} as functionals on the set of $p$-mechanical observables. $p$-Mechanical states come in two equivalent forms: as elements of a Hilbert space and as integration kernels. These states allow us to compute quantum mechanical expectation values and transition amplitudes using solely functions/distributions on the Heisenberg group. The time evolution of both forms of states is defined in Section \ref{subsect:timeevolofstates} and it is shown that the Schr\"odinger and Heisenberg pictures are equivalent in $p$-mechanics. In describing the time evolution of the kernel states we have a close relation between the dynamics of states in classical and quantum mechanics. In Section \ref{sect:cstates} we introduce an overcomplete system of  coherent states for $p$-mechanics; these again come in two equivalent forms as elements of a Hilbert space and as integration kernels. We show that the classical limits of these coherent states are the classical pure states. In Section \ref{sect:interactpic} the interaction picture is discussed in the $p$-mechanical context. Using the Hilbert space states the interaction picture takes a similar form to that in quantum mechanics, however when the kernels are used some new and interesting insights are obtained. Relationships between $\ltworn$ and our new Hilbert space are discussed in Section \ref{sect:relbetltwoandhh} --- this shows how $p$-mechanics is related to the usual $\ltworn$ formulation of quantum mechanics. Section \ref{sect:rhs} describes how rigged Hilbert spaces fit into $p$-mechanics. In doing this we show how $p$-mechanics can deal with unbounded operators which possess continuous spectra.
We introduce two forms of functionals since both have their own advantages. The Hilbert space functionals are useful for deriving quantum properties of a system, while the kernels have a clearer time evolution and classical limit.

\section{States} \label{subsect:whatarestates}

\indent

In this section we introduce states to $p$-mechanics --- these are positive linear functionals on the set of $p$-mechanical observables. For each $h \neq 0$ (the quantum case) we give two equivalent forms of states: the first form we give is as elements of a Hilbert space, the second is as integration with an appropriate kernel. For $h=0$ (the classical case) we have only one form of states, that is as integration with an appropriate kernel.

\begin{Defn}

The Hilbert space $\hilbh$\index{$\hilbh$}, $h \in \Space{R}{} \setminus \{ 0 \}$, is the set of functions on $\Heisn$  defined by
\begin{eqnarray} \label{eq:hh}
\hilbh &=& \left\{ e^{2 \pi ihs} f (x,y) : f \in L^2 \left( \Space{R}{2n} \right) \hspace{0.5cm} \right. \\ \nonumber
&& \left. \hspace{2cm} \textrm{and $f$ is differentiable such that} \hspace{0.5cm} E^j_h f = 0 \hspace{0.5cm} \textrm{for} \hspace{0.5cm} 1 \leq j \leq n \right\}
\end{eqnarray}
where the operator $E^j_h$ is
\begin{equation} \label{eq:defofejh}
E^j_h = \pi h (x_j - i y_j)I + \Partial{x_j} -  i\Partial{y_j}.
\end{equation}
The inner product on $\hilbh$ is defined as
\begin{equation}  \label{hhip}
\langle v_1 , v_2 \rangle_{\hilbh} = \left( \frac{4}{h} \right)^{n} \int_{\Space{R}{2n}} v_1 (s,x,y) \overline{v_2 (s,x,y)} \, dx \, dy.
\end{equation}
\end{Defn}
The operator $E^j_h$ is the inverse Fourier transform of the operator $D^j_h$ since the inverse Fourier transform (as we have defined it) intertwines $\Partial{q}$ with multiplication by $-2\pi i x$ and intertwines multiplication by $q$ with $\frac{1}{2\pi i} \Partial{x}$; clearly the same results hold if we interchange $x$ with $y$ and $q$ with $p$. This implies that
\begin{equation} \label{hh}
\hilbh = \left\{ e^{2 \pi ihs} \check{f}(x,y) : f \in F^2(\horb) \right\},
\end{equation}
where $\check{f}$ is the inverse Fourier transform of $f$. Note in equation (\ref{hhip}) there is no integration over the $s$ variable since for any two functions $v_1 = e^{2 \pi ihs} f_1 (x,y)$ and $v_2 = e^{2 \pi ihs} f_2 (x,y)$ in $\hilbh$
\begin{equation} \nonumber
\langle v_1 , v_2 \rangle = \int_{\Space{R}{2n}} e^{ 2 \pi ihs} e^{-2 \pi ihs} f_1 (x,y) \bar{f}_2 (x,y) \, dx \, dy = \int_{\Space{R}{2n}} f_1 (x,y) \bar{f}_2 (x,y) \, dx \, dy
\end{equation}
and hence there is no $s$-dependence. $\hilbh$ has a reproducing kernel\index{reproducing kernel!$\hilbh$}
\begin{equation} \label{eq:earlyhhrepker}
K^{\hilbh}_{(x',y')} (s,x,y) = \exp \left[ 2 \pi ihs + \frac{\pi}{2h} \left( 2(x+iy)(x'-iy') - x^2 - y^2 -x'^2 -y'^2 \right) \right].
\end{equation}
We delay proving that this is
a reproducing kernel until Section \ref{sect:cstates} when we have some more machinery.

Most $p$-mechanical observables when realised as convolution operators will be unbounded operators \cite[Chap. 8]{ReedSimon80} and not defined on the whole of $\hilbh$.
These problems are resolved through  the use of rigged Hilbert spaces\index{rigged Hilbert spaces} as was discussed in Subsection \ref{subsect:obsvinpmech}. Section \ref{sect:rhs} contains a discussion of a suitable rigged Hilbert space associated to $\hilbh$. For example if a $p$-mechanical observable, $B$, is a distribution then for $B*v$ to be defined we need $v$ to be in the test space for $B$ (see (\ref{eq:convoftwodistsforgeng})).

We define a set of states for each $h \neq 0$ using $\hilbh$ (later in this section we will show how these states for $h\neq0$ can be defined using an integration  kernel).
\begin{Defn}
If $B$ is a $p$-mechanical observable and $v\in \hilbh$, the $p$-mechanical state corresponding to $v$ acting as a functional on $B$ is
\begin{equation} \nonumber
\langle B * v , v \rangle_{\hilbh}.
\end{equation}
\end{Defn}
In \cite{Kisil02.1} it is stated that if $A$ is a quantum mechanical observable (that is an operator on $\fock$) the state corresponding to $f \in \fock$ is
\begin{equation} \nonumber
\langle A f,f \rangle_{\fock} .
\end{equation}
We now introduce a map $\statem$ which maps vectors in $\fock$ to vectors in $\hilbh$
\begin{equation} \label{eq:statemapforhnonzero}
\statem (f(q,p)) = e^{2 \pi ihs} \check{f}(x,y),
\end{equation}
where $\check{f}$ is the inverse Fourier transform of $f$. This map is one to one by equation (\ref{hh}) and so will have a well defined inverse
\begin{equation}
\mathcal{S}_h^{-1} (e^{2\pi ihs} f (x,y)) = \hat{f} (q,p).
\end{equation}
We next prove a Theorem which shows that the states corresponding to vectors $f$ and $\statem f$ give the same expectation values for observables $B$ and $\rho_h (B)$ respectively. Before we state and prove this Theorem we present a Lemma on the map $\statem$.
\begin{Lemma} \label{lem:propofstatem}
$\statem$ from equation (\ref{eq:statemapforhnonzero}) is unitary and
\begin{equation} \label{eq:rhohandleftshift}
\rho_h (g) = \statem^{-1} \lambda_l (g) \statem
\end{equation}
for any $g \in \Heisn$.
\end{Lemma}
\begin{proof}
We first show by a direct calculation that $\statem$ is a unitary operator from $\fock$ to $\hilbh$. If $f_1,f_2 \in \fock$
\begin{eqnarray} \nonumber
\langle \statem f_1 , \statem f_2 \rangle_{\hilbh} &=& \int \check{f_1} (x,y) \overline{\check{f_2} (x,y)} \, dx \, dy \\ \label{eq:usedinvftisunitary}
&=&  \int f_1 (q,p) \overline{f_2 (q,p)} \, dq \, dp \\ \nonumber
&=& \langle f_1 , f_2 \rangle_{\fock}.
\end{eqnarray}
At (\ref{eq:usedinvftisunitary}) we have used the fact that the inverse Fourier transform is a unitary operator on $L^2$. The above calculation proves that $\statem$ is a unitary operator. We now verify equation (\ref{eq:usedinvftisunitary}) this again follows by a direct calculation. Let $f \in \fock$ then $(\statem f)(s,x,y) = e^{2\pi ihs} \check{f} (x,y)$ and
\begin{eqnarray} \nonumber
\lambda_l (s',x',y') (\statem f)(s,x,y) &=& (\statem f) \left( s-s'+ \frac{1}{2} (xy'-x'y) , x-x' , y-y' \right) \\ \nonumber
&=& e^{2\pi ih(s-s')} e^{\pi ih(xy'-x'y)} \check{f} (x-x',y-y').
\end{eqnarray}
This implies that
\begin{eqnarray} \nonumber
\lefteqn{(\statem^{-1} \lambda_l (s',x',y') \statem f)(q,p) } \\ \nonumber
&=& e^{-2\pi ih s'} \int_{\Space{R}{2n}} e^{\pi ih (xy'-x'y)} \check{f} (x-x',y-y') e^{-2\pi i (qx+py)} \, dx \, dy \\ \nonumber
&=& e^{-2\pi ih s'} \int_{\Space{R}{2n}} e^{\pi ih [(x+x')y'-x'(y+y')]} \check{f} (x,y) e^{-2\pi i [q(x+x')+p(y+y')]} \, dx \, dy \\ \nonumber
&=& e^{-2\pi i (hs'+qx'+py')} \int_{\Space{R}{2n}} \check{f} (x,y) e^{-2\pi i x \left( q - \frac{h}{2} y' \right)} e^{-2\pi i y \left( p+\frac{h}{2} x' \right) } \, dx \, dy \\ \nonumber
&=& e^{-2\pi i (hs'+qx'+py')} f \left( q - \frac{h}{2} y', p+\frac{h}{2} x' \right),
\end{eqnarray}
the last step follows by the Fourier inversion formula.
\end{proof}
\begin{Thm} \label{hilbh-f2h}
If $B$ is a $p$-mechanical observable and $f_1,f_2 \in \fock$ such that $B* \statem f_1$ is defined then
\begin{equation} \label{eq:statesrelation}
\langle B * \statem f_1 , \statem f_2 \rangle_{\hilbh} = \left\langle \rho_{h} (B) f_1, f_2 \right \rangle_{\fock}.
\end{equation}
\end{Thm}
\begin{proof}
If $B \in \loneh$ then by (\ref{eq:repofafctnonaliegp})
\begin{equation} \nonumber
\langle \rho_h (B) f_1 , f_2 \rangle = \left\langle \int B(g') \rho_h (g') f_1 \, dg' , f_2 \right\rangle.
\end{equation}
Using Fubini's Theorem (Theorem \ref{thm:fubini}) this becomes
\begin{equation} \nonumber
\langle \rho_h (B) f_1 , f_2 \rangle = \int B(g') \langle \rho_h (g') f_1 , f_2 \rangle \, dg'.
\end{equation}
By (\ref{eq:rhohandleftshift})
\begin{equation} \nonumber
\langle \rho_h (B) f_1 , f_2 \rangle = \int B(g') \langle \statem^{-1} \lambda_l (g') \statem f_1 , f_2 \rangle \, dg'.
\end{equation}
Since $\statem$ is a unitary operator (Lemma \ref{lem:propofstatem})
\begin{eqnarray} \nonumber
\langle \rho_h (B) f_1 , f_2 \rangle &=& \int B(g') \langle \lambda_l (g') \statem f_1 , \statem f_2 \rangle \, dg' \\ \nonumber
&=& \langle B* \statem f_1 , \statem f_2 \rangle.
\end{eqnarray}
Now if $B$ is a distribution on the Heisenberg group and $f_1,f_2 \in \fock$ are such that $(\statem f_1)(g)$, $\langle \rho_h (g) f_1 , f_2 \rangle$ are in the test space then by (\ref{eq:repofadistnonaliegp})
\begin{equation} \nonumber
\langle \rho_h (B) f_1 , f_2 \rangle = \int B(g') \langle \rho_h (g') f_1 , f_2 \rangle \, dg'.
\end{equation}
The result will then follow in the same way as the case of $B \in \loneh$.
\end{proof}

Taking $f_1=f_2$ in (\ref{eq:statesrelation}) shows that the states corresponding to $f$ and $\statem f$ will give the same expectation values for $\rho_h (B)$ and $B$ respectively. If we take $B$ to be a time development operator we can get probability amplitudes between states $f_1 \neq f_2$.

\begin{Remark}
\emph{Lemma \ref{lem:propofstatem} implies that the representation $\lambda_l$ on $\hilbh$ is unitarily equivalent to the $\rho_h$ representation on $\fock$. Hence $\lambda_l$ is a unitary irreducible representation of the Heisenberg group.}
\end{Remark}
We now show that each of these states can also be realised by an appropriate integration kernel\index{kernel states}.
\begin{Thm} \label{thm:statesaresame}
If $l(s,x,y)$ is defined to be the kernel
\begin{equation} \label{eq:relationbetweenkernelandvector}
l(s,x,y) = \left(\frac{4}{h}\right)^n \int_{\Space{R}{2n}}  \overline{v((s,x,y)^{-1} (s',x',y'))} v((s',x',y')) \, dx' \, dy'.
\end{equation}
then if $B*v$ is defined we have
\begin{equation} \nonumber
\langle B*v,v \rangle_{\hilbh} = \int_{\Heisn} B(s,x,y) \overline{l(s,x,y)} \, ds \, dx \, dy.
\end{equation}
\end{Thm}
\begin{proof}
If $B \in L^1 (\Heisn)$ using Fubini's theorem (that is, Theorem \ref{thm:fubini})
\begin{eqnarray} \nonumber
\lefteqn{\langle B*v,v \rangle} \\ \nonumber
&=& \left(\frac{4}{h}\right)^n \int_{\Space{R}{2n}} \int_{\Heisn} B((s,x,y)) v((s,x,y)^{-1} (s',x',y')) \\ \nonumber
&& \qquad \qquad \qquad \qquad \times \overline{v((s',x',y'))} \, ds \, dx \, dy \, dx' \, dy' \\ \label{eq:vandlsxy}
&=& \left(\frac{4}{h}\right)^n \int_{\Heisn} B((s,x,y)) \\ \nonumber
&& \qquad \qquad \times \left( \int_{\Space{R}{2n}} v((s,x,y)^{-1} (s',x',y')  )\overline{v((s',x',y'))}  \,dx' \, dy' \right) \,ds \, dx \, dy.
\end{eqnarray}
Note that there is no integration over $s'$ by the definition of the $\hilbh$ inner product. If $B \in \loneh$ we are allowed to use Fubini's theorem since $\langle B * v,v \rangle < \infty$. If $B$ is a distribution then the result follows by Fubini's theorem for distributions, that is Theorem \ref{thm:fubinifordist}. Since $B*v$ is well defined we can take $v((s,x,y)^{-1} (s',x',y')  )$ as a test function on $\Heisn \times \Heisn$.
\end{proof}
\begin{Defn}
We denote the set of kernels corresponding to the elements in $\hilbh$ as $\lkerh$.
\end{Defn}
If $v(s,x,y) = e^{2\pi ihs} \check{f} (x,y)$ then the corresponding element of $\sokh$ is
\begin{eqnarray} \label{eq:formofkernstraightfromfctn}
\lefteqn{l(s,x,y)} \\ \nonumber
&=& \int_{\Space{R}{2n}} \overline{v \left( s'- s + \frac{1}{2} (x'y - xy'), x'-x,y'-y \right)} v(s',x',y') \, dx' \, dy' \\ \nonumber
&=& e^{2\pi ihs} \int_{\Space{R}{2n}} e^{\pi ih (xy'-x'y) } \overline{\check{f} (x'-x, y'-y)} \check{f} (x', y') \, dx' \, dy'.
\end{eqnarray}
Now we introduce $p$-mechanical $(q,p)$ states which correspond to classical states; they are again functionals on the set of $p$-mechanical observables. Pure states in classical mechanics evaluate observables at particular points of phase space; they can be realised as kernels
$\delta(q-a , p-b)$ for fixed $a,b$ in phase space, that is
\begin{equation} \label{eq:classicalpurestateeval}
\int_{\Space{R}{2n}} F(q,p) \delta(q-a , p-b) \, dq \, dp = F(a,b).
\end{equation}
We now give the $p$-mechanical equivalent of pure classical states.
\begin{Defn}
p-Mechanical $(q,p)$ pure states\index{(q,p) pure states} are defined to be the set of functionals, $k_{(0,a,b)}$, for fixed $a,b \in \Space{R}{2n}$ which act on observables by
\begin{equation} \label{eq:pclasspurestates}
k_{(0,a,b)}(B(s,x,y)) = \int_{\Heisn} B(s,x,y) e^{-2\pi i (a.x+b.y)} \, ds \, dx \, dy.
\end{equation}
Each $(q,p)$ pure state $k_{(0,a,b)}$ is defined entirely by its kernel $l_{(0,a,b)}$
\begin{equation}
l_{(0,a,b)} = e^{2\pi i (a.x+b.y)}.
\end{equation}
\end{Defn}

Note that the kernel is $e^{2\pi i (a.x+b.y)}$ rather than $e^{-2\pi i (a.x+b.y)}$ since we are integrating our observables next to the complex conjugate of an integration kernel.
If $B$ is the $p$-mechanisation, (see equation (\ref{eq:pmechmap})), of a classical observable, $f$, then
\begin{equation} \label{eq:pmechanicalpurestateeval}
\int_{\Heisn} B(s,x,y) e^{-2\pi i (a.x+b.y)}\, ds \, dx \, dy = f(a,b).
\end{equation}
Hence when we apply state $k_{(0,a,b)}$ to a $p$-mechanical observable we get the value of its classical counterpart at the point $(a,b)$ of phase space.
We introduce the map $\statemo$ which maps classical pure state kernels to $p$-mechanical $(q,p)$ pure state kernels
\begin{equation} \nonumber
\statemo (\xi(q,p)) = \overline{\hat{\xi} (x,y)}.
\end{equation}
This equation is almost identical to the relation in equation (\ref{eq:statemapforhnonzero}). The kernels $\labo$, are the complex conjugate of the Fourier transforms of the delta functions $\delta(q-a,p-b)$, and hence pure $(q,p)$ states are just the image of pure classical states.

Mixed states\footnote{Mixed states in quantum mechanics may be infinite linear combinations of pure states and are defined using the density matrix \cite{Merzbacher70}.}, as used in statistical mechanics \cite{Honerkamp98}, are finite linear combinations of pure states. In $p$-mechanics $(q,p)$ mixed states are defined in the same way.
\begin{Defn}
Define $\lkero$, to be the space of all finite linear combinations of $(q,p)$ pure state kernels $l_{(0,a,b)}$, that is the set of all kernels corresponding to $(q,p)$ mixed states.
\end{Defn}
The map $\statemo$ exhibits the same relations on mixed states as pure states due to the linearity of the Fourier transform.

\section{Time Evolution of States} \label{subsect:timeevolofstates}

\indent

We now go on to show how $p$-mechanical states evolve with time. We first show how the elements of $\hilbh$ evolve with time and prove that they agree with the Schr\"odinger picture of motion in quantum  mechanics. We then show how the elements of $\lkerh$, for all $h\in\Space{R}{}$, evolve with time and that this time evolution agrees with the time evolution of $p$-observables. In doing this we show that for the particular case of $\lkero$ the time evolution is the same as classical states under the Liouville equation. Since the kernel states and $\hilbh$ states are equivalent we get a relation between the time evolution of classical states and the time evolution of quantum states.

Before we can do any of this we need to give the definition of a self adjoint $p$-mechanical observable.
\begin{Defn} \label{def:self-adj}
We call a bounded $p$-mechanical observable B self adjoint if and only if for any $v_1,v_2 \in \hilbh$
\begin{equation} \nonumber
\langle B*v_1 , v_2 \rangle = \langle v_1 , B* v_2 \rangle.
\end{equation}
\end{Defn}
For any $p$-mechanical observable, $B$ we denote by $B'$ the $p$-mechanical observable which satisfies
\begin{equation} \nonumber
\langle B*v_1 , v_2 \rangle = \langle v_1 , B' * v_2 \rangle.
\end{equation}
When dealing with unbounded operators the definition of self adjointness is more involved -- this is described in $\cite{ReedSimon80}$.

Now we show how the vectors in $\hilbh$ evolve with time. Initially we extend our definition of $\antid$ which was initially introduced in equation (\ref{eq:defnofantid}).
\begin{Defn}
$\antid$ can also be defined as an operator on each $\hilbh$, $h \in \Space{R}{} \setminus \{ 0 \}$, $\antid:\hilbh \mapsto \hilbh$ by
\begin{equation} \label{eq:newdefofantid}
\antid v = \frac{2\pi}{ih} v.
\end{equation}
\end{Defn}
The following Lemma follows directly from the definition of $\antid$ on $\hilbh$.
\begin{Lemma} \label{lem:propofantidtwo}
If $\antid$ and $\Partial{s}$ are operators on $\hilbh$ then:
\begin{enumerate}
\item The adjoint of $\antid$ is $-\antid$ on each $\hilbh$, $h \in \Space{R}{} \setminus \{ 0 \}$.
\item $\antid \Partial{s} = \Partial{s} \antid = 4\pi^2 I$.
\item $\antid$ commutes with left convolution by a $p$-mechanical observable, that is $B*\antid v = \antid B*v$.
\end{enumerate}
\end{Lemma}
\begin{Defn}
If we have a system with energy $B_H$ then an arbitrary vector $v \in \hilbh$ evolves under the equation
\begin{equation} \label{eq:timevolinhh}
\Fracdiffl{v}{t} = B_H * \antid v.
\end{equation}
\end{Defn}
The operation of left convolution preserves each $\hilbh$ so this time evolution is well defined. Also by Lemma \ref{lem:propofantidtwo} we have
\begin{equation} \nonumber
\frac{dv}{dt} = \antid B_H * v.
\end{equation}
Equation (\ref{eq:timevolinhh}) implies that if we have $B_H$ time-independent and self-adjoint then for any $v \in \hilbh$
\begin{equation} \nonumber
v(t;s,x,y) = e^{t B_H \antid} v(0;s,x,y)
\end{equation}
where $e^{ B_H \antid } $ is the exponential of the operator of  applying $\antid$ and then applying the left convolution of $B_H$ --- this operator is defined using\footnote{This can be done since $B_H \antid$ is an anti-self-adjoint operator. This follows since $\antid$ is anti-self-adjoint, $B_H$ is self-adjoint and $\antid$ commutes with convolution.}  Stone's theorem \cite[Sect. 8.4]{ReedSimon80}. Also by Stone's Theorem we have that $v(t;s,x,y)$ will satisfy (\ref{eq:timevolinhh}) and is differentiable with respect to $t$.
\begin{Thm} \label{thm:schroheis}
If we have a system with energy $B_H$ (assumed to be self-adjoint) then for any state $v \in \hilbh$ and any observable $B$
\begin{equation} \nonumber
\frac{d}{dt} \langle B * v, v \rangle = \langle \ub{B}{B_H}{} * v , v \rangle.
\end{equation}
\end{Thm}
\begin{proof}
The result follows from the direct calculation:
\begin{eqnarray} \nonumber
\frac{d}{dt} \langle B * v (t) , v (t) \rangle
&=& \langle B * \frac{d}{dt} v , v \rangle + \langle B * v, \frac{d}{dt} v \rangle \\ \nonumber
&=& \langle B * B_H * \antid v  , v \rangle + \langle B * v, B_H * \antid v \rangle  \\ \nonumber
&=& \langle B * B_H * \antid v  , v \rangle + \langle B * v, \antid B_H *v \rangle  \\ \label{eq:usedadjofantid}
&=& \langle B* B_H *  \antid v, v \rangle - \langle \antid B * v , B_H *v \rangle \\ \label{eq:usedhermitian}
&=& \langle B * B_H * \antid v  , v \rangle - \langle B_H * \antid B * v, v \rangle  \\ \nonumber
&=& \langle B * B_H * \antid v  , v \rangle - \langle B_H * B * \antid v, v \rangle  \\ \nonumber
&=& \langle \ub{B}{B_H}{} *v , v \rangle.
\end{eqnarray}
Equation (\ref{eq:usedadjofantid}) follows since $\antid$ is skew-adjoint in $\hilbh$. At (\ref{eq:usedhermitian}) we have used the fact that $B_H$ is self-adjoint.
\end{proof}
This Theorem proves that the time evolution of states in $\hilbh$ coincides with the time evolution of observables as described in equation (\ref{eq:pdyneqn}). We now give a corollary to show that the time evolution of $p$-mechanical states in $\hilbh$, $h \in \Space{R}{} \setminus \{ 0 \} $ is the same as the time evolution of quantum states.
\begin{Corollary}
If we have a system with energy $B_H$ (assumed to be self-adjoint) and an arbitrary state $v = \statem f = e^{2 \pi ihs} \check{f} (x,y)$ (assuming $h \neq 0$) then for any $p$-mechanical observable $B$
\begin{equation} \nonumber
\Diffl{t} \langle B * v (t) , v (t) \rangle_{\hilbh} = \Diffl{t} \langle \rho_h (B) f(t) , f(t) \rangle_{\fock},
\end{equation}
where $\Fracdiffl{f}{t} = \frac{1}{i\hbar} \rho_{h} (B_H) f$ (this is just the usual Schr\"odinger equation).
\end{Corollary}
\begin{proof}
From Theorem \ref{thm:schroheis} we have
\begin{eqnarray} \nonumber
\frac{d}{dt} \langle  B * v, v \rangle &=&  \langle \ub{B}{B_H}{} * v , v \rangle \\ \nonumber
&=& \langle (B * B_H - B_H * B ) * \antid v, v \rangle \\ \nonumber
&=& \frac{2 \pi}{ih} \langle (B * B_H - B_H * B ) * v, v \rangle \\ \nonumber
&=& \frac{1}{i \hbar} ( \langle B * B_H *v,v \rangle - \langle B * v, B_H * v \rangle)
\end{eqnarray}
The last step follows since $B_H$ is self-adjoint. Using equation (\ref{eq:statesrelation}), the above equation becomes,
\begin{eqnarray} \nonumber
\frac{d}{dt} \langle B * v, v \rangle &=& \frac{1}{i\hbar} (\langle \rho_h (B) \rho_h (B_H)f,f \rangle_{\fock} - \langle \rho_h (B) f , \rho_h (B_H )f \rangle_{\fock} )\\ \nonumber
&=& \frac{d}{dt} \langle \rho_h (B)f, f \rangle_{\fock},
\end{eqnarray}
which completes the proof.
\end{proof}
Hence the time development in $\hilbh$ for $h \neq 0$ gives the same time development as in $\fock$.
We now look at the time evolution of the kernel coherent states. Before we can do this we need to introduce two definitions. The first definition is of the $p$-mechanical brackets (that is, universal brackets) of a $p$-mechanical observable and a kernel in the space $\sokh$.
\begin{Defn}
If $B$ is a $p$-mechanical observable and $l \in \sokh$ then $\ub{B}{l}$ is defined in exactly the same way as the $p$-mechanical brackets of two observables (see Definition \ref{def:pmechbrackets}).
\end{Defn}
The above definition makes sense since $\sokh$ can easily be realised as a subset of $\mathcal{S}'(\Heisn)$. The next definition we give is of a kernel self-adjoint $p$-mechanical observable.
\begin{Defn}
A $p$-mechanical observable, $B$, is said to be kernel self-adjoint if the adjoint of the operator $\ub{ \cdot}{B}{}$ on the set of $p$-mechanical observables is the operator $\ub{B}{\cdot}{}$ on the set of kernels (which are functionals on the set of $p$-mechanical observables). This is equivalent to the following equation holding
\begin{equation} \nonumber
\langle \ub{C}{B}{} , l \rangle = \langle C , \ub{B}{l}{} \rangle
\end{equation}
for any $p$-mechanical observable $C$, where the brackets represent $\langle B, l \rangle = \int B(g) \overline{l(g)} \, dg$.
\end{Defn}
The $p$-mechanical position and momentum observables are both kernel self adjoint; so are the $p$-mechanical Hamiltonians for the forced and harmonic oscillators (see later in Chapter \ref{chap:forcedosc}) and hence all the Hamiltonians considered in this thesis are kernel self adjoint.
\begin{Defn}
If we have a system with a kernel self-adjoint $p$-mechanical Hamiltonian, $B_H$,  then an arbitrary kernel $l \in \lkerh$, $h\in \Space{R}{}$, evolves under the equation
\begin{equation} \label{eq:zerostatestimeeq}
\Fracdiffl{l}{t} =  \ub{B_H}{l}{}.
\end{equation}
\end{Defn}
We now show that the time evolution of these kernels coincides with the time evolution of $p$-mechanical observables.
\begin{Thm} \label{Thm:timeevolofkernelisok}
If $l$ is a kernel evolving under equation (\ref{eq:zerostatestimeeq}) then any observable $B$ will satisfy
\begin{equation} \nonumber
\Diffl{t} \int_{\Heisn} B \, \overline{l} \, dg = \int_{\Heisn} \ub{B}{B_H}{} \, \overline{l} \, dg .
\end{equation}
\end{Thm}
\begin{proof}
This follows directly from the definition of kernel self-adjointness.
\end{proof}
If we take the representation $\rho_{(q,p)}$ of equation  (\ref{eq:zerostatestimeeq}) we get the Liouville equation \cite[Eq. 5.42]{Honerkamp98} for a kernel $\statemo^{-1}(l)$ moving in a system with energy $\rho_{(q,p)} (B_H)$. This only holds for elements in $\lkero$ and can be verified by a similar calculation to \cite[Prop. 3.5]{Kisil02}.

If $l(s,x,y) = \left( \frac{4}{h} \right)^n \int_{\Heisn} v((s',x',y')) \overline{v((s',x',y')^{-1} (s,x,y))} \, dx' \, dy'$ then by Theorem \ref{thm:schroheis} and Theorem \ref{Thm:timeevolofkernelisok} we have that
\begin{equation} \label{eq:sametimeevolofkernelandvector}
\Diffl{t} \langle B*v,v \rangle_{\hilbh} = \Diffl{t} \int_{\Heisn} B \, \overline{l} \, dg
\end{equation}
in a system governed by a kernel self-adjoint $p$-mechanical Hamiltonian.

\section{Eigenvalues and Eigenfunctions} \label{sect:eigenvalueeigenfunction}

\indent

In this section we introduce the concept of eigenvalues and eigenfunctions\index{eigenfunctions} for $p$-observables.

\begin{Thm}[Eigenfunctions in $\hilbh$] \label{thm:eigvaluesinhh}

For a $p$-mechanical observable $B$ and $f_{\lambda} \in \fock$, $\rho_h (B) f_{\lambda} = \lambda f_{\lambda}$, if and only if for $v_{\lambda} (s,x,y) = \statem f_{\lambda} = e^{2 \pi ihs} \check{f_{\lambda}} (x,y) \in \hilbh$
\begin{equation} \label{eq:evaluesinhh}
\langle B * v_{\lambda}, v \rangle = \lambda \langle v_{\lambda} , v \rangle
\end{equation}
holds for all $v \in \hilbh$.
\end{Thm}

\begin{proof}

If $v = e^{2\pi ihs} \check{f} (x,y)$ where $f$ is an arbitrary element of $\fock$,
\newline
$\rho_h (B) f_{\lambda} = \lambda f_{\lambda} $ implies that
\begin{equation} \label{eiginf}
\langle \rho_h (B) f_{\lambda} ,f \rangle =  \lambda \langle f_{\lambda} ,f \rangle = \lambda \langle \rho_h (\zerodel) f_{\lambda} , f \rangle
\end{equation}
for any $f \in \fock$. By (\ref{eq:statesrelation}) this gives us
\begin{equation} \label{eigink}
\langle B * v_{\lambda} , v \rangle = \lambda \langle \zerodel * v_{\lambda}, v \rangle = \lambda \langle v_{\lambda} , v \rangle
\end{equation}
for $v=e^{2\pi ihs}\hat{f}$. Since we can choose any $f$ in (\ref{eiginf}), (\ref{eigink}) holds for any $v \in \hilbh$. This proves the argument in one direction. Clearly equations (\ref{eiginf}) and (\ref{eigink}) are equivalent so the converse follows since (\ref{eigink}) holding for any $v \in \hilbh$ is equivalent to (\ref{eiginf}) holding for any $f \in \fock$.
\end{proof}
If $v_{\lambda} \in \hilbh$ satisfies (\ref{eq:evaluesinhh}) then we say $v_{\lambda}$ is an eigenvector\index{eigenvector} (or eigenfunction\index{eigenfunction}) of $B$ with eigenvalue\index{eigenvalue} $\lambda$ --- this is just the usual terminology. Note that if we put in the reproducing kernel (see equation (\ref{eq:earlyhhrepker})) for $v$ in (\ref{eq:evaluesinhh}) we get
\begin{equation} \nonumber
B * \vlam = \lambda \vlam.
\end{equation}
Equation (\ref{eq:evaluesinhh}) implies that
\begin{equation} \label{eq:evalforantid}
\langle \antid B * v_{\lambda} , v \rangle =\frac{2\pi}{ih} \lambda \langle v_{\lambda} , v \rangle.
\end{equation}

\section{Coherent States and \newline Creation/Annihilation Operators} \label{sect:cstates}

\indent

In this section we introduce an overcomplete system of vectors in $\hilbh$ by a representation of $\Heisn$. The states which correspond to these vectors are an overcomplete system of coherent states\index{coherent states!$\hilbh$} for each $h \neq 0$. We then show that these vectors correspond to a system of kernels in $\lkerh$, whose limit is the $(q,p)$ pure state kernels. Before we introduce the $p$-mechanical coherent states we give a little history of coherent states.

Coherent states were discovered by Schr\"odinger in 1926. He introduced them as a system of nonorthogonal wave functions which described nonspreading wave packets for the quantum harmonic oscillator. For nearly forty years these states were largely ignored. However in the 1960s a lot of interest in these states was ignited by figures such as Klauder, Glauber, Segal, Berezin, Bargmann, Perelomov and many others \cite{Bargmann61,Berezin75,Perelomov86}. In this section we introduce standard coherent states into $p$-mechanics. More general coherent states have also been considered \cite{Klauder96,Perelomov86} --- their role in $p$-mechanics is discussed in Chapter \ref{chap:keplercoulomb}. The definition of an overcomplete system of coherent states was given in Definition \ref{def:overcompcs}.

Initially we need to introduce a vacuum vector in $\hilbh$. For this we take the vector in $\hilbh$ corresponding to the ground state of the harmonic oscillator with classical Hamiltonian $\frac{1}{2} (m \omega^2 q^2 + \frac{1}{m}p^2)$ where $\omega$ is the constant frequency and $m$ is the constant mass. The vector in $\fock$ corresponding to the ground state is \cite[Eq 2.18]{Kisil02.1}
\begin{equation} \nonumber
f_0 (q,p) = \exp \left( -\frac{2\pi}{h} (\omega m q^2 + (\omega m)^{-1} p^2 ) \right), \hspace{1cm} h>0 .
\end{equation}
The image of this under $\statem$ is
\begin{equation} \nonumber
e^{2 \pi ihs} (\fort^{-1} ( f_0 ))(x,y) = e^{2 \pi ihs} \int_{\Space{R}{2n}}  e^{-\frac{2 \pi}{h} (m \omega q^2 + (m \omega )^{-1} p^2)} e^{2 \pi i(qx+py)} \, dq \, dp.
\end{equation}
Using formula (\ref{eq:waveletwithx}) we get
\begin{equation} \nonumber
\statem (f_0 )(s,x,y) =  \left( \frac{h}{2} \right)^n  \exp \left(  2 \pi ihs - \frac{\pi h}{2} \left( \frac{x^2}{\omega m} + y^2 \omega m \right) \right),
\end{equation}
which is the element of $\hilbh$ corresponding to the ground state. For the rest of this section we assume that $\omega$ and $m$ are equal to unity. In doing this we make the calculations less technical without losing any generality.
\begin{Defn}
Define the vacuum vector in $\hilbh$ as
\begin{equation} \nonumber
\voo (s,x,y) = \left( \frac{h}{2} \right)^n \exp \left( 2\pi i hs - \frac{\pi h}{2} \left( x^2 + y^2 \right) \right).
\end{equation}
\end{Defn}
To generate a system of coherent states in $\hilbh$ we need an irreducible representation of the Heisenberg group on $\hilbh$. The representation we use is\index{$\zeta^h$}
\begin{eqnarray} \label{eq:zetarep}
(\zeta^h_{(r,a,b)} v) (s,x,y) &=& v\left( \left( -\frac{r}{h^2} , -\frac{b}{h} , \frac{a}{h} \right)^{-1} (s,x,y) \right) \\ \nonumber
&=& v \left( \left( \frac{r}{h^2} , \frac{b}{h} , - \frac{a}{h} \right)(s,x,y) \right) \\ \nonumber
&=& v \left( s + \frac{r}{h^2} + \frac{1}{2h} (by+ax) , x + \frac{b}{h}, y - \frac{a}{h} \right).
\end{eqnarray}
\begin{Lemma}
$\zeta^h$ is an irreducible representation of the Heisenberg group.
\end{Lemma}
\begin{proof}
A direct calculation shows that $\zeta^h$ satisfies the group homomorphism property. Irreducibility follows since left shifts are irreducible in $\hilbh$ as explained in Section \ref{subsect:whatarestates}.
\end{proof}
Since $\zeta^h_{(r,0,0)} : v(s,x,y) \mapsto v \left( s + \frac{r}{h^2} , x, y \right) = e^{2\pi i \frac{r}{h}} v(s,x,y)$ we see that $\zeta^h$ satisfies condition 2 of \cite[Defn. 2.2]{Kisil99} with $H$ as the centre of $\Heisn$ (that is $\{ (r,0,0):r \in \Space{R}{} \}$). Since $\zeta^h$ is an irreducible representation of the Heisenberg group by \cite[Thm. 2.11]{Kisil99} the system of vectors given by
\begin{equation} \nonumber
v_{(h,a,b)} = \zeta^h_{(o,a,b)} v_{(h,0,0)}
\end{equation}
is a system of square integrable coherent states \cite{AliAntGaz00,AliAntGazMueller95}. By a direct calculation
\begin{eqnarray} \label{eq:hhcs}
\lefteqn{v_{(h,a,b)} (s,x,y)} \\ \nonumber
&& = v_{(h,0,0)} \left( s + \frac{1}{2h} (by+ax) , x+\frac{b}{h}, y - \frac{a}{h} \right) \\ \nonumber
&& = \exp \left(2\pi ihs + \pi i (by+ax) -\frac{\pi h}{2} \left(x+\frac{b}{h} \right)^2 -\frac{\pi h}{2} \left(y-\frac{a}{h} \right)^2 \right).
\end{eqnarray}
To prepare for later calculations we present two rearrangements of (\ref{eq:hhcs})
\begin{eqnarray} \nonumber
\lefteqn{v_{(h,a,b)} } \\ \nonumber
&=& \exp \left( 2\pi ihs + \pi x (ia-b) + \pi y ( ib + a ) -\frac{\pi h}{2} (x^2 + y^2) - \frac{\pi}{2h}(a^2 + b^2 ) \right) \\ \nonumber
&=& \exp \left( 2\pi ihs + \pi a (ix+y) + \pi b (iy - x ) -\frac{\pi h}{2} (x^2 + y^2) - \frac{\pi}{2h} (a^2 + b^2)  \right).
\end{eqnarray}
Now we have an overcomplete system of coherent sates for $\hilbh$ we can prove the validity of equation (\ref{eq:earlyhhrepker}).
\begin{Lemma}
\begin{equation} \label{eq:repkerforhh}
K^{\hilbh}_{(a',b')} (s,a,b) = \exp \left[ 2\pi ihs + \frac{\pi}{2h} \left( 2(a+ib)(a'-ib') - a^2 - b^2 - a'^2 - b'^2 \right) \right]
\end{equation}
is a reproducing kernel for $\hilbh$\index{reproducing kernel!$\hilbh$}.
\end{Lemma}
\begin{proof}
Since $\{ v_{(h,a,b)} : a,b \in \Space{R}{n} \}$ are an overcomplete system of coherent states,
\begin{equation}
K^{\hilbh}_{(a',b')} (s,a,b) = e^{2\pi ihs} \langle v_{(h,a,b)} , v_{(h,a',b')} \rangle_{\hilbh}
\end{equation}
is a reproducing kernel for $\hilbh$ \cite[Eq. 1.2.13]{Perelomov86}.
Now
\begin{eqnarray}
\lefteqn{ \langle v_{(h,a,b)} , v_{(h,a',b')} \rangle_{\hilbh} } \\ \nonumber
&& =\int_{\Space{R}{2n}} \exp \left( \pi x (ia-b-ia'-b') + \pi y (ib+a-ib'+a') \right)  \\ \nonumber
&& \hspace{2cm} \times \exp \left( -\pi h (x^2 + y^2) \right) \, dx \, dy \, \\ \nonumber
&& \hspace{2.5cm} \times \exp \left( -\frac{\pi}{2h} \left( a^2 + b^2 + a'^2 + b'^2 \right) \right).
\end{eqnarray}
Using (\ref{eq:waveletwithx}) this becomes
\begin{eqnarray} \nonumber
\lefteqn{\exp \left( \frac{\pi}{4h} [ (ia-b)^2 - 2(ia-b)(ia'+b') + (ia'-b')^2  \right. } \\ \nonumber
&& \left. \hspace{2cm}+ (ib+a)^2 + 2(ib+a)(-ib'+a') + (a'-ib')^2 ] \right) \\ \nonumber
&& \hspace{3.5cm} \times \exp \left( -\frac{\pi}{2h} (a^2 + b^2 +a'^2 + b'^2) \right) \\ \nonumber
&=& \exp \left( \frac{\pi}{2h} ( 2(a+ib)(a'-ib') -a^2 -b^2 -a'^2 -b'^2) \right).
\end{eqnarray}
\end{proof}
Also since this system of coherent states is square integrable we can take coherent state expansions \cite{AliAntGaz00} of any element, $v \in \hilbh$, that is
\begin{equation} \label{eq:wecantakecsexpansions}
v = \int_{\Space{R}{2n}} \langle v , \vab \rangle \vab \, da \, db.
\end{equation}
If we map the $\hilbh$ coherent states from equation (\ref{eq:hhcs}) into $\sokh$ we get a system of coherent states\index{coherent states!kernels} realised as kernels.
\begin{Lemma}
The kernel coherent state $\lab$ corresponding to $\vab$ is
\begin{equation}
\lab = \exp \left( 2\pi ihs + 2\pi i (ax+by) -\frac{\pi h}{2}(x^2 + y^2) \right)
\end{equation}
\end{Lemma}
\begin{proof}
By equation (\ref{eq:formofkernstraightfromfctn})
\begin{eqnarray} \nonumber
\lab = e^{2\pi ihs} \int_{\Space{R}{2n}} e^{\pi ih (xy' -x'y)} \overline{\tilde{v} (x'-x,y'-y)} \tilde{v}(x',y') \, dx' \, dy'
\end{eqnarray}
where
\begin{equation} \nonumber
\tilde{v} (x,y) = \exp \left( \pi x (ia-b) + \pi y (ib+a) -\frac{\pi h}{2} (x^2 + y^2) - \frac{\pi}{2h} (a^2 + b^2) \right).
\end{equation}
So
\begin{eqnarray}
\lefteqn{\lab = \exp (2\pi ihs)} \\ \nonumber
&& \times \int_{\Space{R}{2n}} \exp \left( \pi ih (xy' -x'y) + \pi x'(-ia-b) - \pi x (-ia-b) + \pi y' (-ib+a) \right. \\ \nonumber
&& \left. \hspace{2cm} - \pi y (-ib+a) + \pi x'(ia-b) + \pi y' (ib+a) \right. \\ \nonumber
&& \left. \hspace{2cm} -\frac{\pi h}{2} [ (x'-x)^2 + (y'-y)^2 + x'^2 + y'^2]  -\frac{\pi}{h} (a^2 + b^2) \right) \, dx' \, dy'
\end{eqnarray}
Using (\ref{eq:waveletwithx}) this becomes
\begin{eqnarray} \nonumber
\lefteqn{\exp \left( 2\pi ihs - \pi x (-ia-b) - \pi y (a-ib) - \frac{\pi h}{2} (x^2 + y^2) - \frac{\pi}{h} (a^2 + b^2) \right.} \\ \nonumber
&& \left. \hspace{2cm} + \frac{\pi}{4h} \left[ (hx-ihy-2b)^2 + (hy+ihx+2a)^2 \right] \right) \\ \nonumber
&& = \exp \left( 2\pi ihs - \pi x (-ia-b) - \pi y (a-ib) - \frac{\pi h}{2} (x^2 + y^2) - \frac{\pi}{h} (a^2 + b^2) \right. \\ \nonumber
&& \left. \hspace{2cm} + \frac{\pi}{4h} [ 4ah(y+ix) - 4bh(x-iy)] +\frac{\pi}{h} (a^2 + b^2) \right) \\ \nonumber
&& = \exp \left( 2 \pi ihs - \pi x (-ia-b) - \pi y (a-ib) -\frac{\pi h}{2} (x^2 + y^2)  \right. \\ \nonumber
&& \left. \hspace{2cm} + \pi a (y+ix) -\pi b (x-iy) \right) \\ \nonumber
&& = \exp \left( 2\pi ihs + 2\pi i(ax+by) -\frac{\pi h}{2} (x^2 + y^2) \right).
\end{eqnarray}
\end{proof}
\begin{Defn}
For $h \in \Space{R}{} \setminus \{ 0 \}$ and $(a,b) \in \Space{R}{2n}$ define the system of coherent states $k_{(h,a,b)}$ by
\begin{equation} \nonumber
k_{(h,a,b)} (B) = \langle B* \vab,\vab \rangle = \int_{\Heisn} B(g) \overline{\lab (g)} dg.
\end{equation}
\end{Defn}
It is clear that the limit as $h \rightarrow 0$ of the kernels $\lab$ will just be the kernels  $\labo = e^{2\pi i (ax+by)}$. This proves that the system of coherent states we have constructed have the $(q,p)$ pure states, $k_{(0,a,b)}$, from equation (\ref{eq:pclasspurestates}), as their limit as $h \rightarrow 0$.
\begin{Thm}
If we have any $p$-observable $B \in L^1 (\Heisn)$ which is the $p$-mechanisation (see equation (\ref{eq:pmechmap})) of a classical observable, $f$, then
\begin{equation} \nonumber
\lim_{h \rightarrow 0} k_{(h,a,b)} (B) = k_{(0,a,b)} (B) = f(a,b).
\end{equation}
\end{Thm}
\begin{proof}
By the discussion prior to this theorem we clearly have pointwise convergence. Since $|B (s,x,y) \overline{l_{(h,a,b)} (s,x,y)} | \leq B(s,x,y) \overline{l_{(0,a,b)} (s,x,y)}$ for all $h$ the result follows by Lebesgue's dominated convergence theorem \cite[Thm. 1.11]{ReedSimon80}.
\end{proof}
We have used $p$-mechanics to rigorously prove, in a simpler way to previous attempts \cite{Hepp74}, the classical limit of coherent states.

Now we introduce the $p$-mechanical creation\index{creation operator} and annihilation operators\index{annihilation operator}. These operators are of great use in Chapters \ref{chap:forcedosc} and \ref{chap:cantransf}.
\begin{Defn} \label{def:creatandannihil}
The $p$-mechanical creation, $A_j^+$, and annihilation, $A_j^-$, distributions are defined as
\begin{eqnarray} \label{eq:defaplus}
A^+_j &=& \frac{1}{2\pi i} \left(\Partial{x_j} \zerodel - i  \Partial{y_j} \zerodel \right), \\ \label{eq:defaminus}
A^-_j &=& \frac{1}{2 \pi i} \left(\Partial{x_j} \zerodel + i  \Partial{y_j} \zerodel \right).
\end{eqnarray}
The $p$-mechanical creation and annihilation operators are left convolution by the creation and annihilation distributions respectively.
\end{Defn}
The creation and annihilation operators are the $p$-mechanisation of $q-ip$ and $q+ip$ respectively.
\begin{Lemma} \label{lem:csiseigfctnforcreat}
$\vab$ is an eigenfunction for $A_j^-$ with eigenvalue $(a_j+ib_j)$, that is
\begin{equation} \nonumber
A^-_j * \vab = (a_j+ib_j) \vab.
\end{equation}
\end{Lemma}
\begin{proof}
By equations (\ref{eq:deltageninvvfxl}) and (\ref{eq:deltageninvvfyl})
\begin{eqnarray} \nonumber
\lefteqn{ A_j^- * \vab } \\ \nonumber
&=&  \left( \mathfrak{X}^r_j + i \mathfrak{Y}_j^r \right) \vab \\ \nonumber
&=& \frac{1}{2 \pi i}\left( \Partial{x_j} + \frac{y_j}{2} \Partial{s} + i\Partial{y_j} - i \frac{x_j}{2} \Partial{s} \right) \vab \\ \nonumber
&=& \frac{1}{2\pi i} (\pi (ia_j - b_j) - \pi h x_j + \pi ih y_j + i \pi (ib_j + a_j) - \pi ih y_j + \pi h x_j) \vab \\ \nonumber
&=& \frac{1}{2\pi i} (2\pi i a_j - 2 \pi b_j) \\ \nonumber
&=& (a_j+ib_j)
\end{eqnarray}
\end{proof}

\begin{Remark}
\emph{Since the right invariant vector fields generate left shifts \cite{Taylor86} the $\hilbh$ coherent states could also be generated by the operator
$e^{(-bX+aY) \antid}$. From Theorems \ref{thm:schroheis} and \ref{Thm:timeevolofkernelisok} the corresponding operator for the kernel coherent states is
$e^{-b\ub{X}{\cdot} + a\ub{Y}{\cdot}}$.}
\end{Remark}
\section{The Interaction Picture} \label{sect:interactpic}

\indent

In the Schr\"odinger picture, time evolution is governed by the states and their equations $\frac{d v}{dt} = B_H * \antid v$; $\Fracdiffl{l}{t}= \ub{B_H}{l}{}$. In the Heisenberg picture, time evolution is governed by the observables and the equation $\frac{d B}{dt} = \ub{B}{B_H}{}$. In the interaction picture\index{interaction picture} we divide the time dependence between the states and the observables. This is suitable for systems with a Hamiltonian of the form $B_H = B_{H_0} + B_{H_1}$ where $B_{H_0}$ is time independent. The interaction picture has many uses in perturbation theory \cite[Sect. 14.4]{Kurunoglu62}.

Let a $p$-mechanical system have the Hamiltonian $B_H = B_{H_0} + B_{H_1}$ where $B_{H_0}$ is time independent and self-adjoint (see Definition \ref{def:self-adj}). We first describe the interaction picture for elements of $\hilbh$. Define  $\exp (t B_{H_0} \antid)$ as the operator on $\hilbh$ which is the exponential of the operator of applying $\antid$ then taking the convolution with $t B_{H_0}$ --- this is defined using Stone's theorem \cite[Sect. 8.4]{ReedSimon80}. Also by Stone's theorem we have $\frac{d}{dt} e^{t \antid B_{H_0}} v = B_{H_0} * \antid e^{t \antid B_{H_0}} v$. Now if $B$ is an observable let
\begin{equation} \label{eq:obsevolveininter}
\tilde{B} = \exp(-t B_{H_0} \antid) B \exp(+ t B_{H_0} \antid)
\end{equation}
then
\begin{eqnarray} \nonumber
\Fracdiffl{\tilde{B}}{t} &=& - B_{H_0}*\tilde{B} \antid + \tilde{B}*B_{H_0} \antid \\ \nonumber
&=& \ub{\tilde{B}}{B_{H_0}}.
\end{eqnarray}
If $v \in \hilbh$, define $\tilde{v}(t) = (\exp (- t B_{H_0} \antid )) v(t)$. Note for $\tilde{v}$ there is time dependence in both $v$ and $\exp (-t B_{H_0} \antid)$ so when differentiating with respect with to $t$ we get
\begin{eqnarray} \label{eq:gentimeevolofinter}
\frac{d}{dt} \tilde{v} &=& \frac{d}{dt} ( \exp(-t B_{H_0} \antid) v ) \\ \nonumber
&=& - B_{H_0} * \antid \tilde{v} + \exp(-t B_{H_0} \antid) ( (B_{H_0} + B_{H_1}) * \antid v) \\ \nonumber
&=& - B_{H_0} * \antid \tilde{v} + B_{H_0} * \antid \exp(-t B_{H_0}\antid) v + \exp(-t B_{H_0} \antid)  B_{H_1}*\antid v \\ \nonumber
&=& (\exp(-t B_{H_0} \antid) B_{H_1} * \antid \exp(t B_{H_0} \antid ))(\tilde{v}) \\ \nonumber
&=& (\exp(-t B_{H_0} \antid) \antid B_{H_1} * \exp(t B_{H_0} \antid ))(\tilde{v}).
\end{eqnarray}
Now we describe the interaction picture for a state defined by a kernel $l$. Define
\begin{equation} \nonumber
\tilde{l}(t) = e^{- \ub{B_{H_0}}{\cdot}{}t} l (t) =  \exp(-t B_{H_0} \antid ) l(t) \exp(t B_{H_0} \antid)
\end{equation}
so conversely
\begin{equation}
l = \exp (t B_{H_0} \antid ) \tilde{l} \exp (-t  B_{H_0} \antid ).
\end{equation}
Differentiating with respect to $t$ gives us
\begin{eqnarray} \nonumber
\Fracdiffl{\tilde{l}}{t} &=& -B_{H_0} * \antid \tilde{l} + \exp(-t B_{H_0} \antid) \ub{B_{H_0} + B_{H_1}}{l}{} \exp(t B_{H_0} \antid) + \tilde{l} * B_{H_0} \antid \\ \nonumber
&=& -B_{H_0} * \antid \exp (-t B_{H_0} \antid) l \exp (t B_{H_0} \antid) \\ \nonumber
&& \hspace{0.75cm} + \exp (-t B_{H_0} \antid) \ub{B_{H_0} + B_{H_1}}{l} \exp (t B_{H_0} \antid) \\ \nonumber
&& \hspace{1.5cm} + \exp (-t B_{H_0} \antid) l \exp (t B_{H_0} \antid) * B_{H_0} \antid \\ \nonumber
&=& \exp(-t B_{H_0} \antid) \ub{B_{H_1}}{l}{} \exp(t B_{H_0} \antid) \\ \nonumber
&=& \exp(-t B_{H_0} \antid)  B_{H_1} *  \exp(t B_{H_0} \antid) \tilde{l} \exp (-t  B_{H_0} \antid) \exp(t B_{H_0} \antid) \antid \\ \nonumber
&& - \exp(-t B_{H_0} \antid) \exp (t B_{H_0} \antid) \tilde{l} \exp(-t B_{H_0} \antid) * B_{H_1}  \exp(t B_{H_0}\antid )  \antid \\ \label{eq:kernelsatesininteract}
&=& \ub{\exp(-t B_{H_0} \antid) B_{H_1} \exp(t B_{H_0} \antid)} {\tilde{l}}{}.
\end{eqnarray}
This shows us how interaction states evolve with time. Note that if we take $B_{H_0}=B_H$ we have the Heisenberg picture, while if we take $B_{H_1} = B_H$ we have the Schr\"odinger picture. The $p$-mechanical interaction picture here in its abstract form seems very dry, but in Section \ref{sect:interpictofforcedosc} we will see that it is extremely useful in studying the forced oscillator. Also in Section \ref{sect:interpictofforcedosc} we will see how the $p$-mechanical interaction picture can produce simpler calculations than those given by the usual quantum interaction picture.

\section{Relationships Between $\ltworn$ and $\hilbh$} \label{sect:relbetltwoandhh}

\indent

In this subsection we present a kernel which will map an element of $\ltworn$ into an element of $\hilbh$. The standard mathematical formulation of quantum mechanics is given by operators on the Hilbert space $\ltworn$. If we look at relations between $\hilbh$ and $\ltworn$ we will get relations between $p$-mechanics and the standard formulation of quantum mechanics.
\begin{Thm}
$\ltworn$ can be mapped into $\hilbh$ by
\begin{equation}
\psi \mapsto e^{2\pi ihs} \int \psi (\xi) K_{I}^{\hilbh} (x,y,\xi) \, d\xi
\end{equation}
where
\begin{equation} \nonumber
K_I^{\hilbh} (x,y,\xi) = \exp \left( -2\pi \xi (y+ix) -h\pi y^2 + \pi ih xy - \frac{\pi}{h} \xi^2 \right).
\end{equation}
\end{Thm}
\begin{proof}
Theorem \ref{thm:fockandltwointertwine} shows that $\ltworn$ is mapped into $\fock$ by the kernel
\begin{equation}
K_I^{F} (q,p,\xi) = \left( \frac{2}{h} \right)^{n/4} e^{\frac{4\pi i}{h} (p\xi + qp)} e^{-\frac{\pi}{h}(\xi + 2q)^2}.
\end{equation}
Furthermore equation (\ref{hh}) shows us that the inverse Fourier transform followed by multiplication by $e^{2\pi ihs}$ maps $\fock$ into $\hilbh$. So it is clear that the combination of integration next to $K_I^{F}$ and the inverse Fourier transform followed by multiplication by $e^{2\pi ihs}$ will give us a map from $\ltworn$ to $\hilbh$. So if $\psi$ is in $\ltworn$ then
\begin{equation} \label{eq:startofltwotohhkern}
e^{2\pi ihs} \left( \frac{2}{h} \right)^{n/4} \int \psi (\xi) e^{\frac{4\pi i}{h} (p\xi + qp)} e^{-\frac{\pi}{h}(\xi + 2q)^2} e^{2\pi i (q.x+p.y)} \, d\xi \, dq \, dp
\end{equation}
is the associated element of $\hilbh$. The function we are integrating is integrable since we are taking the Fourier transform of an $L^2 (\Space{R}{2n})$ function (see Theorem \ref{thm:tissquareint}) so we can use Fubini's Theorem (see Theorem \ref{thm:fubini}) to interchange the order of integration.
Using Fubini's Theorem the integral in (\ref{eq:startofltwotohhkern}) becomes
\begin{eqnarray*}
\lefteqn{\int \psi (\xi) \exp \left( \frac{4\pi i}{h} p \xi -\frac{\pi}{h} \xi^2 + 2\pi i py \right)} \\ \nonumber
&& \hspace{2cm} \times \left( \int \exp \left( q \left( \frac{4\pi i}{h} p - \frac{4\pi}{h} \xi + 2\pi i x \right) - \frac{4\pi}{h} q^2 \right) \, dq \right) \, dp \, d\xi.
\end{eqnarray*}
Using equation (\ref{eq:waveletwithx}) the above formula becomes
\begin{eqnarray*}
\lefteqn{\int \psi (\xi) \exp \left( \frac{4\pi i}{h} p \xi -\frac{\pi}{h} \xi^2 + 2\pi i py \right) \exp \left( \frac{\pi}{h} \left( ip-\xi + i \frac{h}{2} x \right)^2 \right) \, dp \, d \xi } \\ \nonumber
&=& \int \psi (\xi) \exp \left( -\frac{\pi}{h}\xi^2 + \frac{\pi}{h} \left( \xi - i \frac{h}{2}x \right)^2 \right) \\ \nonumber
&& \hspace{1.5cm} \times \left( \int \exp \left( p \left( \frac{2\pi i}{h} \xi + 2\pi i y - \pi x \right) - \frac{\pi}{h} p^2 \right) \, dp \right) \, d \xi \\ \nonumber
&=& \int \psi (\xi) \exp \left( -\frac{\pi}{h}\xi^2 + \frac{\pi}{h} \left( \xi - i \frac{h}{2}x \right)^2 \right) \exp \left( -h \pi \left( \frac{\xi}{h} + y + i\frac{x}{2} \right)^2 \right) \, d\xi \\
&=& \int \psi (\xi) \exp \left( -2\pi i \xi x - \frac{\pi}{h} \xi^2 - 2\pi \xi y - h \pi y^2 + ih \pi xy \right) \, d\xi \\
&=& \int \psi (\xi) \exp \left( -2 \pi \xi (y+ix) - h\pi y^2 + ih\pi xy - \frac{\pi}{h} \xi^2 \right) \, d\xi.
\end{eqnarray*}
\end{proof}
By equation (\ref{eq:relationbetweenkernelandvector}) we have a map from $\hilbh$ to $\sokh$ so combining this with the above construction we have a map from $\ltworn$ into the space of kernels.

\section{The Rigged Hilbert Spaces Associated \newline with $\hilbh$ and $\fock$} \label{sect:rhs}

\indent

Rigged Hilbert spaces\index{rigged Hilbert spaces} (also known as Gel'fand triples\index{Gel'fand triples} \cite{GelfandVilenkin77}) were introduced in quantum mechanics to help deal with problems which arose from the presence of unbounded operators. Gel'fand and his collaborators discovered rigged Hilbert spaces as a tool for dealing with operators on infinite dimensional vector spaces \cite[Chap. 1, Sect. 4]{GelfandVilenkin77}. Roberts \cite{Roberts66}, Bohm \cite{Bohm01} and Antoine \cite{AntoineVause81} in the 1960s realised that rigged Hilbert spaces could be used to rigorously define Dirac's "bra and ket" formulation of quantum mechanics.

In quantum mechanics the position and momentum observables have continuous spectra. From Example \ref{exam:classicalposition} the $p$-mechanisation of the classical position and momentum observables are the distributions $\frac{1}{2\pi i} \Partial{x} \zerodel$ and $\frac{1}{2\pi i} \Partial{y} \zerodel$ respectively. When realised as operators of convolution on $\hilbh$ they are the following operators
\begin{eqnarray} \label{eq:posonhh}
\mathcal{P} (q) * v &=& \frac{1}{2\pi i} \left( \Partial{x} + \pi ih y \right) v \\ \label{eq:momonhh}
\mathcal{P} (p) * v &=& \frac{1}{2\pi i} \left( \Partial{y} - \pi ih x\right) v.
\end{eqnarray}
These operators clearly are not defined on the whole of $\hilbh$, and do not have any eigenfunctions in $\hilbh$, and so as was mentioned before we need the concept of rigged Hilbert spaces. The idea of a rigged Hilbert space is to start with the original Hilbert space, $H$, then choose a subset, $\Phi$, on which the operator is defined. After this we must also consider the dual space to $\Phi$, denoted $\Phi'$, which will contain the original space. This gives us a triple of vector spaces
\begin{equation} \label{eq:genriggedhs}
\Phi \subset H \subset \Phi'.
\end{equation}
A rigged Hilbert space is a triple as in (\ref{eq:genriggedhs}) where the space $\Phi$ is nuclear \cite[Chap. 1 Sect. 3]{GelfandVilenkin77}.
Suppose $A$ is an operator on the Hilbert space in question then a generalised eigenfunction\index{generalised eigenfunction} of $A$ with eigenvalue $\lambda$ is an element $\psi \in \Phi'$ such that
\begin{equation} \label{eq:geneigenfctn}
\langle A \phi , \psi \rangle = \lambda \langle \phi , \psi \rangle
\end{equation}
for any $\phi \in \Phi$. The brackets $\langle , \rangle$ in the above equation denote the evaluation of an element of $\Phi$ on the left by a functional in $\Phi'$ on the right.
\begin{Thm} \cite[Chap. 1, Sect. 4.5, Thm 5]{GelfandVilenkin77}
A self-adjoint operator in a rigged Hilbert space has a complete system of generalised eigenvectors corresponding to real eigenvalues.
\end{Thm}
In the $\ltworn$ formulation of quantum mechanics the chosen triple is
\begin{equation} \label{eq:riggedforltworn}
\mathcal{S}(\Space{R}{n}) \subset \ltworn \subset \mathcal{S}'(\Space{R}{n})
\end{equation}
where $\mathcal{S}(\Space{R}{n})$ and $\mathcal{S}'(\Space{R}{n})$ are defined in Appendix \ref{app:distributions}. For $\fock$ an associated rigged Hilbert space is
\begin{equation}\index{$Exp_f$}\index{$\focko$}
Exp_f \subset \fock \subset \focko
\end{equation}
where
\begin{equation} \nonumber
\focko = \{ f \in C^{\infty} (\Space{R}{2n}) \, : \, D_j^h f = 0 \hspace{0.3cm} \textrm{for}  \hspace{0.3cm} j=1, \cdots ,n \}
\end{equation}
(the operator $D_j^h$ is defined in equation (\ref{eq:ftwoohpolarz})) and
\begin{equation} \nonumber
Exp_f = \left\{ f \in \focko \, : \, \exists c,a \in \Space{R}{} \hspace{0.3cm} \textrm{such that} \hspace{0.3cm} | f(q,p)| \leq ce^{a \sqrt{q^2 +p^2}} \hspace{0.2cm} \forall q,p \in \Space{R}{n}  \right\}.
\end{equation}
It can be shown that $\focko$ and $Exp_f$ are both nuclear and duals of each other\footnote{The function $f \in Exp_f$ associated with the functional $\mu \in \focko'$ is given by $f(a,b) = \left\langle \mu (q,p) , e^{\langle (a,b)\cdot(q,p)\rangle} \right\rangle$ where $\cdot$ is the dot product on $\Space{R}{2n}$. This map is known as either the Fourier-Laplace transform or the Fourier-Borel transform \cite{Treves67}.} \cite{AntoineVause81,Antoine99}. Similarly a rigged Hilbert space for $\hilbh$ is
\begin{equation} \nonumber
Exp_h \subset \hilbh \subset \hilbho
\end{equation}
where\index{$\hilbho$}
\begin{equation}
\hilbho = \{ e^{2\pi ihs} f(x,y): E_j^h f = 0 \hspace{0.3cm} \textrm{for} \hspace{0.3cm} j = 1 , \cdots , n \}
\end{equation}
(the operator $E_j^h$ is defined in equation (\ref{eq:defofejh})) and\index{$Exp_h$}
\begin{eqnarray} \nonumber
\lefteqn{Exp_h = \left\{ v(s,x,y) \in \hilbho :  \right.} \\ \nonumber
&&  \left. \hspace{2.2cm} \exists c,a \in \Space{R}{} \hspace{0.3cm} \textrm{such that} \hspace{0.3cm} | v(s,x,y)| \leq ce^{a \sqrt{x^2 +y^2}} \hspace{0.2cm} \forall x,y \in \Space{R}{n}  \right\}.
\end{eqnarray}
Now we can find the generalised eigenfunctions for position and momentum in $\hilbho$. The generalised eigenfunctions for position are
\begin{equation}
\exp \left( 2\pi ihs + 2\pi \xi (y+ix) - \pi h y^2 - \pi ih xy - \frac{\pi}{h} \xi^2 \right)
\end{equation}
with eigenvalue $\xi$ -- there is one of these eigenfunctions for every $\xi \in \Space{R}{}$. The generalised eigenfunctions for momentum are
\begin{equation} \nonumber
\exp \left( 2\pi ihs + 2\pi \xi (x+iy) - \pi h x^2 + \pi ih xy - \frac{\pi}{h} \xi^2 \right)
\end{equation}
also with eigenvalue $\xi$ -- again there is one of these eigenfunctions for every $\xi \in \Space{R}{}$. It can be easily verified that both of these functions are in $\hilbho$. It is clear that both of these operators have the continuous spectrum $\Space{R}{}$ --- which is what is required.

\chapter{Examples: The Harmonic Oscillator and the Forced Oscillator} \label{chap:forcedosc}

\indent

In this chapter we look at two examples: the harmonic oscillator\index{harmonic oscillator} and the forced oscillator\index{forced oscillator}. The $p$-mechanical harmonic oscillator has already been discussed in \cite{Kisil02} and \cite{Kisil02.1}. In Section \ref{sect:harmosc} we present a slightly different approach to the problem and develop some new insights. For the rest of the chapter we apply the theory from Chapters \ref{chap:pmechandhn} and \ref{chap:staesandpic}  to the example of the forced oscillator. It is shown that both the quantum and classical pictures are derived from the same source.

The classical forced oscillator has been studied in great depth for a long time --- for a description of this see \cite{Goldstein80} and \cite{Jose98}. The quantum case has also been heavily researched --- see for example \cite[Sect 14.6]{Merzbacher70}, \cite{Martinez83}. Of interest in the quantum case has been the use of coherent states -- this is described in \cite{Perelomov86}. Here we extend these approaches to give a unified quantum and classical solution of the problem based on the $p$-mechanical framework.

\section{The Harmonic Oscillator} \label{sect:harmosc}

\indent

Throughout this section we assume that the forced and harmonic oscillators are one dimensional -- the extension to $n$ dimensions is straight forward. The classical Hamiltonian of the harmonic oscillator\index{harmonic oscillator} with frequency $\omega$ and mass $m$ is
\begin{equation} \label{eq:classhoham}
H(q,p) = \frac{1}{2} \left( m \omega^2 q^2 + \frac{1}{m} p^2 \right).
\end{equation}
This is a $C^{\infty}$ function which can be realised as an element of $\mathcal{S}'(\Space{R}{2n})$. The $p$-mechanisation (see Equation (\ref{eq:pmechmap})) of this is the $p$-mechanical harmonic oscillator Hamiltonian\footnote{ $\zerodelxtwo$ is used\index{$\delta (s) \delta^{(1)} (x) \delta (y)$} to denote the distribution $\frac{\partial^2}{\partial x^2} \zerodel$.}
\begin{equation} \label{eq:pmechharmoscham}
-\frac{1}{8\pi^2} \left( m \omega^2 \zerodelxtwo + \frac{1}{m} \zerodelytwo \right)
\end{equation}
which is a distribution in $\mathcal{S}' (\Heisn)$. The $p$-mechanical harmonic oscillator Hamiltonian has the equivalent form
\begin{equation} \nonumber
B_H = \frac{1}{2m} (A^+ * A^- + i \omega m^2 \zerodelsone ).
\end{equation}
The distributions $A^+$ and $A^-$ were defined in equations (\ref{eq:defaplus}) and (\ref{eq:defaminus}); for the purposes of this chapter we give them a slightly different definition
\begin{eqnarray} \nonumber
A^+ &=& \frac{1}{2\pi i} (m \omega \zerodelxone - i \zerodelyone ) \\ \nonumber
A^- &=& \frac{1}{2\pi i}(m\omega \zerodelxone + i \zerodelyone).
\end{eqnarray}
We denote the $p$-mechanical normalised eigenfunction\index{eigenfunctions!harmonic oscillator} with eigenvalue $n$ of the harmonic oscillator by $v_n \in \hilbh$ (note here that the coherent state $v_{(h,0,0)}=v_0$); it has the form
\begin{eqnarray} \nonumber
v_n &=& \left( \frac{1}{n!} \right)^{1/2} (\antid A^+)^n * v_{(h,0,0)} \\ \nonumber
&=& \left( \frac{1}{n!} \right)^{1/2} \left( \frac{h}{2} \right)^n e^{2\pi i hs} (\omega m y+ix)^n  \exp \left( - \frac{\pi h}{2} \left( \frac{x^2}{\omega m} + y^2 \omega m \right) \right).
\end{eqnarray}
It can be shown by a trivial calculation that the creation and annihilation operators (see Definition \ref{def:creatandannihil}) raise and lower the eigenfunctions of the harmonic oscillator respectively. That is
\begin{equation}
A^+ * v_n = v_{n+1} \hspace{1cm} \textrm{and} \hspace{1cm} A^- * v_n = v_{n-1}.
\end{equation}
It is important to note that these states are orthogonal under the $\hilbh$ inner product defined in equation (\ref{hhip}).

In \cite[Eq. 4.14]{Kisil02.1} it is shown that the $p$-dynamic equation for an arbitary $p$-mechanical observable in this system is
\begin{eqnarray*} \nonumber
\Fracdiffl{B}{t} &=& \ub{B}{B_H}{} \\
&=& \omega^2 m y \Fracpartial{B}{x} - \frac{x}{m} \Fracpartial{B}{y}
\end{eqnarray*}
which has solution
\begin{equation} \label{eq:timeevolofharmoscinpm}
B(t;s,x,y) = B_0 \left( s, x \cos (\omega t) + m\omega y \sin (\omega t), -\frac{1}{m\omega} x \sin (\omega t)  + y \cos (\omega t) \right).
\end{equation}

\section{The $p$-Mechanical Forced Oscillator: The Solution and Relation to Classical Mechanics} \label{sect:forcedoscclass}

\indent

The classical Hamiltonian for an oscillator of frequency $\omega$ and mass $m$ being forced by a real function of a real variable $z(t)$ is
\begin{equation} \nonumber
H(t,q,p) = \frac{1}{2} \left( m \omega^2 q^2 + \frac{1}{m} p^2 \right) - z(t) q.
\end{equation}
 Then for any observable $f \in C^{\infty} (\Space{R}{2n})$ the dynamic equation in classical mechanics is
\begin{eqnarray} \nonumber
\Fracdiffl{f}{t} &=& \{ f,H \} \\ \label{eq:classdyneqnfo}
&=& \frac{p}{m} \Fracpartial{f}{q} - \omega^2 m q \Fracpartial{f}{p} + z(t) \Fracpartial{f}{p}.
\end{eqnarray}
Through the procedure of $p$-mechanisation (see (\ref{eq:pmechmap})) we get the $p$-mechanical forced oscillator Hamiltonian to be
\begin{eqnarray} \nonumber
B_H (t;s,x,y) &=& -\frac{1}{8 \pi^2} \left( m \omega^2 \zerodelxtwo + \frac{1}{m} \zerodelytwo \right) \\ \label{eq:forcedoscpmham}
&& \qquad- \frac{z(t)}{2 \pi i} \zerodelxone.
\end{eqnarray}
From equations (\ref{eq:deltageninvvfxl}), (\ref{eq:deltageninvvfyl}) and (\ref{eq:pdyneqn}) the $p$-dynamic equation for an arbitrary $p$-observable $B$ is
\begin{eqnarray} \nonumber
\Fracdiffl{B}{t} &=& \ub{B}{B_H}{}\\ \nonumber
&=& (B*B_H - B_H * B) \antid \\ \nonumber
&=& \left[-\frac{1}{8\pi^2} \left( m \omega^2 ((\mathfrak{X}^l)^2 - (\mathfrak{X}^r)^2) + \frac{1}{m} ((\mathfrak{Y}^l)^2 - (\mathfrak{Y}^r)^2) \right) \right. \\ \nonumber
&& \left. \hspace{2cm} - \frac{z(t)}{2\pi i} (\mathfrak{X}^l - \mathfrak{X}^r) \right] \antid B \\ \nonumber
&=& \left[ -\frac{1}{8\pi^2} \left( m \omega^2 \left( \left(\Partial{x} - \frac{y}{2} \Partial{s}\right)^2 - \left(\Partial{x} + \frac{y}{2} \Partial{s}\right)^2 \right) \right. \right. \\ \nonumber
&& \left. \left. \hspace{2.5cm} + \frac{1}{m} \left(\left(\Partial{y} + \frac{x}{2} \Partial{s}\right)^2 - \left(\Partial{y} - \frac{x}{2} \Partial{s}\right)^2\right) \right) \right. \\ \nonumber
&& \left. \hspace{1.5cm} -\frac{z(t)}{2\pi i} \left(\Partial{x} - \frac{y}{2} \Partial{s} - \Partial{x} - \frac{y}{2} \Partial{s} \right) \right] \antid B.
\end{eqnarray}
Using the fact that $ \Partial{s} \antid = 4\pi^2 I$ the $p$-dynamic equation for the forced oscillator is
\begin{equation} \label{eq:pdynamiceqnforforcedosc}
\Fracdiffl{B}{t} =  \omega^2 m y \Fracpartial{B}{x} - \frac{x}{m} \Fracpartial{B}{y} - 2\pi i y z(t) B.
\end{equation}
\begin{Thm} \label{thm:forcedoscsoln}
The following expression is a solution of the $p$-dynamic equation for the forced oscillator (\ref{eq:pdynamiceqnforforcedosc})
\begin{eqnarray} \label{eq:solnofpmfo}
\lefteqn{B (t;s,x,y)} \\ \nonumber
&& = \exp \left( -2 \pi i \left( \frac{1}{m \omega} \int_{0}^{t} z(\tau) \sin (\omega \tau ) \, d\tau X(t) + \int_{0}^{t} z(\tau ) \cos (\omega \tau ) \, d \tau Y(t) \right) \right) \\ \nonumber
&& \qquad \qquad \times B (0;s,X(t),Y(t)),
\end{eqnarray}
where
\begin{eqnarray} \nonumber
X(t) &=& x \cos(\omega t) + m \omega y \sin (\omega t), \\ \nonumber
Y(t) &=& -\frac{x}{m\omega} \sin (\omega t) + y \cos (\omega t).
\end{eqnarray}
\end{Thm}
\begin{proof}
We have that
\begin{equation} \label{eq:difftxinforcedosc}
\Fracdiffl{X}{t} =  -\omega x \sin (\omega t) + m \omega^2 y \cos (\omega t) = \omega^2 m y \Fracpartial{X}{x} - \frac{x}{m} \Fracpartial{X}{y}
\end{equation}
\begin{equation} \label{eq:difftyinforcedosc}
\Fracdiffl{Y}{t} =  -\frac{x}{m} \cos (\omega t) - y \omega \sin ( \omega t) = \omega^2 m y \Fracpartial{Y}{x} - \frac{x}{m} \Fracpartial{Y}{y}.
\end{equation}
Equations (\ref{eq:difftxinforcedosc}) and (\ref{eq:difftyinforcedosc}) imply that for any observable $B$
\begin{equation} \label{eq:harmbitofforced}
\Fracdiffl{B(s,X(t),Y(t))}{t} = \omega^2 my \Fracpartial{B(s,X(t),Y(t))}{x} - \frac{x}{m} \Fracpartial{B(s,X(t),Y(t))}{y}.
\end{equation}
Now differentiating expression (\ref{eq:solnofpmfo}) with respect to time gives us
\begin{eqnarray} \nonumber
\lefteqn{\Fracdiffl{ B(t;s,x,y)}{t}} \\ \nonumber
&& = -2\pi i \left[ \frac{1}{m\omega} z(t) \sin (\omega t) X(t) + \frac{1}{m\omega} \int_0^t z(\tau) \sin ( \omega \tau) \, d \tau \Fracdiffl{X}{t} \right. \\ \nonumber
&& \left. \hspace{1cm} + z(t) \cos (\omega t) Y(t) + \int_0^t z(\tau) \cos (\omega \tau) \, d\tau \Fracdiffl{Y}{t} \right] B(t;s,x,y) \\ \nonumber
&& \hspace{1cm} + \exp (F(t,x,y)) \Fracdiffl{B(s,X(t),Y(t))}{t} \\ \nonumber
&& = -2\pi i \left[ y z(t) + \frac{1}{m\omega} \int_0^t z(\tau) \sin ( \omega \tau) \, d \tau \Fracdiffl{X}{t} + \int_0^t z(\tau) \cos (\omega \tau) \, d\tau \Fracdiffl{Y}{t} \right] \\ \nonumber
&& \hspace{3cm} \times B(t;s,x,y) \\ \nonumber
&& \hspace{1.5cm} + \exp (F(t,x,y)) \Fracdiffl{B(s,X(t),Y(t))}{t}
\end{eqnarray}
where
\begin{equation} \nonumber
F(t,x,y) = \left( -2 \pi i \left( \frac{1}{m \omega} \int_{0}^{t} z(\tau) \sin (\omega \tau ) \, d\tau X(t) + \int_{0}^{t} z(\tau ) \cos (\omega \tau ) \, d \tau Y(t) \right) \right).
\end{equation}
Furthermore
\begin{eqnarray} \nonumber
\lefteqn{\Fracpartial{B(t;s,x,y)}{x} =} \\ \nonumber
&& -2\pi i \left( \frac{1}{m\omega} \int_0^t z(\tau) \sin (\omega \tau) \, d\tau \Fracpartial{X}{x} + \int_0^t z(\tau) \cos (\omega \tau) \, d\tau \Fracpartial{Y}{x} \right) B(t;s,x,y)\\ \nonumber
&& \hspace{2cm} + \exp (F(t,x,y)) \Fracpartial{B(s,X(t),Y(t))}{x}
\end{eqnarray}
and
\begin{eqnarray} \nonumber
\lefteqn{\Fracpartial{B(t;s,x,y)}{y} =} \\ \nonumber
 && -2\pi i \left( \frac{1}{m\omega} \int_0^t z(\tau) \sin (\omega \tau) \, d\tau \Fracpartial{X}{y} + \int_0^t z(\tau) \cos (\omega \tau) \, d\tau \Fracpartial{Y}{y} \right) B(t;s,x,y) \\ \nonumber
&& \hspace{3cm} + \exp (F(t,x,y)) \Fracpartial{B(s,X(t),Y(t))}{y}.
\end{eqnarray}
If we substitute this into (\ref{eq:pdynamiceqnforforcedosc}) and equate the coefficients of \newline $\int_0^t z(\tau) \sin (\omega \tau) \, d\tau B(t;s,x,y)$, $\int_0^t z(\tau) \cos (\omega \tau) \, d\tau B(t;s,x,y)$ then using equations  (\ref{eq:difftxinforcedosc}), (\ref{eq:difftyinforcedosc}) and (\ref{eq:harmbitofforced}) we get the required result.
\end{proof}
Now we show that if we take the $p$-mechanisation of a classical observable, $f$, then the one dimensional representation of (\ref{eq:solnofpmfo}) will give the classical flow for the forced oscillator.
\begin{eqnarray} \nonumber
\lefteqn{ f(t;q,p) } \\ \nonumber
&& = \int_{\Space{R}{2n+1}} B (t;s,x,y) e^{-2\pi i (q.x+p.y)} \, ds \, dx \, dy  \\ \nonumber
&& = \int_{\Space{R}{2n+1}} \exp \left( - 2 \pi i \left( \frac{1}{m \omega} \int_{0}^{t} z(\tau) \sin (\omega \tau ) \, d\tau X(t) \right. \right. \\ \nonumber
&& \hspace{4.5cm} \left. \left. + \int_{0}^{t} z(\tau ) \cos (\omega \tau ) \, d \tau Y(t) \right) \right) \\ \nonumber
&& \qquad \qquad \times  \exp (- 2\pi i (q.x +p.y) ) \, B(0;s,X(t),Y(t)) \, ds \, dx \, dy.
\end{eqnarray}
Making the change of variable $u=X(t)$ and $v=Y(t)$ the above equation becomes
\begin{eqnarray} \nonumber
\lefteqn{\int_{\Space{R}{2n+1}} \exp \left( -2 \pi i \left( \frac{1}{m \omega} \int_{0}^{t} z(\tau) \sin (\omega \tau ) \, d\tau u + \int_{0}^{t} z(\tau ) \cos (\omega \tau ) \, d \tau v \right) \right) } \\ \nonumber
&& \times \exp \left(-2\pi i \left( q.(u \cos(\omega t) -vm\omega \sin(\omega t)) +p.\left(\frac{u}{m\omega} \sin(\omega t) +v \cos(\omega t) \right) \right) \right) \\ \nonumber
&& \qquad \qquad \times B (0;s,u,v) \, ds \, du \, dv \\ \nonumber
&& = \int_{\Space{R}{2n+1}} \exp \left( -2\pi i u. \left( q cos(\omega t) + \frac{p}{m \omega} \sin (\omega t) + \frac{1}{m \omega} \int_{0}^{t} z(\tau) \sin (\omega \tau ) \, d\tau \right) \right) \\ \nonumber
&& \qquad \times \exp \left( - 2 \pi i v. \left( - q m\omega \sin (\omega t) +p \cos (\omega t) + \int_{0}^{t} z(\tau ) \cos (\omega \tau ) \, d \tau \right) \right) \\ \nonumber
&& \qquad \qquad \times B (0;s,u,v) \, ds \, du \, dv \\ \nonumber
&& = f \left( 0; q \cos(\omega t) + \frac{p}{m\omega} \sin (\omega t) + \frac{1}{m \omega} \int_{0}^{t} z(\tau) \sin (\omega \tau ) \, d\tau , \right.
\\ \label{eq:classicalflowfo}
&& \left. \qquad \qquad - q m\omega \sin (\omega t) +p \cos (\omega t) + \int_{0}^{t} z(\tau ) \cos (\omega \tau ) \, d \tau \right).
\end{eqnarray}
This flow satisfies the classical dynamic equation (\ref{eq:classdyneqnfo}) for the forced oscillator --- this is shown in \cite{Jose98}. Similarly if we take an infinite dimensional representation of $B(t;s,x,y)$ we will get the quantum observable which is $\rho_h (B(0;s,x,y))$ after spending time $t$ in the forced oscillator system.

\section{A Periodic Force and Resonance} \label{sect:periodandres}

\indent

In classical mechanics the forced oscillator is of particular interest if we take the external force to be $z(t)=Z_0 \cos(\Omega t)$ \cite{Jose98}, that is the oscillator is being driven by a harmonic force of constant frequency $\Omega$ and constant amplitude $Z_0$. First we define the functions $\psi_1 (\omega , \Omega ,t)$ and $\psi_2 (\omega , \Omega ,t)$ as
\begin{eqnarray} \nonumber
\psi_1 (\omega , \Omega ,t) &=& \int_0^t \cos (\Omega \tau) \sin(\omega \tau) \, d \tau , \\ \nonumber
\psi_2 (\omega , \Omega ,t) &=& \int_0^t \cos (\Omega \tau) \cos(\omega \tau) \, d \tau.
\end{eqnarray}
By a simple calculation we have for $\Omega \neq \omega$
\begin{equation} \label{eq:sinint}
\psi_1 (\omega , \Omega ,t) = \frac{2}{(\Omega^2 -\omega^2 )}  [ \Omega \cos (\Omega t) \cos (\omega t) + \omega \sin (\Omega t) \sin (\omega t) - \Omega ]
\end{equation}
\begin{equation} \label{eq:cosint}
\psi_2 (\omega , \Omega ,t) = \frac{2}{(\Omega^2 -\omega^2 ) } [ -\Omega \sin (\Omega t) \cos (\omega t) + \omega \cos (\Omega t) \sin (\omega t) ].
\end{equation}
By substituting these two equations into (\ref{eq:solnofpmfo}) we get the $p$-mechanical solution for the oscillator being forced by a periodic force as
\begin{eqnarray} \nonumber
\lefteqn{B (t;s,x,y)} \\ \label{eq:solofforcedosctwo}
&& = \exp \left( -2 \pi i \left( \frac{1}{m \omega} \psi_1 (\omega , \Omega ,t) X(t) +\psi_2 (\omega , \Omega ,t) Y(t) \right) \right) \\ \nonumber
&& \qquad \qquad \times B (0;s,X(t),Y(t)),
\end{eqnarray}
where $X(t)$ and $Y(t)$ are as defined in Theorem \ref{thm:forcedoscsoln}. We can see that the solution is the flow of the $p$-mechanical unforced oscillator multiplied by an exponential term which is also periodic. However the argument of this exponential term will become infinitely large as $\Omega$ comes close to $\omega$. If we substitute $(\ref{eq:sinint})$ and $(\ref{eq:cosint})$ into $(\ref{eq:classicalflowfo})$ we obtain a classical flow which is periodic but with a singularity as $\Omega$ tends towards $\omega$. These two effects show a correspondence between classical and $p$-mechanics. When $\Omega = \omega$ the functions $\psi_1 (\omega, \Omega , t)$ and $\psi_2 (\omega , \Omega ,t)$ become
\begin{eqnarray} \label{eq:sinint2}
\psi_1 (\omega,\Omega,t) = \int_0^t \cos (\omega \tau) \sin(\omega \tau) \, d \tau &=& \frac{1-\cos(2\omega t)}{4 \omega} \\ \label{eq:cosint2}
\psi_2 (\omega,\Omega,t) = \int_0^t \cos (\omega \tau) \cos(\omega \tau) \, d \tau &=& \frac{t}{2} + \frac{1}{4 \omega} \sin(2\omega t).
\end{eqnarray}
Now when these new values are substituted into (\ref{eq:solofforcedosctwo}) the argument of the exponential term will expand without bound as $t$ becomes large. When (\ref{eq:sinint2}) and (\ref{eq:cosint2}) are substituted into $(\ref{eq:classicalflowfo})$ the classical flow will also expand without bound --- this is the effect of resonance\index{resonance}.

\section{The Interaction Picture of the Forced Oscillator} \label{sect:interpictofforcedosc}

\indent

We now use the interaction picture\index{interaction picture} (see Section \ref{sect:interactpic}) to get a better description of the $p$-mechanical forced oscillator and also to demonstrate some of the quantum effects. The interaction picture has already been used in quantum mechanics \cite[Sect. 14.6]{Merzbacher70} to analyse the forced oscillator; we show in this section how $p$-mechanics can simplify some of these calculations. In $p$-mechanics we get a solution for the problem directly without any need for a time ordering operator \cite[Eq. 14.129]{Merzbacher70}. Taking the infinite dimensional representation of our solution we obtain the quantum interaction picture. This is a more straight forward way of analysing the quantum forced oscillator than is given in the current quantum mechanical literature.

 To simplify the calculations we take the constants $m$ and $\omega$ to be unity throughout this section. To use the interaction picture (see Section \ref{sect:interactpic}) we split the $p$-mechanical Hamiltonian  for the forced oscillator (from equation (\ref{eq:forcedoscpmham})) into two parts $B_{H_0} = -\frac{1}{8 \pi^2} \left( \zerodelxtwo + \zerodelytwo \right)$ and $B_{H_1} = -\frac{z(t)}{2 \pi i} \zerodelxone$. Now by equation (\ref{eq:gentimeevolofinter}) a $\hilbh$ state $\tilde{v} = \exp (-t \antid B_{H_0} ) v$ will evolve by the equation\footnote{Note that $\exp (t B_{H_0} \antid)$ is well defined using Stone's Theorem since $B_{H_0}$ is self-adjoint on $\hilbh$ and $\antid$ is just multiplication by $\frac{2\pi}{ih}$.}
\begin{equation} \nonumber
\Fracdiffl{\tilde{v}}{t} = \exp (-t B_{H_0} \antid ) \antid B_{H_1} \exp (t B_{H_0} \antid) * \tilde{v}.
\end{equation}
Since $B_{H_0}$ is just the Hamiltonian for the harmonic oscillator we have using equation (\ref{eq:timeevolofharmoscinpm}) and Property 3 of Lemma \ref{lem:propofantidtwo}
\begin{eqnarray} \nonumber
\Fracdiffl{\tilde{v}}{t} &=& B_{H_1} (s, x \cos (t) + y \sin (t), -x \sin(t) + y \cos (t)) * \antid \tilde{v} \\ \nonumber
&=& \antid \left( - \frac{z(t)}{2\pi i} \right) \delta (s) \delta^{(1)} (x\cos(t) + y \sin(t) ) \delta (-x\sin(t)+y\cos(t)) * \tilde{v} \\ \nonumber
&=& \antid \left[ - \frac{z(t)}{2\pi i} \int_{\Heisn} \delta (s') \delta^{(1)} (x' \cos (t) + y' \sin (t)) \delta (-x' \sin(t) + y' \cos(t)) \right. \\ \nonumber
&& \left. \hspace{3.4cm} \times \tilde{v} ( (-s',-x'-y').(s,x,y)) \, ds' \, dx' \, dy' \right].
\end{eqnarray}
Since $\tilde{v} \in \hilbh$ we have $\antid \tilde{v} = \frac{2\pi}{ih} \tilde{v}$ then by a change of variable the right hand side of the above equation is equal to
\begin{eqnarray}   \nonumber
\lefteqn{-\frac{2\pi}{ih} \frac{z(t)}{2\pi i} \int_{\Heisn} \delta (s') \delta^{(1)} (x') \delta (y')} \\ \nonumber
&& \hspace{2.3cm} \times \tilde{v} ((-s',-x'\cos(t) +y' \sin (t) , -x' \sin (t) - y' \cos (t) ).(s,x,y))  \\ \nonumber
&=& \frac{z(t)}{h} \Partial{x'} \tilde{v} \left( s-s' + \frac{1}{2} [ (-x' \cos (t) + y' \sin(t) )y +(x'\sin (t) + y' \cos(t))x] , \right. \\ \nonumber
&& \hspace{2.5cm} \left. x- x' \cos (t) + y' \sin (t) , y- x' \sin (t) - y' \cos (t) \right) |_{(s',x',y)=(0,0,0)} \\ \nonumber
&=& \frac{z(t)}{h} \left( \frac{1}{2} (x\sin(t) - y \cos (t) ) \Partial{s} - \cos(t) \Partial{x} -\sin(t) \Partial{y} \right) \tilde{v} \\ \nonumber
&=& \left(\pi i z(t) [x\sin(t) - y \cos(t)] - \frac{z(t)\cos(t)}{h}  \Partial{x} - \frac{z(t) \sin(t)}{h} \Partial{y}\right) \tilde{v}.
\end{eqnarray}
A solution of this equation is
\begin{eqnarray} \label{eq:hhintersolofforcedosc}
\lefteqn{\tilde{v}(t;s,x,y)} \\ \nonumber
&=& \exp \left(-\pi i \int_0^t \int_0^\tau (z ( \tau) \cos (\tau) z (\tau') \sin (\tau') - z(\tau') \cos(\tau') z(\tau) \sin (\tau) \, d\tau' d\tau \right) \\ \nonumber
&& \hspace{1cm} \times v_0 \left( \left( 0, -\frac{1}{h} \int_0^t z(\tau) \cos(\tau) \, d\tau , -\frac{1}{h} \int_0^t z(\tau) \sin(\tau) \, d\tau \right).(s,x,y) \right)  \\ \nonumber
&=& \exp \left(-\pi i \int_0^t \int_0^\tau z ( \tau) z(\tau') \sin (\tau - \tau')  \, d\tau' d\tau \right) \\ \nonumber
&& \hspace{1cm} \times v_0 \left( s + \frac{1}{2h} \left( x \int_0^t z(\tau) \sin(\tau) \, d\tau - y \int_0^t z(\tau) \cos(\tau) \, d\tau \right), \right. \\ \nonumber
&& \left. \hspace{3.5cm} x  -\frac{1}{h} \int_0^t z(\tau) \cos(\tau) \, d\tau, y -\frac{1}{h} \int_0^t z(\tau) \sin(\tau) \, d\tau \right).
\end{eqnarray}
So the time evolution is just a left shift by
\begin{equation}
\left(0,\frac{1}{h} \int_0^t z(\tau) \cos (\tau) \, d\tau ,  \frac{1}{h} \int_0^t z(\tau) \sin (\tau) \, d\tau \right)
\end{equation}
and  multiplication by a numerical phase of modulus $1$. The numerical phase can be ignored when taking expectation values of observables --- it will be canceled out by the complex conjugation in the $\hilbh$ inner product.

If at time $0$ the system is in a coherent state $v_{(h,a,b)}$ (see equation (\ref{eq:hhcs})) that is
\begin{equation}
\tilde{v}(0;s,x,y) = v_{(h,0,0)} \left( \left( 0, \frac{b}{h} , -\frac{a}{h} \right) (s,x,y) \right)
\end{equation}
then by time $t$ the system will be in state
\begin{eqnarray} \nonumber
\lefteqn{\tilde{v}(t;s,x,y)} \\ \nonumber
&=& e^{f_1 (t)} v_{(h,0,0)} \left( \left(0,-\frac{1}{h} \int_0^t z(\tau) \cos (\tau) \, d\tau ,  - \frac{1}{h} \int_0^t z(\tau) \sin (\tau) \, d\tau \right) \right. \\ \nonumber
&& \hspace{4cm} \left. \times \left(0, \frac{b}{h} , -\frac{a}{h} \right) (s,x,y) \right) \\ \nonumber
&=& e^{f_1 (t)} v_{(h,0,0)} \left( \left( \frac{1}{2h^2} \left[ a \int_0^t z(\tau) \cos (\tau) \, d \tau - b \int_0^t z(\tau) \sin (\tau) \, d\tau \right], \right. \right. \\ \nonumber
&& \left. \left. \hspace{2cm} \frac{1}{h} \left( b- \int_0^t z (\tau) \cos (\tau) \, d\tau \right) , -\frac{1}{h} \left( a+ \int_0^t z(\tau) \sin (\tau) \, d \tau \right) \right) \right.
\\ \nonumber
&& \left.\hspace{4cm} .(s,x,y) \right),
\end{eqnarray}
where
\begin{equation} \nonumber
f_1 (t) = -\pi i \int_0^t \int_0^\tau z(\tau ) z(\tau') \sin (\tau - \tau') \, d\tau' d\tau.
\end{equation}
Since any element of $\hilbh$ is of the form $e^{2\pi ihs} \hat{f} (x,y)$
\begin{equation} \label{eq:forcedoscstayscoherent}
\tilde{v}(t;s,x,y) = e^{f_1 (t) + f_2 (t)} v_{\left( h, a+ \int_0^t z(\tau) \sin (\tau) \, d \tau , b- \int_0^t z (\tau) \cos (\tau) \, d\tau \right) } (s,x,y)
\end{equation}
where
\begin{equation}
f_2 (t) = \frac{\pi i}{h} \left[ a \int_0^t z(\tau) \cos (\tau) \, d \tau - b \int_0^t z(\tau) \sin (\tau) \, d\tau \right].
\end{equation}
The $e^{f_1(t) + f_2(t)}$ part is just a numerical phase of modulus $1$ which can be ignored when taking expectation values. So (\ref{eq:forcedoscstayscoherent}) implies that if the system starts in a coherent state then it will always be a coherent state up to a numerical phase. This is a known fact in quantum theory, but we have proved it using much simpler methods than is commonly found in the literature (see \cite[Sect. 14.6]{Merzbacher70}, for example).

We can make these calculations even simpler using the kernel coherent states. Taking\footnote{It should be noted that $B_{H_0}$ is kernel self-adjoint.} $B_{H_0}$ and $B_{H_1}$ the same as we used for the $\hilbh$ interaction picture, the interaction picture kernel coherent states evolve by equation (\ref{eq:kernelsatesininteract})
\begin{equation} \label{eq:timeevolforforcedoscinterker}
\Fracdiffl{\tilde{l}}{t} = \ub{\exp(-t \antid B_{H_0}) B_{H_1} \exp(t \antid B_{H_0})} {\tilde{l}}{}.
\end{equation}
Using the same method as for the $\hilbh$ states the right hand side of (\ref{eq:timeevolforforcedoscinterker}) takes the form
\begin{equation} \nonumber
\ub{-\frac{z(t)}{2\pi i} \delta (s) \delta^{(1)} (x \cos (t) + y \sin (t)) \delta ( - x \sin (t) + y \cos (t) ) }{\tilde{l}}.
\end{equation}
We have that
\begin{eqnarray} \nonumber
\lefteqn{-\frac{z(t)}{2\pi i} \delta (s) \delta^{(1)} (x \cos (t) + y \sin (t)) \delta ( - x \sin (t) + y \cos (t) ) * \tilde{l} }\\ \nonumber
&=& -\frac{z(t)}{2\pi i} \left( \frac{1}{2} (x\sin(t) - y \cos (t) ) \Partial{s} - \cos(t) \Partial{x} -\sin(t) \Partial{y} \right)  \tilde{l}
\end{eqnarray}
and
\begin{eqnarray} \nonumber
\lefteqn{\tilde{l} * \left(-\frac{z(t)}{2\pi i} \delta (s) \delta^{(1)} (x \cos (t) + y \sin (t)) \delta ( - x \sin (t) + y \cos (t) ) \right) }\\ \nonumber
&=& -\frac{z(t)}{2\pi i} \left( \frac{1}{2} ( y \cos (t) - x \sin (t) ) \Partial{s} - \cos(t) \Partial{x} -\sin(t) \Partial{y} \right)  \tilde{l}.
\end{eqnarray}
Hence using the fact that $\Partial{s} \antid = 4 \pi^2 I$ equation (\ref{eq:timeevolforforcedoscinterker}) becomes
\begin{eqnarray} \nonumber
\Fracdiffl{\tilde{l}}{t} = 2\pi i z(t) [ x \sin (t) - y \cos (t) ] \tilde{l}.
\end{eqnarray}
This has the solution
\begin{equation} \nonumber
\tilde{l} (t;s,x,y) = \exp \left( 2\pi i \left( x \int_0^t z(\tau) \sin (\tau) \, d \tau - y \int_0^t z(\tau) \cos (\tau) \, d\tau \right) \right) \tilde{l}(0;s,x,y).
\end{equation}
From this it can be realised that if the system started in the coherent state $\tilde{l}(0;s,x,y) = l_{(h,a,b)}$ then after time $t$ it will be in the coherent state
\begin{equation} \nonumber
\tilde{l} (t;s,x,y) = l_{\left( h, a+ \int_0^t z(\tau) \sin (\tau) \, d \tau , b- \int_0^t z (\tau) \cos (\tau) \, d\tau \right) }.
\end{equation}
There is no numerical phase because the kernels directly evaluate expectation values.
\newline
\begin{Remark}
\emph{The states remaining coherent means if we let $h \rightarrow 0$ we can consider the classical time evolution by evaluating the observables at different points (that is the coordinates given by the coherent state). The observables themselves are moving, but just as they would under the unforced oscillator.}
\end{Remark}

\chapter{Canonical Transformations} \label{chap:cantransf}

\indent

In this chapter we consider the representation of canonical transformations in $p$-mechanics. In doing so we obtain relationships between canonical transformations in classical mechanics and quantum mechanics. Also we give a new method for deriving the representation of non-linear canonical transformations in quantum mechanics.

In Subsection \ref{subsect:ctransincm} we introduce canonical transformations in classical mechanics and describe their uses. The passage of canonical transformations from classical to quantum mechanics is the content of Subsection \ref{subsect:ctinqm}. In Subsection \ref{subsect:ctinpm} we give a summary of the role which canonical transformations have played in $p$-mechanics to date and give some motivation for why we are studying them. In Section \ref{sect:linearct} we calculate the effect of linear canonical transformations on the $p$-mechanical states. We describe the operators on $\fock$ (Subsection \ref{subsect:linctonfock}) and $\hilbh$ (Subsection \ref{subsect:linctonhh}) which correspond to particular linear classical canonical transformations. In Section \ref{sect:nonlincteqns} we use the coherent states defined in equation  (\ref{eq:hhcs}) to generate a system of integral equations which when solved will give the matrix elements of an operator on $\hilbh$ for a particular canonical transformation. In Subsection \ref{subsect:nonlinexample} we solve this equation for a non-linear example which is similar to the time evolution of the forced oscillator.

\section{Canonical Transformations in Classical \newline Mechanics, Quantum Mechanics and \newline $p$-Mechanics} \label{sect:ctsincmqmpm}

\indent

In this section we consider the different roles which canonical transformations play in classical, quantum and $p$-mechanics. We also look at relations between these three sets of transformations.

\subsection{Canonical Transformations in Classical Mechanics} \label{subsect:ctransincm}

\indent

Canonical transformations are at the centre of classical mechanics \cite{Arnold90,Goldstein80,Jose98}. A canonical transformation in classical mechanics is a map $A$ defined on phase space which preserves the symplectic form on $\Space{R}{2n}$. That is $A:\Space{R}{2n} \rightarrow \Space{R}{2n}$ such that
\begin{equation} \label{eq:canonsfcond}
\omega ( A(q,p), A(q',p')) = \omega ( (q,p),(q',p'))
\end{equation}
where $\omega$ is defined as $\omega((q,p),(q',p')) = qp'-q'p$. The effect of a canonical transformation is that it will map the set of coordinates $(q,p)$ into another set of coordinates $(Q,P)$ where $(Q(q,p),P(q,p)) = A(q,p)$. A condition equivalent to (\ref{eq:canonsfcond}) is
\begin{equation} \label{eq:QPcondforct}
\{ Q_i(q,p),P_j(q,p) \} = \delta_{i,j} = \{ q_i , p_j \}.
\end{equation}
For a classical system with Hamiltonian, $H$, the transformed Hamiltonian denoted by $K$ is defined by
\begin{equation} \nonumber
H(q,p) = K(Q(q,p),P(q,p)).
\end{equation}
The equations of motion for the new coordinates $Q,P$ are
\begin{eqnarray} \label{eq:firstaftercteqnofmom}
\Fracdiffl{Q_i}{t} &=& \Fracpartial{K}{P_i} \\ \label{eq:secondaftercteqnofmom}
\Fracdiffl{P_i}{t} &=& - \Fracpartial{K}{Q_i}.
\end{eqnarray}
It can be shown \cite{Goldstein80} that the time evolution in the new coordinates is the same as the time evolution in the old coordinates.

Canonical transformations can be realised as operators on the set of classical mechanical observables. If $f(q,p)$ is a classical mechanical observable then the image of $f$ under the canonical transformation defined by an invertible map $A: \Space{R}{2n} \rightarrow \Space{R}{2n}$ is
\begin{equation} \label{eq:ctmaponclassobsv}
\tilde{f}(Q,P) = f (A^{-1}(Q,P)).
\end{equation}
Alternatively we have
\begin{equation}
\tilde{f}(A(q,p)) = f (q,p).
\end{equation}
If $A$ represents a canonical transformation then for any two classical mechanical observables $f,g$
\begin{equation}
\{ f , g \} = \{ \tilde{f} , \tilde{g} \}.
\end{equation}
All these results are proved in \cite{Goldstein80}.

One of the aims in developing classical canonical transformations is to derive transformations which for particular systems simplify Hamilton's equation (see equation (\ref{eq:dynforclassobsv})). The most advanced applications of canonical transformations in classical mechanics are the Hamilton-Jacobi theory \cite[Chap. 10]{Goldstein80} \cite[Chap. 9]{Arnold90} and the passage to action angle variables \cite[Sect. 6.2]{Jose98} \cite[Chap. 9]{Arnold90}.

\subsection{Canonical Transformations in Quantum Mechanics} \label{subsect:ctinqm}

\indent

The passage of canonical transformations from classical mechanics to quantum mechanics has been a long journey which is still incomplete. The first person to give a clear formulation of quantum canonical transformations was Dirac; this is presented in his book \cite{Dirac47}. Mario Moshinsky along with a variety of collaborators has published a great deal of enlightening papers on the subject \cite{DirlKasperkovitzMoshinsky88,GarciaMoshinsky80,MelloMoshinsky75,MoshinskySeligman78,MoshinskySeligman79}. In these papers the aim is to find an operator $U$, defined on a Hilbert space, which corresponds to the canonical transformation. Moshinsky and his collaborators developed a system of differential equations which when solved gave the matrix elements --- with respect to the eigenfunctions of the position or momentum operator --- of $U$. More recently Arlen Anderson \cite{Anderson94} has published some results on modelling canonical transformations in quantum mechanics using non-unitary operators. Canonical transformations in phase space quantisation are discussed in \cite{ZachosCurtrightFairlie98}.

\subsection{Canonical Transformations in $p$-Mechanics} \label{subsect:ctinpm}

\indent

In this chapter we use $p$-mechanics to exhibit relations between classical and quantum canonical transformations. Canonical transformations in $p$-mechanics have already been mentioned briefly in the papers \cite{KisilBrodlie02,Kisil02.2,Kisil02.1}. It has been shown that if we have a $p$-mechanical observable $f(q,p)$ and its $p$-mechanisation is $B(s,x,y)$, then for any $A \in \Symp{n}$ ($\Symp{n}$ is the group of all linear symplectic transformations on $\Space{R}{n}$ see Definition \ref{def:symplecticgroup}) the $p$-mechanisation\footnote{The matrix $A$ here is in fact the inverse of the matrix which would describe the canonical transformation in equation (\ref{eq:ctmaponclassobsv}). This is only a matter of notation since the Symplectic group is closed under matrix inversion.} of $f(A(q,p))$ is\footnote{Throughout this chapter if $A$ is an $n$ by $n$ matrix then $A^*$ represents the transpose (that is, adjoint) of this matrix.}
\begin{equation} \label{eq:linctonpmechobsv}
B(s,(A^{-1})^* (x,y)).
\end{equation}
In these papers the discussion is restricted to the effect of linear canonical transformation on observables only. In this chapter we consider the effect of both linear and non-linear canonical transformations on the $p$-mechanical states which were introduced in Chapter \ref{chap:staesandpic}. We now present a result on linear canonical transformations in $p$-mechanics..
\begin{Proposition} \label{prop:linctwithdetonedynpres}
Let $A$ be a linear canonical transformation and let $B_1,B_2$ be two $p$-mechanical observables. If $\tilde{B}$ is the $p$-mechanical observable defined as $\tilde{B}(s,x,y) = B (s, A(x,y))$ then
\begin{equation}
\ub{\tilde{B_1}}{\tilde{B_2}}{} = \widetilde{\ub{B_1}{B_2}{}}.
\end{equation}
\end{Proposition}
\begin{proof}
This follows by a direct calculation:
\begin{eqnarray} \nonumber
\lefteqn{\ub{\tilde{B_1}}{\tilde{B_2}}{}} \\ \nonumber
&=& \int \tilde{B_1}(s',x',y') \tilde{B_2}(s-s' + \omega((-x',-y'),(x,y)), x-x',y-y') \, ds' \, dx' \, dy' \\ \nonumber
&& - \int \tilde{B_2}(s',x',y') \tilde{B_1}(s-s'+\omega((-x',-y')(x,y)), x-x',y-y') \, ds' \, dx' \, dy'  \\ \nonumber
&=& \int B_1(s',A(x',y')) \\ \nonumber
&& \hspace{1.3cm} \times B_2 (s-s' + \omega((-x',-y'),(x,y)), A(x-x',y-y')) \, ds' \, dx' \, dy' \\ \nonumber
&& - \int B_2 (s',A(x',y')) \\ \nonumber
&& \hspace{1.3cm} \times B_1 (s-s'+\omega((-x',-y')(x,y)), A(x-x',y-y')) \, ds' \, dx' \, dy'.
\end{eqnarray}
$\omega$ is the symplectic form on $\Space{R}{2n}$ as defined in equation (\ref{eq:sympformonrn}). By a change of variables this becomes
\begin{eqnarray} \nonumber
&& \int B_1(s',x',y') \\ \nonumber
&& \hspace{1.3cm} \times B_2(s-s' + \omega(A^{-1}(-x',-y'),(x,y)),A(x,y) - (x',y')) \, ds' \, dx' \, dy' \\ \nonumber
&& - \int B_2(s',x',y') \\ \nonumber
&& \hspace{1.3cm} \times B_1(s-s'+\omega(A^{-1}(-x',-y')(x,y)), A(x,y) - (x',y')) \, ds' \, dx' \, dy'.
\end{eqnarray}
Since $A$ is a canonical transformation and hence preserves the symplectic form $\omega$ the above expression is equal to
\begin{eqnarray} \nonumber
&& \int B_1 (s',x',y') \\ \nonumber
&& \hspace{1.3cm} \times B_2 (s-s' + \omega((-x',-y'),A(x,y)),A(x,y) - (x',y')) \, ds' \, dx' \, dy' \\ \nonumber
&& - \int B_2 (s',x',y') \\ \nonumber
&& \hspace{1.3cm} \times B_1 (s-s'+\omega((-x',-y'),A(x,y)), A(x,y) - (x',y')) \, ds' \, dx' \, dy' \\ \nonumber
&=& \widetilde{\ub{B_1}{B_2}{}}.
\end{eqnarray}
\end{proof}
Since the symplectic group is closed under inversion and transposition this result implies that a linear canonical transformation will preserve the time evolution of $p$-mechanical observables. We use this result in Subsection \ref{subsect:coupledosc} when considering the example of two coupled oscillators.

Chapter \ref{chap:staesandpic} showed us that we can find out both quantum and classical results using the $p$-mechanical states and observables. If we can apply a canonical transformation in $p$-mechanics we can immediately find out information about both the classical and the quantum system after the canonical transformation has taken place.
In studying $p$-mechanical canonical transformations we show how canonical transformations can be represented in the mathematical framework of both quantum and classical mechanics. It is stated in \cite{Anderson94} that canonical transformations have three important roles in both quantum and classical mechanics:
\begin{itemize}
\item time evolution;
\item physical equivalence of two theories;
\item solving a system.
\end{itemize}
Taking the one and infinite dimensional representations of the $p$-mechanical system will show how these properties are exhibited in classical and quantum mechanics respectively.

There are further benefits of considering canonical transformations in $p$-mechanics. Canonical transformations can represent the symmetries of a classical mechanical system. By looking at the image of canonical transformations in quantum mechanics we can see how these symmetries are represented in quantum mechanics.

Another reason to study $p$-mechanical canonical transformations is the possibility to transform the $p$-dynamic equation. In Chapter \ref{chap:forcedosc} we solved the $p$-dynamic equation (\ref{eq:pdyneqn}) for the forced and harmonic oscillators. In doing so it was made evident that the quantum and classical pictures of the problems were generated from the same source. For more complicated problems the $p$-dynamic equation becomes much more complicated and technical problems are encountered (see Section \ref{sect:pdynforkepcou} for example). In classical mechanics when these problems arise the solution often lies in finding a canonical transformation to a set of coordinates in which Hamilton's equations have a more manageable form. For example the transformation to action-angle variables completely solves the Kepler problem \cite[Sect. 10.8]{Goldstein80}. By studying canonical transformations in $p$-mechanics we have a tool which will transform the $p$-dynamic equation (\ref{eq:pdyneqn}) into possibly a more desirable form.

\section{Linear Canonical Transformations} \label{sect:linearct}

\indent

In this section we just consider linear canonical transformations. Linear canonical transformations are useful in both classical and quantum mechanics. For example the time evolution of the harmonic oscillator --- as discussed in Section \ref{sect:harmosc}--- is a linear canonical transformation. Dirac, in his original treatment of canonical transformations in quantum mechanics, dealt with exclusively linear canonical transformations. Linear canonical transformations are also a good stepping stone towards non-linear canonical transformations. In this section we show how linear canonical transformations affect the $p$-mechanical states. As well as having physical implications this is of interest as an area of pure mathematics because it generates the Metaplectic representation of the Symplectic group \cite[Chap. 4]{Folland89}.

\subsection{The Metaplectic Representation for $\fock$} \label{subsect:linctonfock}
\indent

 The set of all linear canonical transformations can be realised as a subset of $GL(2n,\Space{R}{})$.
\begin{Defn} \label{def:symplecticgroup}
The symplectic group, denoted $\Symp{n}$, is the subgroup of $GL(2n,\Space{R}{})$ which preserves the standard symplectic form (see (\ref{eq:canonsfcond})).
\end{Defn}
See \cite[Prop. 4.1]{Folland89} for alternative definitions of $\Symp{n}$. The symplectic group can also be realised as a  subgroup of the group of automorphisms of the Heisenberg group. The automorphism corresponding to $M\in \Symp{n}$ is $T_M$ defined by
\begin{equation} \nonumber
T_M:(s,x,y) \mapsto (s,M(x,y)).
\end{equation}
By the Stone-von Neumann Theorem the representations $\rho_h$ and $\rho_h \circ T_M$ are unitarily equivalent, hence there exists a unitary operator $\nu(M)$ such that
\begin{equation} \nonumber
\rho_h (s,M(x,y)) = \nu(M) \rho_h (s,x,y) \nu(M)^{-1}.
\end{equation}
This gives a representation\footnote{In fact this gives a double-valued representation of $\Symp{n}$ since it is defined up to a phase factor of $\pm 1$ \cite[Sect. 4.1]{Folland89}.}  of $\Symp{n}$ as operators on the space $\fock$.

We now identify the form of $\mu$ for particular $M \in \Symp{n}$. Initially we need the result that if $M = \left( \begin{array}{cc}
A & B \\
C & D
\end{array} \right) \in \Symp{n}$
then $D=A^{*-1} + A^{*-1} C^* B$ (see \cite[Prop. 4.1e]{Folland89}). Using this we have that any
\newline
$M = \left( \begin{array}{cc}
A & B \\
C & D
\end{array} \right) \in \Symp{n}$ with\footnote{We use the notation $|A|$ to denote the determinant of a matrix.} $|A| \neq 0$ can be expanded out as
\begin{eqnarray} \label{eq:expandofspnr}
\left( \begin{array}{cc}
A & B \\
C & D
\end{array} \right)
&=&
\left( \begin{array}{cc}
I & 0 \\
CA^{-1} & I
\end{array} \right)
\left( \begin{array}{cc}
A & 0 \\
0 & A^{*-1}
\end{array} \right)
\left( \begin{array}{cc}
0 & I \\
-I & 0
\end{array} \right) \\ \nonumber
&& \times  \left( \begin{array}{cc}
I & 0\\
-A^{-1} B & I
\end{array} \right)
\left( \begin{array}{cc}
0 & -I \\
I & 0
\end{array} \right).
\end{eqnarray}
The problem of finding a formula for $\mu(M)$ for any $M = \left( \begin{array}{cc}
A & B \\
C & D
\end{array} \right) \in \Symp{n}$ with $|A| \neq 0$ is reduced to three simpler cases which we tackle in this next theorem. First we state a result from \cite{Folland89} about the Metaplectic representation for the Schr\"odinger representation on $\ltworn$.
\begin{Thm} \label{thm:metapforschro}
If $\psi \in \ltworn$ and $\mu$ is the metaplectic representation on $\ltworn$ then
\begin{eqnarray} \label{eq:firstmetonschro}
\mu \left( \left( \begin{array}{cc}
A & 0 \\
0 & A^{* -1}
\end{array} \right) \right) (\psi)(\xi) &=& |A|^{-1/2} \psi(A^{-1} \xi);
\\ \label{eq:secondmetonschro}
\mu \left( \left( \begin{array}{cc}
I & 0 \\
C & I
\end{array} \right) \right) (\psi)(\xi) &=&  \pm e^{-\pi i x Cx} \psi(\xi), \qquad \textrm{if $C=C^*$};
\\ \label{eq:thirdmetonschro}
\mu \left( \left( \begin{array}{cc}
0 & I \\
-I & 0
\end{array} \right) \right) \psi (\xi) &=& i^{n/2} \int_{\Space{R}{n}} \psi (\xi') e^{2\pi i \xi \xi'} \, d\xi' .
\end{eqnarray}
\end{Thm}
\begin{proof}
A proof of this can be found in \cite[Sect. 4]{Folland89}.
\end{proof}
\begin{Remark}
\emph{Note that in the above theorem all the operators are double valued. This is because we have a double valued representation of the symplectic group. One way of getting a single valued representation is to use the double cover of the symplectic group known as the metaplectic group. This is the reason why this double valued representation is known as the metaplectic representation.}
\end{Remark}
We now give the metaplectic representation for elements of $\fock$.

\begin{Thm} \label{thm:metap}
If $f$ is an element of $\fock$ and $\nu$ is the metaplectic representation on $\fock$ then
\begin{eqnarray} \label{eq:firstmet}
\lefteqn{ \nu \left( \left( \begin{array}{cc}
A & 0 \\
0 & A^{* -1}
\end{array} \right) \right) (f)(q,p) } \\ \nonumber
&=& |A|^{-1/2} \int_{\Space{R}{3n}} f(q',p') e^{-\frac{4\pi i}{h} (p'A^{-1} \xi + q'p')} e^{-\frac{\pi}{h} (A^{-1} \xi + 2q')^2} \, dq' \, dp' \\ \nonumber
&& \hspace{3cm} \times e^{\frac{4\pi i}{h} (p\xi + qp)} e^{-\frac{\pi}{h} (\xi + 2q)^2}  \, d\xi;
\end{eqnarray}
\begin{eqnarray}  \label{eq:secondmet}
\lefteqn{ \nu \left( \left( \begin{array}{cc}
I & 0 \\
C & I
\end{array} \right) \right) (f)(q,p)} \\ \nonumber
&=&  \pm \int_{\Space{R}{3n}} f(q',p') e^{-\frac{4\pi i}{h} (p'\xi + q'p')} e^{-\frac{\pi}{h} (\xi + 2q')^2} \, dq' \, dp' \\ \nonumber
&& \hspace{3cm} \times e^{-\pi i \xi C \xi} e^{\frac{4\pi i}{h} (p\xi + qp)} e^{-\frac{\pi}{h} (\xi + 2q)^2} \, d\xi; \qquad \textrm{if $C=C^*$};
\\ \label{eq:thirdmet}
\lefteqn{ \nu \left( \left( \begin{array}{cc}
0 & I \\
-I & 0
\end{array} \right) \right) (f)(q,p)} \\ \nonumber
&=& i^{n/2} \int_{\Space{R}{4n}} f(q',p') e^{-\frac{4\pi i}{h} (p'\xi' + q'p')} e^{-\frac{\pi}{h} (\xi' + 2q')^2} \, dq' \, dp' \\ \nonumber
&& \hspace{3cm} \times e^{2\pi i \xi \xi'} \, d\xi'  e^{\frac{4\pi i}{h} (p\xi + qp)} e^{-\frac{\pi}{h} (\xi + 2q)^2}  \, d\xi.
\end{eqnarray}
\end{Thm}
\begin{proof}
By Theorem \ref{thm:fockandltwointertwine} we have
\begin{equation} \label{eq:intertwinethmres}
\mathcal{T} \rho_h^S \mathcal{T}^{-1} = \rho_h.
\end{equation}
Furthermore Theorem \ref{thm:metapforschro} gives us the operators $\mu(M)$ for these particular examples in the Schr\"odinger picture
\begin{equation} \label{eq:metapforschrogen}
\rho_h^S (s,M(x,y)) = \mu (M) \rho_h^S (s,x,y) \mu(M)^{-1}.
\end{equation}
So by a direct calculation using (\ref{eq:intertwinethmres}) and (\ref{eq:metapforschrogen}) for any $M \in \Symp{n}$
\begin{eqnarray} \nonumber
\rho_h (s,M(x,y)) &=& \mathcal{T} (\rho_h^S (s,M(x,y))) \mathcal{T}^{-1} \\ \nonumber
&=& \mathcal{T} \mu (M) \rho_h^S (s,x,y) \mu(M)^{-1} \mathcal{T}^{-1} \\ \nonumber
&=& \mathcal{T} \mu (M) \mathcal{T}^{-1} \rho_h (s,x,y) \mathcal{T} \mu (M)^{-1} \mathcal{T}^{-1} \\ \nonumber
&=& ( \mathcal{T} \mu (M) \mathcal{T}^{-1}) \rho_h (s,x,y) (\mathcal{T} \mu (M) \mathcal{T}^{-1})^{-1}.
\end{eqnarray}
This shows us that $\nu(M) = \mathcal{T} \mu(M) \mathcal{T}^{-1}$. We procede to calculate this for each of the three matrices in question. To make the calculations simpler we define the three matrices $M_1, M_2 , M_3$ as
\begin{eqnarray} \nonumber
M_1 &=& \left( \begin{array}{cc}
A & 0 \\
0 & A^{* -1}
\end{array} \right) \\ \nonumber
M_2 &=& \left( \begin{array}{cc}
I & 0 \\
C & I
\end{array} \right) \\ \nonumber
M_3 &=& \left( \begin{array}{cc}
0 & I \\
-I & 0
\end{array} \right)
\end{eqnarray}
By Theorem \ref{thm:inverseoft} for any $f \in \fock$ we have
\begin{equation} \nonumber
(\mathcal{T}^{-1} f) (\xi) = \int_{\Space{R}{2n}} f(q',p') e^{-\frac{4\pi i}{h} (p'\xi + q'p')} e^{-\frac{\pi}{h}(\xi + 2q')^2} \, dq' \, dp'.
\end{equation}
Furthermore by equation (\ref{eq:firstmetonschro})
\begin{equation} \nonumber
(\mu (M_1) \mathcal{T}^{-1} f)(\xi) = |A|^{-1/2} \int_{\Space{R}{2n}} f(q',p') e^{-\frac{4\pi i}{h} (p' A^{-1} \xi + q'p')} e^{-\frac{\pi}{h}(A^{-1} \xi + 2q')^2} \, dq' \, dp'.
\end{equation}
Finally by equation (\ref{eq:defoft})
\begin{eqnarray} \nonumber
(\mathcal{T} \mu (M_1) \mathcal{T}^{-1} f) (q,p)  &=& |A|^{-1/2} \int_{\Space{R}{2n}} f(q',p') e^{-\frac{4\pi i}{h} (p' A^{-1} \xi + q'p')} e^{-\frac{\pi}{h}(A^{-1} \xi + 2q')^2} \, dq' \, dp' \\ \nonumber
&& \hspace{3cm} \times e^{\frac{4\pi i}{h} (p\xi + qp)} e^{-\frac{\pi}{h} (\xi + 2q)^2}  \, d\xi.
\end{eqnarray}
This verifies equation (\ref{eq:firstmet}). We show that this new function will satisfy the polarization $D_j^h$
\begin{eqnarray} \nonumber
\lefteqn{ \Partial{q_j} (\mathcal{T} \mu (M_1) \mathcal{T}^{-1} f)} \\ \nonumber
&=& |A|^{-1/2} \int_{\Space{R}{2n}} f(q',p') e^{-\frac{4\pi i}{h} (p' A^{-1} \xi + q'p')} e^{-\frac{\pi}{h}(A^{-1} \xi + 2q')^2} \, dq' \, dp' \\ \nonumber
&& \hspace{3cm} \times \left( \frac{4\pi i}{h} p_j - \frac{4\pi}{h} (\xi_j + 2q_j) \right) e^{\frac{4\pi i}{h} (p\xi + qp)} e^{-\frac{\pi}{h} (\xi + 2q)^2}  \, d\xi.
\end{eqnarray}
and
\begin{eqnarray} \nonumber
\lefteqn{ \Partial{p_j} (\mathcal{T} \mu (M_1) \mathcal{T}^{-1} f)} \\ \nonumber
&=& |A|^{-1/2} \int_{\Space{R}{2n}} f(q',p') e^{-\frac{4\pi i}{h} (p' A^{-1} \xi + q'p')} e^{-\frac{\pi}{h}(A^{-1} \xi + 2q')^2} \, dq' \, dp' \\ \nonumber
&& \hspace{3cm} \times  \left( \frac{4\pi i}{h} \xi_j + \frac{4\pi i}{h}q_j \right) e^{\frac{4\pi i}{h} (p\xi + qp)} e^{-\frac{\pi}{h} (\xi + 2q)^2}  \, d\xi.
\end{eqnarray}
So
\begin{eqnarray} \nonumber
\lefteqn{ D_j^h ( \nu (M_1) f)} \\ \nonumber
&=& \left( \frac{h}{2}  \left(\Partial{p_j} + i \Partial{q_j} \right) + 2\pi (p_j+iq_j) \right)  (\nu (M_1) f) \\ \nonumber
&=& |A|^{-1/2} \int_{\Space{R}{2n}} f(q',p') e^{-\frac{4\pi i}{h} (p' A^{-1} \xi + q'p')} e^{-\frac{\pi}{h}(A^{-1} \xi + 2q')^2} \, dq' \, dp' \\ \nonumber
&& \hspace{1cm} \times  \left( \frac{h}{2} \left[ \frac{4\pi i}{h} \xi_j + \frac{4\pi i}{h} q_j - \frac{4\pi }{h}p_j - \frac{4\pi i}{h} \xi_j - \frac{8\pi i}{h}q_j \right] +2\pi (p_j+iq_j) \right) \\ \nonumber
&& \hspace{2cm} \times e^{\frac{4\pi i}{h} (p\xi + qp)} e^{-\frac{\pi}{h} (\xi + 2q)^2}  \, d\xi \\ \nonumber
&=& 0.
\end{eqnarray}
We now do a similar calculation to verify equation (\ref{eq:secondmet}). From equation (\ref{eq:secondmetonschro}) we have
\begin{eqnarray} \nonumber
(\mu (M_2) \mathcal{T}^{-1} f)(\xi) = \int_{\Space{R}{2n}} f(q',p') e^{-\frac{4\pi i}{h} (p' \xi + q'p')} e^{-\frac{\pi}{h}( \xi + 2q')^2} \, dq' \, dp' e^{-\pi i \xi C \xi}.
\end{eqnarray}
Applying $\mathcal{T}$ to this will give us equation (\ref{eq:secondmet}). It can be shown to satisfy the polarization by a similar calculation to that for $\nu (M_1)$. Similarly by (\ref{eq:thirdmetonschro}) we have
\begin{eqnarray} \nonumber
(\mu (M_3) \mathcal{T}^{-1} f)(\xi) = i^{n/2} \int_{\Space{R}{2n}} f(q',p') e^{-\frac{4\pi i}{h} (p' \xi + q'p')} e^{-\frac{\pi}{h}( \xi + 2q')^2} \, dq' \, dp' e^{2\pi i \xi \xi'} \, d\xi'.
\end{eqnarray}
By applying $\mathcal{T}$ to the above equation we get (\ref{eq:thirdmet}).
\end{proof}
Note that by expanding any matrix in $\Symp{n}$ by (\ref{eq:expandofspnr}), it is a product of the above types of matrices -- this is true since $(CA^{-1})^* = CA^{-1}$ and $(A^{-1} B)^* = A^{-1} B$ by  the properties of the symplectic group (see \cite[Prop. 4.1e,f]{Folland89}).

Now if we have an observable $B(s,x,y)$ the effect of a canonical transformation on this observable by equation (\ref{eq:linctonpmechobsv}) is $B \mapsto \tilde{B}$ where
\begin{equation}
\tilde{B} (s,x,y) = B ( s , (M^{-1})^* (x,y)).
\end{equation}
So the effect of this on a state $f \in \fock$ is given by
\begin{eqnarray} \nonumber
\rho_h ( \tilde{B})f &=& \int_{\Heisn} \tilde{B} (g) \rho_h (g) \, dg \, f \\ \nonumber
&=& \int_{\Heisn} B (s,(M^{-1})^* (x,y)) \rho_h (s,x,y) \, ds \, dx \, dy  \, f \\ \nonumber
&=& |M| \int_{\Heisn} B (s,x,y) \rho_h (s, M^* (x,y)) \, ds \, dx \, dy \, f \\ \nonumber
&=& |M| \int_{\Heisn} B(s,x,y) \nu (M^*) \rho_h (s,x,y) \nu (M^*)^{-1} \, ds \, dx \, dy \, f \\ \nonumber
&=& |M| \nu(M^*) \int_{\Heisn} B(s,x,y) \rho_h (s,x,y) \, ds \, dx \, dy  \nu (M^*)^{-1} \, f \\ \label{eq:effectonfockstates}
&=& |M| \nu (M^*) \rho_h (B) \nu(M^*)^{-1} f.
\end{eqnarray}
We now show how states in $\hilbh$ will be affected by a canonical transformation. By equation (\ref{eq:rhohandleftshift}) we have
\begin{equation} \nonumber
\lambda_l (s,x,y) v = \statem \rho_h (s,x,y) \statem^{-1} v
\end{equation}
so
\begin{eqnarray} \nonumber
\lambda (s,M(x,y)) &=& \statem (\rho_h (s,M(x,y))) \statem^{-1} \\ \nonumber
&=& \statem \nu (M) \rho_h(s,x,y) \nu(M)^{-1} \statem^{-1} \\ \nonumber
&=& \statem \nu (M) \statem^{-1} \lambda_l (s,x,y) \statem \nu (M)^{-1} \statem^{-1} \\ \nonumber
&=& ( \statem \nu (M) \statem^{-1}) \lambda_l (s,x,y) (\statem \nu (M) \statem^{-1})^{-1}.
\end{eqnarray}
So if we let $\tilde{\nu} (M) = \mathcal{S}_h \nu (M) \mathcal{S}_h^{-1}$ then
\begin{equation} \nonumber
\lambda_l (s, M(x,y)) = \tilde{\nu} (M) \lambda_l (s,x,y) \tilde{\nu}(M)^{-1}.
\end{equation}
By a direct calculation
\begin{eqnarray} \nonumber
\tilde{B} * v  &=& \int_{\Heisn} \tilde{B} (g) \lambda_l (g) \, dg \, v \\ \nonumber
&=& \int_{\Heisn} B (s,(M^{-1})^* (x,y)) \lambda_l (s,x,y) \, ds \, dx \, dy  \, v \\ \nonumber
&=& |M| \int_{\Heisn} B (s,x,y) \lambda_l (s, M^* (x,y)) \, ds \, dx \, dy \, v \\ \nonumber
&=& |M| \int_{\Heisn} B(s,x,y) \tilde{\nu} (M^*) \lambda_l (s,x,y) \tilde{\nu} (M^*)^{-1} \, ds \, dx \, dy \, v \\ \nonumber
&=& |M| \tilde{\nu} (M^*) \int_{\Heisn} B(s,x,y) \lambda_l (s,x,y) \, ds \, dx \, dy  \tilde{\nu} (M^*)^{-1} \, v \\ \label{eq:effectonfockstates}
&=& |M| \tilde{\nu} (M^*) B* \tilde{\nu}(M^*)^{-1} v.
\end{eqnarray}
This formula shows us how a $\hilbh$ state will transform under a canonical transformation.

\subsection{Linear Canonical Transformations for States Represented by Kernels} \label{sect:linctforkerns}

\indent

We now show how under certain conditions a linear canonical transformation will affect a kernel state. If $A\in \Symp{n}$, $B$ is a $p$-mechanical observable and $l$ is a kernel such that $B(s,x,y) l(s,x,y) \in \loneh$ then we have
\begin{eqnarray} \nonumber
\lefteqn{\int_{\Heisn} B(s,(A^{-1})^*(x,y) ) \overline{l(s,x,y)} \, ds \, dx \, dy} \\ \nonumber
&=& \int_{\Heisn} B(s,x,y) \overline{l(s,((A^{-1})^*)^{-1} (x,y))} \, ds \, dx \, dy \\ \nonumber
&=& \int_{\Heisn} B(s,x,y ) \overline{l(s,A^*(x,y))} \, ds \, dx \, dy.
\end{eqnarray}
So hence under these conditions a linear canonical transformation will map $l(s,x,y) \mapsto \tilde{l} (s,x,y) = l (s, A^* (x,y))$.

\subsection{Coupled Oscillators: An Application of \newline $p$-Mechanical Linear Canonical Transformations} \label{subsect:coupledosc}

In this subsection we apply the theory of linear canonical transformations to solve the $p$-dynamic equation for the system of two coupled oscillators. The problem of analysing two coupled oscillators is an important one in both classical and quantum mechanics \cite{CocolicchioViggiano00,HanKimNoz95,HanKimNozYeh93}.
The classical Hamiltonian for a system of two coupled oscillators both with mass unity is
\begin{equation} \label{eq:classcoupledham}
H = \frac{1}{2} (p_1^2 + p_2^2 ) + \frac{A}{2} q_1^2 + \frac{B}{2} q_2^2 + \frac{C}{2} q_1 q_2
\end{equation}
where $A,B$ and $C$ are constants such that $A>0$, $B>0$ and $4AB-C^2>0$. The $p$-mechanisation (see equation (\ref{eq:pmechmap})) of this is
\begin{equation} \label{eq:pcoupledham}
B_H = - \frac{1}{8\pi^2} \left[ \left( \Partialtwo{y_1} + \Partialtwo{y_2} \right) + A \Partialtwo{x_1} + B \Partialtwo{x_2} + C \frac{\partial^2}{\partial x_1 \partial x_2} \right] \delta (s,x_1,x_2,y_1,y_2).
\end{equation}
The canonical transformation
\begin{eqnarray} \label{eq:wherethemmatrixcomesfrom}
\left( \begin{array}{c}
q_1 \\ q_2 \\ p_1 \\ p_2
\end{array} \right)
=
\left(\begin{array}{cccc}
\cos \left( \frac{\alpha}{2} \right) & \sin \left( \frac{\alpha}{2} \right) & 0 & 0 \\
-\sin \left( \frac{\alpha}{2} \right) & \cos \left( \frac{\alpha}{2} \right) & 0 & 0 \\
0 & 0 & \cos \left( \frac{\alpha}{2} \right) & \sin \left( \frac{\alpha}{2} \right) \\
0 & 0 & - \sin \left( \frac{\alpha}{2} \right) &  \cos \left( \frac{\alpha}{2} \right) \\
\end{array} \right)
\left( \begin{array}{c}
Q_1 \\ Q_2 \\ P_1 \\ P_2
\end{array} \right)
\end{eqnarray}
where
\begin{equation} \label{eq:valueofalpha}
\alpha = \tan^{-1} \left( \frac{C}{B-A} \right)
\end{equation}
has been shown to be of use in studying the classical coupled oscillator \cite{HanKimNoz95}. We use $M$ to denote the matrix in equation (\ref{eq:wherethemmatrixcomesfrom}). Since $(M^{-1})^* = M$, by (\ref{eq:linctonpmechobsv}) the image of this transformation on the set of $p$-mechanical observables is
\begin{eqnarray*}
&& B(s,x_1,x_2,y_1,y_2) \\
&& \mapsto B \left( s,x_1 \cos \left( \frac{\alpha}{2} \right) + x_2 \sin \left( \frac{\alpha}{2} \right) , - x_1 \sin \left( \frac{\alpha}{2} \right) + x_2  \cos \left( \frac{\alpha}{2} \right), \right. \\
&& \hspace{3cm} \left. y_1 \cos \left( \frac{\alpha}{2} \right) + y_2 \sin \left( \frac{\alpha}{2} \right) , - y_1 \sin \left( \frac{\alpha}{2} \right) + y_2  \cos \left( \frac{\alpha}{2} \right)\right).
\end{eqnarray*}
Hence under this canonical transformation the $p$-mechanical Hamiltonian will be transformed into
\begin{eqnarray*}
&& -\frac{1}{8\pi^2} \left[ \left( \Partialtwo{y_1} + \Partialtwo{y_2} \right) + A \Partialtwo{x_1} + B \Partialtwo{x_2} + C \frac{\partial^2}{\partial x_1 \partial x_2} \right] \\
&& \hspace{1cm} \delta \left( s,x_1 \cos \left( \frac{\alpha}{2} \right) + x_2 \sin \left( \frac{\alpha}{2} \right) , - x_1 \sin \left( \frac{\alpha}{2} \right) + x_2  \cos \left( \frac{\alpha}{2} \right), \right. \\
&& \hspace{2.5cm}  \left. y_1 \cos \left( \frac{\alpha}{2} \right) + y_2 \sin \left( \frac{\alpha}{2} \right) , - y_1 \sin \left( \frac{\alpha}{2} \right) + y_2  \cos \left( \frac{\alpha}{2} \right) \right).
\end{eqnarray*}
This distribution is equal to
\begin{eqnarray} \nonumber
\lefteqn{ -\frac{1}{8\pi^2} \left\{ \left[ A \cos^2 \left( \frac{\alpha}{2} \right) + B  \sin^2 \left( \frac{\alpha}{2} \right) - C \sin \left( \frac{\alpha}{2} \right)  \cos \left( \frac{\alpha}{2} \right) \right] \Partialtwo{x_1} \right.} \\ \nonumber
&& \left. + \left[ A \sin^2 \left( \frac{\alpha}{2} \right) + B  \cos^2 \left( \frac{\alpha}{2} \right) + C  \sin \left( \frac{\alpha}{2} \right)  \cos \left( \frac{\alpha}{2} \right) \right] \Partialtwo{x_2} \right. \\ \nonumber
&& \left. + \left[ 2(A-B) \sin \left( \frac{\alpha}{2} \right) \cos \left( \frac{\alpha}{2} \right) + C \left( \cos^2 \left( \frac{\alpha}{2} \right) - \sin^2 \left( \frac{\alpha}{2} \right) \right) \right] \frac{\partial^2}{\partial x_1 \partial x_2} \right. \\ \nonumber
&& \left. \hspace{0.5cm} + \left[ \cos^2 \left( \frac{\alpha}{2} \right) + \sin^2 \left( \frac{\alpha}{2} \right) \right] \Partialtwo{y_1} + \left[ \cos^2 \left( \frac{\alpha}{2} \right) + \sin^2 \left( \frac{\alpha}{2} \right) \right] \Partialtwo{y_2} \right. \\ \nonumber
&& \left. \hspace{0.5cm} + \left[ 2 \cos \left( \frac{\alpha}{2} \right)  \sin \left( \frac{\alpha}{2} \right) - 2 \cos \left( \frac{\alpha}{2} \right)  \sin \left( \frac{\alpha}{2} \right) \right] \frac{\partial^2}{\partial y_1 \partial y_2 }  \right\} \\ \label{eq:thedistafterct}
&& \hspace{8cm} \delta (s,x_1,x_2,y_1,y_2).
\end{eqnarray}
Since $\tan(\alpha) = \frac{C}{B-A}$ we have
\begin{equation} \nonumber
\sin (\alpha ) = \frac{C}{\sqrt{C^2 + (B-A)^2}} \hspace{2cm} \cos (\alpha ) = \frac{B-A}{\sqrt{C^2 + (B-A)^2}}.
\end{equation}
By the trigonometric identity
\begin{equation} \nonumber
\sin (\alpha) = 2 \sin \left( \frac{\alpha}{2} \right) \cos \left( \frac{\alpha}{2} \right)
\end{equation}
we have
\begin{equation} \label{eq:firsttrigid}
\sin \left( \frac{\alpha}{2} \right) \cos \left( \frac{\alpha}{2} \right) = \frac{C}{2 \sqrt{C^2 + (B-A)^2}}.
\end{equation}
Furthermore the trigonometric identity
\begin{equation} \nonumber
\cos (\alpha) = \cos^2 \left( \frac{\alpha}{2} \right) - \sin^2 \left( \frac{\alpha}{2} \right)
\end{equation}
implies that
\begin{equation} \label{eq:secondtrigid}
\cos^2 \left( \frac{\alpha}{2} \right) - \sin^2 \left( \frac{\alpha}{2} \right) = \frac{B-A}{\sqrt{C^2 + (B-A)^2}}.
\end{equation}
If we substitute equations (\ref{eq:firsttrigid}) and (\ref{eq:secondtrigid}) into (\ref{eq:thedistafterct}) we see that the coefficient of $\frac{\partial^2}{\partial x_1 \partial x_2}$ disappears. To simplify matters we define $W_1$ and $W_2$ as
\begin{eqnarray*}
W_1 &=& A \cos^2 \left( \frac{\alpha}{2} \right) + B \sin^2 \left( \frac{\alpha}{2} \right) - C \sin \left( \frac{\alpha}{2} \right)  \cos \left( \frac{\alpha}{2} \right) \\
W_2 &=& A \sin^2 \left( \frac{\alpha}{2} \right) + B  \cos^2 \left( \frac{\alpha}{2} \right) + C  \sin \left( \frac{\alpha}{2} \right)  \cos \left( \frac{\alpha}{2} \right).
\end{eqnarray*}
Distribution (\ref{eq:thedistafterct}) now becomes
\begin{equation} \nonumber
\tilde{B_H} = -\frac{1}{8\pi^2} \left[ W_1 \Partialtwo{x_1} + W_2 \Partialtwo{x_2} + \left( \Partialtwo{y_1} + \Partialtwo{y_2} \right) \right] \delta (s,x_1,x_2,y_1,y_2).
\end{equation}
Hence by this canonical transformation we have managed to decouple the oscillators. By Proposition \ref{prop:linctwithdetonedynpres} the dynamics of this observable are the same after this canonical transformation. If $B$ is an arbitary $p$-mechanical observable whose image after the canonical transformation is $\tilde{B}$, then in the coupled oscillator system the dynamics will be given by
\begin{equation} \label{eq:timeevolforcoupled}
\Fracdiffl{\tilde{B}}{t} = \ub{\tilde{B}}{\tilde{B_H}}{}.
\end{equation}
Using the commutation of left and right invariant vector fields on $\Heisn$ equation (\ref{eq:timeevolforcoupled}) becomes
\begin{eqnarray} \nonumber
\Fracdiffl{\tilde{B}}{t} = \left( y_1 \Partial{x_1} + y_2 \Partial{x_2}  - W_1 x_1 \Partial{y_1} - W_2 x_2 \Partial{y_2} \right) \tilde{B}.
\end{eqnarray}
A solution of this is
\begin{eqnarray*} \nonumber
\lefteqn{\tilde{B}(t;s,x_1,x_2,y_1,y_2)} \\
&=& \tilde{B_0} \left[ s, x_1 \cos \left( \sqrt{W_1} t \right) + \frac{y_1}{\sqrt{W_1}} \sin \left( \sqrt{W_1} t \right) , \right. \\
&& \left. \hspace{2cm} x_2 \cos \left( \sqrt{W_2} t \right) + \frac{y_2}{\sqrt{W_2}} \sin \left( \sqrt{W_2} t \right) , \right.\\
&& \left. \hspace{2.2cm} - \sqrt{W_1} x_1 \sin \left( \sqrt{W_1} t \right) + y_1 \cos \left( \sqrt{W_1} t \right) , \right. \\
&& \left. \hspace{2.4cm} - x_2 \sqrt{W_2} \sin \left( \sqrt{W_2} t \right) + y_2 \cos \left( \sqrt{W_2}t \right) \right].
\end{eqnarray*}
By applying this linear canonical transformation we have simplified the $p$-mechanical dynamics for the coupled oscillator. By taking the one and infinite dimensional representations of this flow we would get the classical and quantum dynamics in the new coordinates. To return to the usual coordinates we would just need to take the inverse of this canonical transformation -- this is just the inverse of the matrix $M$.

\section{Non-Linear Canonical Transformations} \label{sect:nonlincteqns}

\indent

Unfortunately the majority of canonical transformations which are physically useful are non-linear. For example the passage to action angle variables \cite{CooperPellegrini99} for the one dimensional harmonic oscillator is a non-linear canonical transformation. In this section we look at ways of modelling non-linear canonical transformations in $p$-mechanics. We follow an approach which is an enhancement of a method pioneered by Mario Moshinsky and a variety of collaborators \cite{MelloMoshinsky75,MoshinskySeligman78,MoshinskySeligman79,GarciaMoshinsky80,DirlKasperkovitzMoshinsky88}. Moshinsky and his collaborators attempted to find operators on the Hilbert space of quantum mechanical states which correspond to particular classical canonical transformations. To do this they generated a system of differential equations which when solved gave the matrix elements --- with respect to the position or momentum eigenfunctions --- of this operator.

In our approach we use the $\hilbh$ coherent states (see Section \ref{sect:cstates}) to generate a system of integral equations which when solved will give the coherent state expansion of the operator on $\hilbh$ which corresponds to the classical canonical transformation. In this paper we are looking at general $p$-mechanical observables and states as opposed to just quantum mechanical observables and states. This means that our equations have both quantum and classical realisations.

\subsection{Equations for Non-Linear Transformations Involving $\hilbh$ States} \label{subsect:nonlintranseq}

\indent

This method starts with the observation that a canonical transformation in classical mechanics described by $2n$ independent relations
\begin{eqnarray} \label{eq:oldinitialconrelq}
q_i \rightarrow Q_i(q,p) \\ \label{eq:oldinitialconrelp}
p_i \rightarrow P_i(q,p)
\end{eqnarray}
$i=1 \hdots n$ where $\{ Q_i , P_j \}_{q,p}= \delta_{ij}$ can be realised implicitly by $2n$ functional relations
\begin{eqnarray} \label{eq:newinitialconrelq}
f_i (q,p) &=& F_i (Q,P) \\ \label{eq:newinitialconrelp}
g_i (q,p) &=& G_i (Q,P)
\end{eqnarray}
for $i=1 \hdots n$. We cannot just choose any sets of functions, they need to satisfy a certain property, this is the content of the following proposition. The one dimensional version of the following proposition appears in \cite{MoshinskySeligman78}; we extend it to $n$ dimensions.
\begin{Proposition} \label{prop:implicitctok}
If $Q_i, P_i, f_i, g_i, F_i, G_i$ for $i=1, \cdots , n$ are all differentiable and invertible functions on $\Space{R}{2n}$ then we have the following relation.
$\{ Q_i(q,p), P_j(q,p) \} = \delta_{i,j}$ for all $i,j=1, \cdots , n$ if and only if $\{f_i, g_j \}_{q,p} = \{ F_i , G_j \}_{Q,P}$ for all $i,j=1, \cdots , n$.
\end{Proposition}
\begin{proof}
By the $n$-dimensional chain rule and equations (\ref{eq:newinitialconrelq}) for any $i,j = 1, \cdots , n$
\begin{eqnarray} \nonumber
\Partial{q_j} f_i(Q(q,p),P(q,p)) &=& \sum_{k=1}^n \left(\Fracpartial{f_i}{Q_k} \Fracpartial{Q_k}{q_j} + \Fracpartial{f_i}{P_k} \Fracpartial{P_k}{q_j}\right) \\ \label{eq:ndimchruleforctimplicit}
&=& \sum_{k=1}^n \left(\Fracpartial{F_i}{Q_k} \Fracpartial{Q_k}{q_j} + \Fracpartial{F_i}{P_k} \Fracpartial{P_k}{q_j}\right).
\end{eqnarray}
Identical relations hold if we replace $F$ with $G$ and $q$ with $p$. Using (\ref{eq:ndimchruleforctimplicit}) we get
\begin{eqnarray*}
\lefteqn{\{ f_i , g_j \}_{q,p} =} \\
&& \sum_{k=1}^n \left\{ \left[ \sum_{m=1}^n  \left(\Fracpartial{F_i}{Q_m} \Fracpartial{Q_m}{q_k} + \Fracpartial{F_i}{P_m} \Fracpartial{P_m}{q_k} \right) \right] \left[ \sum_{m'=1}^n  \left(\Fracpartial{G_j}{Q_{m'}} \Fracpartial{Q_{m'}}{p_k} + \Fracpartial{G_j}{P_{m'}} \Fracpartial{P_{m'}}{p_k} \right) \right] \right. \\
&& \left. \hspace{1cm} - \left[ \sum_{r=1}^n  \left(\Fracpartial{F_i}{Q_r} \Fracpartial{Q_r}{p_k} + \Fracpartial{F_i}{P_r} \Fracpartial{P_r}{p_k} \right) \right] \left[ \sum_{r'=1}^n  \left(\Fracpartial{G_j}{Q_{r'}} \Fracpartial{Q_{r'}}{q_k} + \Fracpartial{G_j}{P_{r'}} \Fracpartial{P_{r'}}{q_k} \right) \right] \right\} \\
&=& \sum_{k,m,m'=1}^n \left( \Fracpartial{F_i}{Q_m}  \Fracpartial{G_j}{Q_{m'}} \Fracpartial{Q_m}{q_k} \Fracpartial{Q_{m'}}{p_k} \right) -  \sum_{k,r,r' =1}^n \left(\Fracpartial{F_i}{Q_r} \Fracpartial{G_j}{Q_{r'}} \Fracpartial{Q_r}{p_k}  \Fracpartial{Q_{r'}}{q_k} \right) \\
&&  + \sum_{k,m,m'=1}^n \left( \Fracpartial{F_i}{Q_m}  \Fracpartial{G_j}{P_{m'}} \Fracpartial{Q_m}{q_k} \Fracpartial{P_{m'}}{p_k} \right) -  \sum_{k,r,r' =1}^n \left(\Fracpartial{F_i}{Q_r} \Fracpartial{G_j}{P_{r'}} \Fracpartial{Q_r}{p_k}  \Fracpartial{P_{r'}}{q_k} \right) \\
&&  + \sum_{k,m,m'=1}^n \left( \Fracpartial{F_i}{P_m}  \Fracpartial{G_j}{Q_{m'}} \Fracpartial{P_m}{q_k} \Fracpartial{Q_{m'}}{p_k} \right) -  \sum_{k,r,r' =1}^n \left(\Fracpartial{F_i}{P_r} \Fracpartial{G_j}{Q_{r'}} \Fracpartial{P_r}{p_k}  \Fracpartial{Q_{r'}}{q_k} \right) \\
&&  + \sum_{k,m,m'=1}^n \left( \Fracpartial{F_i}{P_m}  \Fracpartial{G_j}{P_{m'}} \Fracpartial{P_m}{q_k} \Fracpartial{P_{m'}}{p_k} \right) -  \sum_{k,r,r' =1}^n \left(\Fracpartial{F_i}{P_r} \Fracpartial{G_j}{P_{r'}} \Fracpartial{P_r}{p_k}  \Fracpartial{P_{r'}}{q_k} \right). \\
\end{eqnarray*}
In the above expression the first and second terms are equal and so cancel each other out; the same applies for the seventh and eight terms. Hence this becomes
\begin{eqnarray} \nonumber
\lefteqn{ \sum_{k,m,m'=1}^n \left( \Fracpartial{F_i}{Q_m}  \Fracpartial{G_j}{P_{m'}} \Fracpartial{Q_m}{q_k} \Fracpartial{P_{m'}}{p_k} \right) -  \sum_{k,r,r' =1}^n \left(\Fracpartial{F_i}{Q_r} \Fracpartial{G_j}{P_{r'}} \Fracpartial{Q_r}{p_k}  \Fracpartial{P_{r'}}{q_k} \right) } \\ \nonumber
&& + \sum_{k,m,m'=1}^n \left( \Fracpartial{F_i}{P_m}  \Fracpartial{G_j}{Q_{m'}} \Fracpartial{P_m}{q_k} \Fracpartial{Q_{m'}}{p_k} \right) -  \sum_{k,r,r' =1}^n \left(\Fracpartial{F_i}{P_r} \Fracpartial{G_j}{Q_{r'}} \Fracpartial{P_r}{p_k}  \Fracpartial{Q_{r'}}{q_k} \right) \\ \nonumber
&=& \sum_{m,m'=1}^n \left\{ \Fracpartial{F_i}{Q_m}  \Fracpartial{G_j}{P_{m'}} \left[ \sum_{k=1}^n \left( \Fracpartial{Q_m}{q_k} \Fracpartial{P_{m'}}{p_k} - \Fracpartial{Q_m}{p_k} \Fracpartial{P_{m'}}{q_k} \right) \right] \right\} \\ \nonumber
&& + \sum_{m,m'=1}^n \left\{ \Fracpartial{F_i}{P_m}  \Fracpartial{G_j}{Q_{m'}} \left[ \sum_{k=1}^n \left( \Fracpartial{P_m}{q_k} \Fracpartial{Q_{m'}}{p_k} - \Fracpartial{P_m}{p_k} \Fracpartial{Q_{m'}}{q_k} \right) \right] \right\} \\ \label{eq:finalofqpctisok}
&=& \sum_{m,m'=1}^n  \Fracpartial{F_i}{Q_m}  \Fracpartial{G_j}{P_{m'}} \{ Q_m , P_{m'} \}_{q,p} + \sum_{m,m'=1}^n  \Fracpartial{F_i}{P_m}  \Fracpartial{G_j}{Q_{m'}} \{ P_m , Q_{m'} \}_{q,p}.
\end{eqnarray}
If we assume $\{ Q_i(q,p), P_j(q,p) \}_{q,p} = \delta_{i,j}$ for all $i,j=1, \cdots n$ then the above expression becomes
\begin{eqnarray*}
\lefteqn{ \sum_{m,m'=1}^n  \Fracpartial{F_i}{Q_m}  \Fracpartial{G_j}{P_{m'}} \delta_{m,m'} - \sum_{m,m'=1}^n  \Fracpartial{F_i}{P_m}  \Fracpartial{G_j}{Q_{m'}} \delta_{m,m'} } \\
&=& \sum_{m=1}^n \Fracpartial{F_i}{Q_m}  \Fracpartial{G_j}{P_{m}}  - \sum_{m=1}^n  \Fracpartial{F_i}{P_m}  \Fracpartial{G_j}{Q_{m}}  \\
&=& \{ F_i , G_j \}_{Q,P}.
\end{eqnarray*}
Since this holds for any $i,j=1, \cdots , n$ one direction of the argument has been proved. The inverse follows since if expression (\ref{eq:finalofqpctisok}) is equal to $\{ F_i , G_j \}$ for all $i,j=1, \cdots , n$ then $\{ Q_i , P_j \} = \delta_{i,j}$ for all $i,j=1, \cdots , n$.
\end{proof}
By (\ref{eq:QPcondforct}) and Proposition \ref{prop:implicitctok} we can see that $\{ f_i,g_j\}_{q,p} = \{ F_i,G_j \}_{Q,P}$ for all $i,j=1, \cdots , n$ is a necessary and sufficient condition for equations (\ref{eq:newinitialconrelq}), (\ref{eq:newinitialconrelp}) to describe a canonical transformation.

The advantage of describing the canonical transformation implicitly is that the $p$-mechanisation (\ref{eq:pmechmapwc}) of the functions in (\ref{eq:newinitialconrelq}), (\ref{eq:newinitialconrelp}) may be easier to define than the functions on the right hand side of equations (\ref{eq:oldinitialconrelq}), (\ref{eq:oldinitialconrelp}). We assume throughout the chapter that the above functions of $q$ and $p$ are $C^{\infty}$ and when integrated next to an element of $\mathcal{S}(\Space{R}{2n})$ will be finite. This means they can always be realised as elements of $\mathcal{S}'(\Space{R}{2n})$.

We now derive an equation which will give us a clear form of an operator $U$ on $\hilbh$ corresponding to a canonical transformation. This equation will supply us with the matrix elements of the operator $U$ with respect to the overcomplete set of coherent states, that is  it will give us $\langle U v_{(h,a,b)} , v_{(h,a',b')} \rangle$ for all $a,b,a',b' \in \Space{R}{n}$.

In Dirac's original treatment of quantum canonical transformations \cite{Dirac47} he proposed that the canonical transformation from equations (\ref{eq:oldinitialconrelq}) and (\ref{eq:oldinitialconrelp}) should be represented in quantum mechanics by an unitary operator $U$ on a Hilbert space such that
\begin{equation} \nonumber
\tilde{Q_i}= U\tilde{q_i} U^{-1} \hspace{1cm} \textrm{and} \hspace{1cm} \tilde{P_i} = U \tilde{p_i} U^{-1}
\end{equation}
$i=1, \cdots n$. Here $\tilde{Q_i},\tilde{P_i},\tilde{q_i}, \tilde{p_i}$ are the quantum mechanical observables corresponding to the classical mechanical observables $Q_i,P_i,q_i,p_i$ respectively.

In \cite{MelloMoshinsky75} Mello and Moshinsky suggested that in some circumstances it is easier to define the operator $U$ by the equations
\begin{equation} \nonumber
\tilde{F} U = U\tilde{f}  \hspace{1cm} \textrm{and} \hspace{1cm} \tilde{G} U = U \tilde{g}
\end{equation}
where $\tilde{F}, \tilde{G}, \tilde{f}, \tilde{g}$ are the operators corresponding to the classical observables $F,G,f,g$ from equations (\ref{eq:newinitialconrelq}) and (\ref{eq:newinitialconrelp}). The actual definition of this operator $U$ will depend on the example in question. In \cite{MoshinskySeligman78} there is a lot of discussion on defining this operator for nonbijective transformations.

We proceed to transfer this approach into $p$-mechanics. We first fix a set of functions $f,g,F,G$ which define the canonical transformation in question and have a clear $p$-mechanisation. We want to understand the operator $U$ on $\hilbh$ which is defined by the equations
\begin{eqnarray} \label{eq:beginofourmmeqn}
U \proj ( f_i (q,p) ) * v &=& \proj (F_i(Q,P)) * U v \\ \label{eq:beginofourmmeqntwo}
U \proj ( g_i (q,p) ) * v &=& \proj (G_i(Q,P)) * U v
\end{eqnarray}
where $\proj$ is the map of $p$-mechanisation (\ref{eq:pmechmapwc}) and $v$ is any element of $\hilbh$.

We will now divert from deriving the general equation by giving an example to illuminate these ideas (the example we give is a linear transformation but it must be stressed that this work holds for non-linear transformations too).
\begin{Example} \label{ex:flipexample}
Consider the linear canonical transformation
\begin{equation} \nonumber
q \rightarrow -P \hspace{2cm} p \rightarrow Q
\end{equation}
which has already been discussed in Section \ref{sect:linearct}. This can be realised by the two equations
\begin{eqnarray} \label{eq:goodformofflipctone}
q+ip &=& -P+iQ \\ \label{eq:goodformofflipcttwo}
q-ip &=& -P-iQ.
\end{eqnarray}
The $p$-mechanisation of these two equations is
\begin{eqnarray} \label{eq:flipincreaandanihilone}
a^- &=& i A^-  \\ \label{eq:flipincreaandanihiltwo}
a^+ &=& -i A^+
\end{eqnarray}
where $a^-$ and $a^+$ are defined in equations (\ref{eq:defaplus}) and (\ref{eq:defaminus}). This may seem like we have made the equations more complicated, but we will see later in this chaper that we have got them into a more manageable form.
\end{Example}

We now continue to derive the system of equations which will help us understand the operator $U$. We begin by taking the matrix elements of equation (\ref{eq:beginofourmmeqn}) with respect to the coherent states defined in equation (\ref{eq:hhcs})
\begin{eqnarray} \label{eq:matrixeltofcteqn}
\langle U \proj ( f_i) *  \vab , \vabd \rangle &=& \langle \proj (F_i) * U \vab, \vabd \rangle , \\ \label{eq:matrixeltofcteqntwo}
\langle U \proj ( g_i) *  \vab , \vabd \rangle &=& \langle \proj (G_i) * U \vab, \vabd \rangle.
\end{eqnarray}
We can expand $U \vab$ using our system of coherent states (see (\ref{eq:wecantakecsexpansions}))
\begin{equation} \nonumber
U \vab = \int_{\Space{R}{2n}} \langle U \vab , \vabdd \rangle \vabdd \, da'' \, db''.
\end{equation}
The right hand sides of equations (\ref{eq:matrixeltofcteqn}), (\ref{eq:matrixeltofcteqntwo}) now become
\begin{eqnarray} \nonumber
\int_{\Space{R}{2n}} \langle U \vab , \vabdd \rangle \langle \proj (F_i) * \vabdd, \vabd \rangle \, da'' \, db'', \\ \nonumber
\int_{\Space{R}{2n}} \langle U \vab , \vabdd \rangle \langle \proj (G_i) * \vabdd, \vabd \rangle \, da'' \, db''.
\end{eqnarray}
Similarly we expand $\proj (f_i) * \vab$ as
\begin{equation} \nonumber
\proj (f_i) * \vab = \int_{\Space{R}{2n}} \langle  \proj (f_i) * \vab, \vabdd \rangle \vabdd \, da'' \,  db''.
\end{equation}
so the left hand sides of equations (\ref{eq:matrixeltofcteqn}), (\ref{eq:matrixeltofcteqntwo}) become
\begin{eqnarray*}
&& \int_{\Space{R}{2n}} \langle U \vabdd, \vabd \rangle \langle \proj (f_i) *\vab, \vabdd \rangle \, da'' \, db'' , \\
&& \int_{\Space{R}{2n}} \langle U \vabdd, \vabd \rangle \langle \proj (g_i) *\vab, \vabdd \rangle \, da'' \, db''.
\end{eqnarray*}

Hence if we set $m(a,b,c,d) = \langle U v_{(h,a,b)} , v_{(h,c,d)} \rangle$ equations (\ref{eq:matrixeltofcteqn}), (\ref{eq:matrixeltofcteqntwo})  become
\begin{eqnarray} \label{eq:thefinaleqnforkernsinctsof}
\lefteqn{\int_{\Space{R}{2n}} m(a'',b'',a',b') \langle \proj (f_i) * \vab, \vabdd \rangle \, da'' \, db''} \\ \nonumber
&& = \int_{\Space{R}{2n}} m(a,b,a'',b'') \langle \proj (F_i) *\vabdd, \vabd \rangle \, da'' \, db'', \\ \label{eq:thefinaleqnforkernsinctsoftwo}
\lefteqn{\int_{\Space{R}{2n}} m(a'',b'',a',b') \langle \proj (g_i) * \vab, \vabdd \rangle \, da'' \, db''} \\ \nonumber
&& = \int_{\Space{R}{2n}} m(a,b,a'',b'') \langle \proj (G_i) *\vabdd, \vabd \rangle \, da'' \, db''.
\end{eqnarray}

If we can solve this integral equation for $m$ then we can understand the effect of $U$ on any element $v$ of $\hilbh$ since
\begin{equation} \nonumber
v=\int_{\Space{R}{2n}} \langle v, \vab \rangle \vab \, da \, db
\end{equation}
and
\begin{equation} \nonumber
Uv = \int_{\Space{R}{2n}} \langle U v, \vabd \rangle \vabd \, da' \, db'
\end{equation}
which together give us
\begin{eqnarray} \nonumber
Uv &=& \int_{\Space{R}{2n}} \langle U \left(\int_{\Space{R}{2n}} \langle v, \vab \rangle \vab \, da \, db \right) , \vabd \rangle \vabd \, da' \, db' \\ \nonumber
&=& \int_{\Space{R}{2n}} \int_{\Space{R}{2n}} \langle U \vab, \vabd \rangle \langle v,\vab \rangle \vabd \, da \, db \, da' \, db' \\ \label{eq:waveletexpofsoltomabcd}
&=& \int_{\Space{R}{2n}} \int_{\Space{R}{2n}} m(a,b,a',b') \langle v,\vab \rangle \vabd \, da \, db \, da' \, db'.
\end{eqnarray}
\begin{Remark}
\emph{ The existence and uniqueness of a solution $m$ for the system (\ref{eq:thefinaleqnforkernsinctsof}), (\ref{eq:thefinaleqnforkernsinctsoftwo}) will depend on the canonical transformation in question. For complex examples this would involve some delicate use of the theory of integral equations.}
\end{Remark}
Since for many functions $f,g$, $\langle \proj (f) *\vab,\vabd \rangle$ and \newline $\langle \proj (g) *\vab , \vabd \rangle$ are manageable functions of $a,b,a',b'$, equations (\ref{eq:thefinaleqnforkernsinctsof}) will take a simple form for a variety of examples. For example consider the distributions involved in equations (\ref{eq:flipincreaandanihilone}) and (\ref{eq:flipincreaandanihiltwo}). Since $\vab$ is an eigenfunction of the annihilation operator $a^-$ with eigenvalue $(a+ib)$ (see Lemma \ref{lem:csiseigfctnforcreat}) we have
\begin{equation} \nonumber
\langle a^- * \vab, \vabd \rangle = (a+ib) \langle \vab, \vabd \rangle,
\end{equation}
and hence
\begin{equation} \label{eq:actiofcreatonctsone}
\langle \proj (q+ip) * \vab,\vabd \rangle = (a+ib) \langle \vab, \vabd \rangle.
\end{equation}
Furthermore by Lemma \ref{eq:posandmomincsmatrixelements} (which we state and prove later) we have
\begin{equation} \label{eq:actiofanihilonctsone}
\langle \proj (q-ip) * \vab , \vabd \rangle = (a'-ib') \langle \vab, \vabd \rangle.
\end{equation}
We are now in a position to present equations (\ref{eq:thefinaleqnforkernsinctsof}), (\ref{eq:thefinaleqnforkernsinctsoftwo}) for the canonical transformation
\begin{equation} \label{eq:statementoflineqfornonlineqn}
q \rightarrow -P \hspace{2cm} p \rightarrow Q.
\end{equation}
Using equations (\ref{eq:goodformofflipctone}), (\ref{eq:goodformofflipcttwo}), (\ref{eq:actiofcreatonctsone}) and  (\ref{eq:actiofanihilonctsone}) we can see that equations (\ref{eq:thefinaleqnforkernsinctsof}) and (\ref{eq:thefinaleqnforkernsinctsoftwo}) must take the form
\begin{eqnarray} \nonumber
\lefteqn{(a+ib) \int_{\Space{R}{2n}} m(a'',b'',a',b') \langle \vab, \vabdd \rangle \, da'' \, db''} \\ \nonumber
&& = \int_{\Space{R}{2n}} m(a,b,a'',b'') i(a'' + ib'') \langle \vabdd, \vabd \rangle \, da'' \, db'',
\end{eqnarray}

\begin{eqnarray} \nonumber
\lefteqn{(a-ib) \int_{\Space{R}{2n}} m(a'',b'',a',b') \langle \vab, \vabdd \rangle \, da'' \, db''} \\ \nonumber
&& = \int_{\Space{R}{2n}} m(a,b,a'',b'') (-i)(a'' -i b'') \langle \vabdd, \vabd \rangle \, da'' \, db''.
\end{eqnarray}

Using equation (\ref{eq:repkerforhh}) the above system becomes
\begin{eqnarray} \nonumber
\lefteqn{(a+ib) \int_{\Space{R}{2n}} m(a'',b'',a',b') } \\ \nonumber
&& \times \exp \left[ \frac{\pi}{2h} \left( 2(a+ib)(a''-ib'') - a^2 - b^2 - a''^2 - b''^2 \right) \right]\, da'' \, db'' \\ \nonumber
&& = \int_{\Space{R}{2n}} m(a,b,a'',b'') (ia'' - b'') \\ \nonumber
&& \times \exp \left[ \frac{\pi}{2h} \left( 2(a''+ib'')(a'-ib') - a''^2 - b''^2 - a'^2 - b'^2 \right) \right] \, da'' \, db''.
\end{eqnarray}

\begin{eqnarray} \nonumber
\lefteqn{ (a-ib) \int_{\Space{R}{2n}} m(a'',b'',a',b') } \\ \nonumber
&& \times \exp \left[ \frac{\pi}{2h} \left( 2(a+ib)(a''-ib'') - a^2 - b^2 - a''^2 - b''^2 \right) \right] , da'' \, db''  \\ \nonumber
&& = \int_{\Space{R}{2n}} m(a,b,a'',b'') (-ia'' + b'')   \\ \nonumber
&& \times \exp \left[ \frac{\pi}{2h} \left( 2(a''+ib'')(a'-ib') - a''^2 - b''^2 - a'^2 - b'^2 \right) \right] da'' \, db''.
\end{eqnarray}
The function
\begin{equation} \label{eq:soloflinexofnonlineqn}
m(a,b,a',b') = \exp \left( \frac{\pi}{h}(a+ib)(-ia'-b')-\frac{\pi}{2h}(a^2 + b^2 + a'^2 + b'^2) \right)
\end{equation}
can be shown to satisfy these equations through the repeated use of formulae (\ref{eq:waveletwith}) and (\ref{eq:waveletwithx}). Another verification of this is given in Corollary \ref{cor:linisrightbynonlinex} which appears later in the thesis. Using formula (\ref{eq:waveletexpofsoltomabcd}) we can obtain the integral operator corresponding to this transformation.

\subsection{A Non-Linear Example} \label{subsect:nonlinexample}

\indent

We now go through an example of a non-linear transformation in detail to demonstrate how equations (\ref{eq:thefinaleqnforkernsinctsof}), (\ref{eq:thefinaleqnforkernsinctsoftwo}) can be used for modelling non-linear canonical transformations. The example we discuss is the canonical transformation given by the following equations
\begin{eqnarray} \label{eq:qeqnforstartofnonlinex}
Q &=& q \cos (t) + p \sin (t) - C \\ \label{eq:peqnforstartofnonlinex}
P &=& -q\sin (t) + p \cos (t) - C
\end{eqnarray}
where $C$ is a constant. This is similar to the canonical transformation which generates the time evolution for the classical forced oscillator (see equation (\ref{eq:classicalflowfo})). This is a relatively straightforward non-linear canonical transformation. To apply this method to more complicated non-linear examples numerical methods would need to be used to solve equations (\ref{eq:thefinaleqnforkernsinctsof}),(\ref{eq:thefinaleqnforkernsinctsoftwo}). Before we set out to form and solve the system of equations (\ref{eq:thefinaleqnforkernsinctsof}),(\ref{eq:thefinaleqnforkernsinctsoftwo}) for this example we present some preliminary results which will help us along the way.
\begin{Lemma}
The following relations hold
\begin{eqnarray} \label{eq:xandcs}
\langle x \, \vab , \vabd \rangle &=& \frac{1}{2h} (ia-b -ia' -b') \langle \vab, \vabd \rangle \\ \label{eq:yandcs}
\langle y \, \vab , \vabd \rangle &=& \frac{1}{2h} (ib+a -ib' +a') \langle \vab, \vabd \rangle.
\end{eqnarray}
\end{Lemma}
\begin{proof}
Since
\begin{eqnarray} \nonumber
\lefteqn{ \langle x v_{(h,a,b)} , v_{(h,a',b')} \rangle } \\ \nonumber
&& =\int_{\Space{R}{2n}} x \exp \left( \pi x (ia-b-ia'-b') + \pi y (ib+a-ib'+a') \right)  \\ \nonumber
&& \hspace{2cm} \times \exp \left( -\pi h (x^2 + y^2) \right) \, dx \, dy \, \\ \nonumber
&& \hspace{2.5cm} \times \exp \left( -\frac{\pi}{2h} \left( a^2 + b^2 + a'^2 + b'^2 \right) \right),
\end{eqnarray}
(\ref{eq:xandcs}) follows through a direct application of equation (\ref{eq:waveletwith}). Similarly since
\begin{eqnarray} \nonumber
\lefteqn{ \langle y v_{(h,a,b)} , v_{(h,a',b')} \rangle } \\ \nonumber
&& =\int_{\Space{R}{2n}} y \exp \left( \pi x (ia-b-ia'-b') + \pi y (ib+a-ib'+a') \right)  \\ \nonumber
&& \hspace{2cm} \times \exp \left( -\pi h (x^2 + y^2) \right) \, dx \, dy \, \\ \nonumber
&& \hspace{2.5cm} \times \exp \left( -\frac{\pi}{2h} \left( a^2 + b^2 + a'^2 + b'^2 \right) \right)
\end{eqnarray}
we can obtain (\ref{eq:yandcs}) using equation (\ref{eq:waveletwith}).
\end{proof}
\begin{Lemma} \label{eq:posandmomincsmatrixelements}
We have the following relations
\begin{eqnarray} \nonumber
\lefteqn{\langle \proj (q) * \vab , \vabd \rangle} \\ \label{eq:projqandcs}
&& \hspace{2cm} = \frac{1}{2} [(a+ib) + (a'-ib')] \langle \vab , \vabd \rangle \\ \nonumber
\lefteqn{ \langle \proj (p) * \vab , \vabd \rangle} \\ \label{eq:projpandcs}
&& \hspace{2cm} = \frac{1}{2} [(b-ia) + (b'+ia')] \langle \vab , \vabd \rangle.
\end{eqnarray}
\end{Lemma}
\begin{proof}
By (\ref{eq:posonhh})
\begin{eqnarray} \nonumber
\proj (q) * \vab &=& \frac{1}{2\pi i} \left( \Partial{x} + \frac{y}{2} \Partial{s} \right) \vab \\ \nonumber
&=& \frac{1}{2\pi i} (\pi i a - \pi b - \pi hx + \pi ihy) \vab \\ \nonumber
&=& \frac{1}{2} ( a+ib + ihx + hy) \vab.
\end{eqnarray}
So
\begin{eqnarray} \nonumber
\lefteqn{\langle \proj (q) * \vab , \vabd \rangle = \frac{1}{2} (a+ib) \langle \vab , \vabd \rangle} \\ \nonumber
&& \hspace{5cm}  + \frac{h}{2} \langle (y+ix) \vab , \vabd \rangle.
\end{eqnarray}
Using equations (\ref{eq:xandcs}) and (\ref{eq:yandcs}) we get
\begin{eqnarray*}
\lefteqn{\langle \proj (q) * \vab , \vabd \rangle} \\
&=& \left( \frac{1}{2} (a+ib) + \frac{h}{2} \left[ \frac{1}{2h} (ib+a - ib' + a') + \frac{i}{2h} (ia-b-ia'-b') \right] \right) \\
&& \hspace{2cm} \times \langle \vab , \vabd \rangle \\
&=& \frac{1}{2} (a+ib + a'-ib') \langle \vab , \vabd \rangle
\end{eqnarray*}
which is relation (\ref{eq:projqandcs}). (\ref{eq:projpandcs}) follows similarly, by (\ref{eq:momonhh})
\begin{eqnarray} \nonumber
\proj (p) * \vab &=& \frac{1}{2\pi i} \left( \Partial{y} - \frac{x}{2} \Partial{s} \right) \vab \\ \nonumber
&=& \frac{1}{2\pi i} (\pi i b + \pi a - \pi hy - \pi ihx) \vab \\ \nonumber
&=& \frac{1}{2} ( b-ia + ihy - hx) \vab.
\end{eqnarray}
Using equations (\ref{eq:xandcs}) and (\ref{eq:yandcs})
\begin{eqnarray} \nonumber
\langle \proj (p) * \vab , \vabd \rangle = \frac{1}{2} (b-ia + b' + ia') \langle \vab , \vabd \rangle.
\end{eqnarray}
\end{proof}
We are now in a position to create system (\ref{eq:thefinaleqnforkernsinctsof}), (\ref{eq:thefinaleqnforkernsinctsoftwo}) for this example. We are aiming to find the coherent state expansion of the operator $U$ defined by
the equations
\begin{eqnarray} \nonumber
U \proj ( q \cos (t) + p \sin (t) - C ) * v &=& \proj (Q) * U v \\ \nonumber
U \proj (-q \sin (t) + p \cos (t) - C) * v &=& \proj (P) * U v.
\end{eqnarray}
Now if $m(a,b,c,d) = \langle U \vab , v_{(h,c,d)} \rangle$ then by equations (\ref{eq:thefinaleqnforkernsinctsof}),(\ref{eq:thefinaleqnforkernsinctsoftwo}) we have
\begin{eqnarray} \label{eq:bigqinnonlinex}
\lefteqn{\int_{\Space{R}{2n}} m(a, b,a'',b'') \langle \proj (Q) * \vabdd , \vabd \rangle \, da'' \, db''} \\ \nonumber
&=& \cos (t) \int_{\Space{R}{2n}} m (a'', b'',a',b') \langle \proj (q) * \vab , \vabdd \rangle \, da'' \, db'' \\ \nonumber
&& \hspace{1cm} + \sin (t) \int_{\Space{R}{2n}} m (a'',b'',a',b') \langle \proj (p) * \vab , \vabdd \rangle \, da'' \, db'' \\ \nonumber
&& \hspace{1cm} - C \int_{\Space{R}{2n}} m (a'',b'',a',b') \langle \vab , \vabdd \rangle \, da'' \, db'',
\end{eqnarray}
and
\begin{eqnarray} \label{eq:bigpinnonlinex}
\lefteqn{\int_{\Space{R}{2n}} m(a, b,a'',b'') \langle \proj (P) * \vabdd , \vabd \rangle \, da'' \, db''} \\ \nonumber
&=& -\sin (t) \int_{\Space{R}{2n}} m (a'', b'',a',b') \langle \proj (q) * \vab , \vabdd \rangle \, da'' \, db'' \\ \nonumber
&& \hspace{1cm} + \cos (t) \int_{\Space{R}{2n}} m (a'',b'',a',b') \langle \proj (p) * \vab , \vabdd \rangle \, da'' \, db'' \\ \nonumber
&& \hspace{1cm} - C \int_{\Space{R}{2n}} m (a'',b'',a',b') \langle \vab , \vabdd \rangle \, da'' \, db''.
\end{eqnarray}
\begin{Thm} \label{thm:nonlinexamplesoln}
The expression
\begin{eqnarray} \label{eq:solnofnonlinex}
\lefteqn{m(a,b,a',b')} \\ \nonumber
&=& \exp \left( \frac{\pi}{h} [(a+ib)(\cos(t) - i \sin(t))(a'-ib') + C (\cos(t) - i \sin (t) )(a+ib)] \right) \\ \nonumber
&& \hspace{0.5cm} \times \exp \left( -\frac{\pi}{2h} (a^2 + b^2 + a'^2 + b'^2) \right)
\end{eqnarray}
satisfies equations (\ref{eq:bigqinnonlinex}) and (\ref{eq:bigpinnonlinex}).
\end{Thm}
\begin{proof}
We start to prove this by directly substituting (\ref{eq:solnofnonlinex}) into (\ref{eq:bigqinnonlinex}). If we let
\begin{eqnarray} \nonumber
A &=& m(a,b,a'',b'') \langle \vabdd , \vabd \rangle \\ \nonumber
B &=& m (a'',b'' ,a',b') \langle \vab , \vabdd \rangle,
\end{eqnarray}
then using equations (\ref{eq:projqandcs}) and (\ref{eq:projpandcs}) equation (\ref{eq:bigqinnonlinex}) becomes
\begin{eqnarray} \label{eq:midtablesystem}
&& (a'-ib') \int_{\Space{R}{2n}} A \, da'' db'' + \int_{\Space{R}{2n}} (a'' +ib'') A \, da'' \, db'' \\ \nonumber
&& \hspace{0,5cm} = \cos (t) (a+ib) \int_{\Space{R}{2n}} B \, da'' \, db'' + \cos(t)  \int_{\Space{R}{2n}} (a''-ib'') B \, da'' \, db'' \\ \nonumber
&& \hspace{1.5cm} + \sin (t) (b-ia) \int_{\Space{R}{2n}} B \, da'' \, db'' + \sin (t) \int_{\Space{R}{2n}} (b'' + i a'') B \, da'' \, db'' \\ \nonumber
&& \hspace{1.5cm} - C \int_{\Space{R}{2n}} B \, da'' \, db''.
\end{eqnarray}
This is equivalent to
\begin{eqnarray} \label{eq:lastbeforethelemma}
\lefteqn{ (a'-ib') \int_{\Space{R}{2n}} A \, da'' db'' + \int_{\Space{R}{2n}} (a'' +ib'') A \, da'' \, db''} \\ \nonumber
&=& (\cos (t) -i \sin(t) )(a+ib) \int_{\Space{R}{2n}} B \, da'' \, db'' \\ \nonumber
&& \hspace{1.5cm} + (\cos(t) + i \sin(t)) \int_{\Space{R}{2n}} (a''-ib'') B \, da'' \, db'' - C \int_{\Space{R}{2n}} B \, da'' \, db''.
\end{eqnarray}
We now require two results -- Lemma \ref{lem:helpstosolvenonlin} and Lemma \ref{lem:intanadintbaresame} -- which are stated and proved after this proof. By applying Lemma \ref{lem:helpstosolvenonlin}, equation (\ref{eq:lastbeforethelemma}) becomes
\begin{eqnarray}
\lefteqn{ \left[ (a'-ib') + \frac{1}{2} [ (a+ib)(\cos(t)-i \sin(t) ) + (a'-ib')  \right. } \\ \nonumber
&& \hspace{2cm} \left. + (a+ib) (\cos(t) - i \sin(t)) + (-a'+ib') ] \right] \int A \, da'' \, db'' \\ \nonumber
&=& \left[ (\cos(t) -i \sin(t) )(a+ib) + (\cos(t) + i \sin (t))\right. \\ \nonumber
&&  \hspace{0.5cm} \times \frac{1}{2} [(\cos(t) - i \sin (t))(a'-ib'+C)  +(a+ib) \\ \nonumber
&& \left. \hspace{1.5cm} + (\cos (t) -  i \sin (t))(a'-ib'+C) + (-a-ib)] -C \right] \int B \, da'' \, db''
\end{eqnarray}
which  is equivalent to
\begin{eqnarray}
\lefteqn{[a'-ib' + (a+ib)(\cos(t) - i \sin(t))] \int A \, da'' \, db'' } \\ \nonumber
&=& [ (\cos(t)-i \sin (t))(a+ib) + a'-ib' + C -C] \int B \, da'' \, db''.
\end{eqnarray}
Then Lemma \ref{lem:intanadintbaresame} tells us that (\ref{eq:solnofnonlinex}) is a solution of (\ref{eq:bigqinnonlinex}). A similar calculation will show us that (\ref{eq:solnofnonlinex}) also satisfies (\ref{eq:bigpinnonlinex}).
\end{proof}

\begin{Lemma} \label{lem:helpstosolvenonlin}
We have the following results
\begin{eqnarray} \nonumber
\lefteqn{\int a'' A \, da'' \, db''} \\ \nonumber
&=& \frac{1}{2}[(a+ib)(\cos (t) - i \sin (t)) + (a'-ib')] \int A \, da'' \, db'' \\ \nonumber
\lefteqn{\int b'' A \, da'' \, db''} \\ \nonumber
&=& \frac{1}{2}[(a+ib)(-\sin (t) - i \cos (t)) + (ia'+b')] \int A \, da'' \, db'' \\ \nonumber
\lefteqn{\int a'' B \, da'' \, db''} \\ \nonumber
&=& \frac{1}{2}[(\cos (t) - i \sin (t))(a'-ib') + C(\cos(t) -i\sin(t))+(a+ib)] \int B \, da'' \, db'' \\ \nonumber
\lefteqn{\int b'' B \, da'' \, db''} \\ \nonumber
&=& \frac{1}{2}[(\sin (t) + i \cos (t))(a'-ib') + C(i\cos(t) + \sin(t)) + (b-ia)] \int B \, da'' \, db''.
\end{eqnarray}
In the above equations all the integrals are over $\Space{R}{2n}$.
\end{Lemma}
\begin{proof}
Since
\begin{eqnarray} \nonumber
\lefteqn{A=\exp \left( \frac{\pi}{h} [(a+ib)(\cos(t) -i\sin(t))(a''-ib'')\right)} \\ \nonumber
&& \hspace{0.5cm} \times \exp \left( \frac{\pi}{h}[ C (\cos(t) - i \sin (t) )(a+ib) + (a''+ib'')(a'-ib')] \right) \\ \nonumber
&& \hspace{1cm} \times \exp \left( \frac{\pi}{2h} [-2a''^2 -2b''^2 -a'^2 -b'^2-a^2 -b^2] \right),
\end{eqnarray}
by directly applying equation (\ref{eq:waveletwith}) we get the first two results of the Lemma. Also
\begin{eqnarray} \nonumber
\lefteqn{B=\exp \left( \frac{\pi}{h} [(a''+ib'')(\cos(t) -i\sin(t))(a'-ib') \right) } \\ \nonumber
&& \hspace{0.5cm} \times \exp \left( \frac{\pi}{h} [C (\cos(t) - i \sin (t) )(a''+ib'') + (a+ib)(a''-ib'') ] \right) \\ \nonumber
&& \hspace{1cm} \times \exp \left( \frac{\pi}{2h} [-2a''^2 -2b''^2 -a'^2 -b'^2-a^2 -b^2] \right).
\end{eqnarray}
From this the final two results of the Lemma follow by again directly applying equation (\ref{eq:waveletwith}).
\end{proof}
\begin{Lemma} \label{lem:intanadintbaresame}
We have the following relation
\begin{equation}
\int_{\Space{R}{2n}} A \, da'' \, db'' = \int_{\Space{R}{2n}} B \, da'' \, db''.
\end{equation}
\end{Lemma}
\begin{proof}
By equation (\ref{eq:waveletwithx}) we have
\begin{eqnarray} \nonumber
\lefteqn{ \int A \, da'' \, db'' = \exp \left( \frac{\pi^2/h^2}{4\pi/h} \{ [ (a+ib)(\cos(t) - i \sin (t) ) + (a'-ib')]^2 \right.} \\ \nonumber
&& \left. \hspace{4.5cm} + [-i (a+ib) (\cos(t) - i \sin (t) ) + i (a'-ib') ]^2 \} \right) \\ \nonumber
&& \hspace{1.7cm} \times \exp \left( \frac{\pi}{h}[ C (\cos(t) - i \sin (t) )(a+ib)] - \frac{\pi}{2h} (a^2 + b^2 + a'^2 + b'^2 ) \right) \\ \nonumber
&=& \exp \left( \frac{\pi}{h} \{ (a+ib)(\cos(t) -i \sin (t) )(a'-ib') + C (\cos(t) -i \sin(t))(a+ib) \} \right. \\ \nonumber
&& \left. \hspace{1.8cm} - \frac{\pi}{2h} (a^2 + b^2 + a'^2 + b'^2 ) \right) \\ \nonumber
&=& \exp \left( \frac{\pi}{h}  (a+ib)[(\cos(t) -i \sin (t) )(a'-ib') + C (\cos(t) -i \sin(t))]  \right. \\ \label{eq:aandbarethesameone}
&& \left. \hspace{1.8cm} - \frac{\pi}{2h} (a^2 + b^2 + a'^2 + b'^2 ) \right).
\end{eqnarray}
Also using (\ref{eq:waveletwithx}) we get
\begin{eqnarray} \nonumber
\lefteqn{\int B \, da'' \, db''} \\ \nonumber
&=& \exp \left( \frac{\pi}{4h} \{ [ (\cos(t) - i \sin(t))(a'-ib') + C (\cos(t) - i\sin(t) ) + (a+ib) ]^2 \right. \\ \nonumber
&& \left. \hspace{1.3cm} + [i(\cos(t) -i\sin(t) )(a'-ib') + i C ( \cos(t) - i \sin(t)) - i (a+ib)]^2 \} \right) \\ \nonumber
&& \times \exp \left( - \frac{\pi}{2h} (a^2 + b^2 + a'^2 + b'^2 ) \right) \\ \nonumber
&=& \exp \left( \frac{\pi}{h} (a+ib)[(\cos(t)-i\sin(t))(a'-ib') + C (\cos(t) - i \sin(t))] \right. \\ \label{eq:andbarethesametwo}
&& \left. \hspace{1.5cm} - \frac{\pi}{2h} (a^2 + b^2 + a'^2 + b'^2 ) \right).
\end{eqnarray}
By comparing (\ref{eq:aandbarethesameone}) and (\ref{eq:andbarethesametwo}) we obtain the desired result.
\end{proof}

We now present the integration kernel which corresponds to the canonical transformation given by (\ref{eq:qeqnforstartofnonlinex}) and (\ref{eq:peqnforstartofnonlinex}).
\begin{Corollary}
The  operator on $\hilbh$ corresponding to the integration kernel
\begin{eqnarray} \nonumber
&& \exp \left\{ 2\pi ih (s-s') + \pi h (\cos (t) - i \sin (t))\left( ix+y + \frac{C}{h} \right) (y'-ix') \right.  \\ \nonumber
&& \left. \hspace{2.5cm} - \frac{\pi h}{2} (x^2 + y^2 + x'^2 + y'^2) \right\}
\end{eqnarray}
models the canonical transformation given by equations (\ref{eq:qeqnforstartofnonlinex}) and (\ref{eq:peqnforstartofnonlinex}).
\end{Corollary}
\begin{proof}
Let $v$ be an arbitrary element of $\hilbh$ and let $U$ be the operator on $\hilbh$ corresponding to the canonical transformation given by ((\ref{eq:qeqnforstartofnonlinex}), (\ref{eq:peqnforstartofnonlinex})). By equation (\ref{eq:waveletexpofsoltomabcd})
\begin{eqnarray} \nonumber
\lefteqn{(Uv)(s,x,y) = \int_{\Space{R}{7}} m(a,b,a',b') v(s',x',y') \overline{\vab (s',x',y')} } \\ \nonumber
&& \hspace{4.5cm} \times \vabd (s,x,y) \, da \, db \, ds \, dx' \, dy' \, da' \, db'
\end{eqnarray}
where $m(a,b,a',b') = \langle U \vab , \vabd \rangle$. So the integration kernel corresponding to the operator $U$ is
\begin{eqnarray} \nonumber
\lefteqn{K(s,x,y,s',x',y')} \\ \nonumber
&&  = \int_{\Space{R}{4}} m(a,b,a',b') \overline{\vab (s',x',y') } \vabd (s,x,y) \, da \, db \, da' \, db'.
\end{eqnarray}
For this example $m(a,b,a',b')$ is given in Theorem \ref{thm:nonlinexamplesoln}, so
\begin{eqnarray*}
\lefteqn{K(s,x,y,s',x',y')} \\
&=& \int_{\Space{R}{4}} \exp \left\{ \frac{\pi}{h} [(a+ib)(\cos(t) - i \sin(t))(a'-ib') \right.\\
&& \left. \hspace{3cm} + C (\cos(t) - i \sin (t) )(a+ib)] \right\} \\
&& \hspace{0.5cm} \times \exp \left( -\frac{\pi}{2h} (a^2 + b^2 + a'^2 + b'^2) \right) \\
&& \times \exp \left[ 2\pi ih (s-s') + \pi a (-ix'+y') + \pi b (-iy'-x') \right. \\
&& \left. \hspace{3cm} + \pi a' (ix+y) + \pi b' (iy-x) \right] \\
&& \times \exp \left\{ -\frac{\pi h}{2} (x^2 + y^2 +x'^2 + y'^2) -\frac{\pi}{2h} (a^2 + b^2 + a'^2 + b'^2 ) \right\} \, da \, db \, da' \, db' \\
\end{eqnarray*}
\begin{eqnarray*}
&=& \int_{\Space{R}{4}} \exp \left\{ 2\pi ih (s-s') + a' \left[ \frac{\pi}{h} (a+ib) (\cos(t) - i \sin (t)) + \pi (ix+y) \right] \right. \\
&& \left. \hspace{3.4cm} + b' \left[ -i \frac{\pi}{h} (a+ib) (\cos(t) - i \sin (t)) + i \pi (ix+y) \right]  \right\} \\
&& \times \exp \left\{ -\frac{\pi}{h} (a^2 + b^2 + a'^2 + b'^2) - \frac{\pi h}{2} (x^2 + y^2 + x'^2 + y'^2 ) \right\} \\
&& \times \exp \left\{ \frac{\pi}{h} C (\cos(t) - i \sin (t)) (a+ib) + \pi a(-ix'+y') \right. \\
&& \left. \hspace{3cm} + \pi b (-iy'-x') \right\} \, da' \, db' \, da \, db.
\end{eqnarray*}
Using equation (\ref{eq:waveletwithx}) this becomes
\begin{eqnarray*}
\lefteqn{\int_{\Space{R}{2}} \exp \left\{ 2\pi ih (s-s') + \pi (a+ib)(\cos (t) - i \sin (t))(ix+y) \right\} } \\
&& \times \exp \left\{ -\frac{\pi}{h} (a^2+b^2) -\frac{\pi h}{2} (x^2 + y^2 + x'^2 + y'^2) \right\} \\
&& \times \exp \left\{ C(\cos(t) - i\sin(t))(a+ib) + \pi (a-ib)(-ix'+y') \right\} \, da \, db  \\
\end{eqnarray*}
\begin{eqnarray*}
&=& \exp \left\{ 2\pi ih(s-s') + a \left[ \pi (\cos(t) - i \sin(t))\left( ix+y+ \frac{C}{h} \right) + \pi (y'-ix') \right] \right. \\
&& \left. \hspace{2.5cm} + b \left[ i \pi (\cos(t) - i \sin (t))\left(ix+y+\frac{C}{h} \right) - i \pi (y'-ix') \right] \right\} \\
&& \times \exp \left\{ -\frac{\pi}{h} (a^2+b^2) - \frac{\pi h}{2} (x^2 + y^2 + x'^2 + y'^2 ) \right\}.
\end{eqnarray*}
Another application of (\ref{eq:waveletwithx}) reduces this to
\begin{eqnarray*}
&& \exp \left\{ 2\pi ih (s-s') + \pi h (\cos(t)-i \sin (t))\left(ix+y + \frac{C}{h} \right)(y'-ix') \right\}  \\
&& \hspace{0.5cm} \times \exp \left\{ -\frac{\pi h}{2} (x^2 + y^2 + x'^2 + y'^2 ) \right\}
\end{eqnarray*}
which gives us the required result.
\end{proof}
We now give another corollary of Theorem \ref{thm:nonlinexamplesoln}. This corollary gives further verification of our solution to the linear example presented at the end of Subsection \ref{subsect:nonlintranseq}.
\begin{Corollary} \label{cor:linisrightbynonlinex}
Expression (\ref{eq:soloflinexofnonlineqn}) gives the correct matrix elements of the operator corresponding to canonical transformation (\ref{eq:statementoflineqfornonlineqn}).
\end{Corollary}
\begin{proof}
This follows by putting $t= \frac{\pi}{2}$ and $C=0$ into (\ref{eq:solnofnonlinex}).
\end{proof}

\chapter{The Kepler/Coulomb Problem} \label{chap:keplercoulomb}

\indent

In this chapter we consider modelling the Kepler/Coulomb\index{Kepler/Coulomb problem} problem in classical and quantum mechanics. The Kepler/Coulomb problem is the three dimensional classical system governed by the $\frac{1}{\sqrt{q_1^2 + q_2^2 + q_3^2}}$ potential and the associated quantum system. We use the name Kepler/Coulomb problem since we are looking at both the classical and the quantum problems. The associated classical problem is often referred to as the Kepler\index{Kepler problem} problem \cite{GuilleminSternberg90} since it was studied in great depth by Kepler in the early 1600s. The classical problem also gave birth to classical analytic mechanics in the works of Newton. The problem of quantising this system is closely related to the fundamental problem of mathematically modelling the Hydrogen atom\index{hydrogen atom}. The $\frac{1}{\sqrt{q_1^2 + q_2^2 + q_3^2}}$ potential in the quantum mechanics literature is usually referred to as the Coulomb potential.

In the 1970s some important and interesting work was done on the classical Kepler problem by Moser \cite{Moser70} and Souriau \cite{Souriau74} --- this work is summarised in \cite{GuilleminSternberg90}. The work involved showing that the classical flow of the Kepler problem was equivalent to the geodesic flow on the four dimensional sphere. The quantum system has been of interest to physicists since the very birth of the subject due to its close relation to the Hydrogen atom \cite{Bohm01}. The standard treatment of the Coulomb potential in the quantum mechanics literature \cite[Sect. 12.6]{Merzbacher70}  \cite[Chap. 11]{Messiah61} involves finding the eigenvalues and eigenfunctions of the Kepler/Coulomb Hamiltonian using spherical harmonics \cite{Vilenkin68} and associated Laguerre polynomials \cite{Vilenkin68}. The geometric quantisation of the Kepler problem was described by Simms in the papers \cite{Simms73,Simms74}.

In Section \ref{sect:kepcouprobstart} we look at the $p$-mechanisation of the Kepler/Coulomb problem. We present the $p$-mechanisation of the Kepler/Coulomb Hamiltonian along with the $p$-mechanisation of its constants of motion (that is, the angular momentum vector and the Laplace--Runge--Lenz vector). In Section \ref{sect:pdynforkepcou} we present the $p$-dynamic equation for the Kepler/Coulomb problem and decribe its non-trivial nature. The limitations of the $L^2 (\Space{R}{3})$ and $\fock$ spaces in analysing the Kepler/Coulomb problem are discussed in Section \ref{sect:kepcouinltwoandfock}.
In Section \ref{sect:spherpolinpmech} we develop a new form of the infinite dimensional unitary irreducible representation of the Heisenberg group using spherical polar coordinates. The purpose of Section \ref{sect:transfposspace} is to generalise this to any transformation of position space. In Section \ref{sect:klaudcs} we describe Klauder's coherent states for the hydrogen atom. These coherent states are used to define a new Hilbert space in Section \ref{sect:klaudrep}. This Hilbert space is shown to be very useful for analysing the Kepler/Coulomb problem. In Section \ref{sect:genofkepcou} we extend this approach to more general systems.

\section{The $p$-Mechanisation of the \newline Kepler/Coulomb Problem} \label{sect:kepcouprobstart}

\indent

In this section we derive the $p$-mechanisation of the Hamiltonian for the Kepler/Coulomb problem. The Kepler/Coulomb Hamiltonian in three dimensional classical mechanics is\footnote{Here we have taken all constants equal to one to reduce the technicalities in the calculations.}
\begin{equation} \label{eq:ckepham}
H(q,p) = \frac{\|p\|^2}{2} - \frac{1}{\|q\|}.
\end{equation}
All the norms in the above equation are the $2$-norm on $\Space{R}{3}$ (that is $\|x\| = \sqrt{x_1^2 + x_2^2 + x_3^2}$ ). We wish to obtain the $p$-mechanisation of this Hamiltonian. In order to do this we need a known result on the inverse Fourier transform of $\frac{1}{\|q\|}$.
\begin{Lemma}
The inverse Fourier transform (see (\ref{eq:ftonschwp})) of $\frac{1}{\|q\|}$ is the element of $\mathcal{S}'(\Space{R}{3})$, $\frac{1}{\pi \|x\|^2}$, that is
\begin{equation} \nonumber
\int_{\Space{R}{3}} \frac{1}{\|q\|} \int_{\Space{R}{3}} \phi (x) e^{-2\pi i qx} \, dx \, dq = \int_{\Space{R}{3}} \frac{1}{\pi \|x\|^2} \phi (x) \, dx,
\end{equation}
for any $\phi \in \mathcal{S}'(\Space{R}{3})$.
\end{Lemma}
\begin{proof}
See for example \cite[Chap 2, Sect 3.3]{GelfandShilov77} for a proof. A slight change by a factor of $2\pi$ is needed to get the result into the form we require.
\end{proof}
The $p$-mechanisation (see (\ref{eq:pmechmap})) of $H$ is
\begin{equation} \label{eq:pkepham}
B_H(s,x,y) = \left( -\frac{1}{8\pi^2} \zerodelytwo - \delta (s) \frac{1}{\pi \|x\|^2} \delta (y) \right).
\end{equation}
This is a distribution in the space $\mathcal{S}'(\Space{H}{3})$.
 $\delta^{(2)} (y)$ is notation for the distribution $\left( \Partialtwo{y_1} + \Partialtwo{y_2} + \Partialtwo{y_3} \right) \delta (y)$.

Three classical constants of the motion are the components of the classical angular momentum vector\index{angular momentum}
\begin{equation} \label{eq:cangmomvec}
l=q \times p
\end{equation}
where $\times$ here denotes the cross product of two vectors. Using summation convention the $i$th component of the classical angular momentum vector can be written as
\begin{equation} \label{eq:cangmomi}
l_i = \epsilon_{ijk} q_j p_k ,
\end{equation}
where
\begin{displaymath}
\epsilon_{ijk} =
\left\{
\begin{array}{ll}
1 & \textrm{if $(ijk)$ is an even permutation of (123)} \\
-1 & \textrm{if $(ijk)$ is an odd permutation of (123)} \\
0 & \textrm{otherwise.}
\end{array} \right.
\end{displaymath}
The $p$-mechanisation of the $i$th component of angular momentum is
\begin{equation} \label{eq:pmechofangmom}
L_i = -\frac{1}{4\pi^2} \epsilon_{ijk} \delta (s) \delta_{(j)}^{(1)}(x) \delta_{(k)}^{(1)} (y).
\end{equation}
$\delta_{(j)}^{(1)}(x)$ represents the distribution $\Partial{x_j} \delta (x_1, x_2, x_3)$. The total angular momentum $l^2 = \sum_{j=1}^3 l_j^2$ is another constant of the motion. The $p$-mechanisation of $l^2$ is
\begin{equation} \nonumber
L^2 = \sum_{j=1}^3 L_j * L_j
\end{equation}
where $*$ represents the noncommutative convolution on the Heisenberg group.

Three more constants of the classical motion are the three components of the classical Laplace--Runge--Lenz vector\index{Laplace--Runge--Lenz vector}
\begin{equation} \label{eq:clenzvec}
f=l \times p + \frac{q}{r}.
\end{equation}
Again using summation convention the $i$th component of the Laplace--Runge--Lenz vector can be written as
\begin{equation} \label{eq:clenzi}
f_i = \epsilon_{ijk} l_j p_k + \frac{q_i}{r}.
\end{equation}
The $p$-mechanisation of this observable is
\begin{eqnarray} \label{eq:pmechoflenz}
\lefteqn{F_i = \frac{1}{2\pi i} \epsilon_{ijk} L_j * \zerodelky } \\ \nonumber
&& \hspace{2.5cm} + \frac{1}{2\pi i} \zerodelix * \delta (s) \frac{1}{\pi \| x \|^2} \delta (y).
\end{eqnarray}
\begin{Remark}
\emph{The Hamiltonian along with both the angular momentum and the Lenz vector are shown to satisfy an $o(4)$ symmetry\index{o(4) symmetry} \cite{GuilleminSternberg90} under both the Poisson brackets and the quantum commutator. Using the commutation of the left and right invariant vector fields along with the results
\begin{equation} \nonumber
\sum_{j=1}^3 \Partial{x_j} x_j \frac{1}{\|x\|^2} = \frac{1}{\|x\|^2}
\end{equation}
\begin{equation} \nonumber
x_i \Partial{x_j} \frac{1}{\|x\|^2} = -\frac{2 x_i x_j}{\| x \|^4 } = x_j \Partial{x_i} \frac{1}{\|x \|^2}
\end{equation}
we get the same $o(4)$ symmetry under the universal brackets (see equation (\ref{eq:pmechbrackets})). This means that if $\xi$ and $\eta$ are elements of $\Space{R}{3}$ then
\begin{eqnarray} \label{eq:doto4bracket1}
\ub{L.\xi}{L.\eta}{} &=& L.(\xi \times \eta) \\ \label{eq:doto4bracket2}
\ub{L.\xi}{F.\eta}{} &=& F.(\xi \times \eta) \\ \label{eq:doto4bracket3}
\ub{F.\xi}{F.\eta}{} &=& -2 H * L.(\xi \times \eta).
\end{eqnarray}
}
\end{Remark}

\section{The $p$-Dynamic Equation for the \newline  Kepler/Coulomb Problem} \label{sect:pdynforkepcou}

\indent

In Chapter \ref{chap:forcedosc} we solved the $p$-dynamic equation (see (\ref{eq:pdyneqn})) for the harmonic and forced oscillators. This showed us that the classical and quantum dynamics were generated from the same source. We would like to do the same for the Kepler/Coulomb problem. The Kepler/Coulomb $p$-dynamic equation\index{p-dynamic equation!Kepler/Coulomb problem} for an arbitrary $p$-mechanical observable, $B$, takes the form:
\begin{eqnarray*}
\lefteqn{ \Fracdiffl{B}{t} = - \sum_{j=1}^3 y_j \Fracpartial{B}{x_j} + \frac{1}{\pi} \int_{\Space{R}{3}} \frac{1}{\|x-x'\|^2}\left[ B \left( s+\frac{1}{2}y(x-x'), x' ,y \right) \right. } \\
&& \left. \hspace{7cm} - B \left( s+\frac{1}{2}y(x'-x), x' ,y \right) \right] dx'.
\end{eqnarray*}
This equation is very hard to analyse due to being the mixture of a differential equation and an integral equation. This shows that taking this approach to obtain relations between classical and quantum mechanics is  not suitable for this system. This leads us to look at new representations of the Heisenberg group --- this is the main focus for the rest of this chapter.

\section{The Kepler/Coulomb Problem in $L^2 (\Space{R}{3})$ and $\fock$} \label{sect:kepcouinltwoandfock}

\indent

We now prove two Lemmas which show the limitations of both the \newline Schr\"odinger\index{Schr\"odinger representation} representation on $L^2(\Space{R}{3})$ and the $\rho_h$\index{$\rho_h$} representation on $\fock$\index{$\fock$} when dealing with the Kepler/Coulomb problem.
\begin{Lemma}

The representation of the distribution $B_H$ (from equation (\ref{eq:pkepham})) using the Schr\"odinger representation (see (\ref{eq:schrorep})) on\footnote{This operator is not defined on the whole of $L^2(\Space{R}{3})$ --- it is defined on the space $\mathcal{S}(\Space{R}{3})$ as discussed in Section \ref{sect:rhs}.} $L^2(\Space{R}{3})$
is\footnote{Throughout this chapter we use $\eta$ to denote an element of $L^2 (\Space{R}{3})$ since we reserve $\psi$ for an element of the space $\fsp$ which we introduce later.}
\begin{equation} \nonumber
(\rho_h^S (B_H) \eta )(\xi) = \left( -\frac{h^2}{8\pi^2} \nabla^2 + \frac{1}{\|\xi\|} \right) \eta (\xi).
\end{equation}
\end{Lemma}

\begin{proof}

From (\ref{eq:repofadistnonaliegp}) we have the definition for the representation $\rho_h^S$ of a distribution $B_H \in \mathcal{S}'(\Space{H}{3})$ is

\begin{equation} \label{eq:firstltworepofkep}
\langle \rho_h^S (B_H) \eta_1 , \eta_2 \rangle = \langle \langle \rho_h^S (s,x,y) \eta_1 , \eta_2 \rangle , B_H (s,x,y) \rangle.
\end{equation}
$\eta_1$ and $\eta_2$ are elements of $L^2 (\Space{R}{3})$ such that $\langle \rho_h^S (s,x,y) \eta_1 , \eta_2 \rangle$ is in $\mathcal{S}(\Space{H}{3})$. The right hand side of this equation is equal to
\begin{eqnarray*} \nonumber
&& \int_{\Space{R}{7}} \int_{\Space{R}{3}} e^{-2\pi ihs + 2\pi ix \xi + \pi ihxy} \eta_1 (\xi+hy) \overline{\eta_2 (\xi )} \, d\xi \\
&& \hspace{2.5cm} \times \left( -\frac{1}{8 \pi^2} \zerodelytwo + \kephamwkp \right) \, ds \, dx \, dy \, \\
&& = - \frac{1}{8 \pi^2} \int_{\Space{R}{7}} \int_{\Space{R}{3}}  e^{-2\pi ihs + 2\pi ix\xi + \pi ihxy} \eta_1 (\xi+hy) \overline{\eta_2 (\xi) } \, d\xi \, \zerodelytwo \, ds \, dx \, dy \\
&& \hspace{1cm} + \frac{1}{\pi} \int_{\Space{R}{3}} \int_{\Space{R}{3}} e^{2\pi i x \xi } \eta_1 (\xi) \overline{ \eta_2 (\xi) } \, d\xi \frac{1}{\|x\|^2} \, dx \\
&& = \left\langle \left( - \frac{h^2}{8 \pi^2} \nabla^2  + \frac{1}{\|\xi\|}  \right) \eta_1 , \eta_2 \right\rangle.
\end{eqnarray*}
At the final step we used the fact that the Fourier transform of $\frac{1}{\pi \|x\|^2}$ is $\frac{1}{\|\xi\|}$. So equation (\ref{eq:firstltworepofkep}) becomes
\begin{equation} \nonumber
\langle \rho_h^S (B_H) \eta_1 , \eta_2 \rangle = \left\langle \left( -\frac{h^2}{8 \pi^2} \nabla^2 + \frac{1}{\|\xi\|}  \right) \eta_1 , \eta_2 \right\rangle.
\end{equation}
\end{proof}
Under this representation we have an operator on $L^2(\Space{R}{3})$ associated with the Kepler/Coulomb Hamiltonian. Unfortunately both the Schr\"odinger and Heisenberg equations of motion are hard to study using this operator \cite[Chap. 6]{Bohm01}.

In the forced and harmonic oscillator examples we saw that the $\fock$ representation made the problem much easier to solve. We now work out the representation of the Kepler/Coulomb Hamiltonian in the $\fock$ representation.
\begin{Lemma}
The $\rho_h$ representation of the Kepler/Coulomb Hamiltonian, $B_H$, applied to $f \in \fock$ is
\begin{eqnarray} \nonumber
\lefteqn{\rho_h (B_H) f(q,p) = -\frac{1}{8\pi^2}  \sum_{j=1}^{3} \left(-2\pi i p_j - \frac{h}{2} \Partial{q_j}\right)^2 f(q,p) } \\ \nonumber
&& \hspace{5cm} + \int_{\Space{R}{3}} e^{-2\pi i q x } \frac{1}{\|x\|^2} f \left( q,p + \frac{h}{2} x \right) \, dx .
\end{eqnarray}
\end{Lemma}
\begin{proof}
Equation (\ref{eq:firstltworepofkep}) holds for this representation with $\rho_h^S$ replaced by $\rho_h$ and $\psi_1, \psi_2$ being replaced by functions $f_1,f_2 \in \fock$ such that $\langle \rho_h (g) f_1 , f_2 \rangle \in \mathcal{S}(\Space{H}{3})$. The right hand side of this is
\begin{eqnarray*} \nonumber
&& \int_{\Space{R}{7}} \int_{\Space{R}{6}} e^{-2\pi i (hs+qx+py)} f_1 \left( q- \frac{h}{2} y , p + \frac{h}{2} x \right) \overline{ f_2 (q,p) } \, dq \, dp \\
&& \hspace{1.5cm} \times \left( -\frac{1}{8\pi^2} \zerodelytwo + \kephamwkp \right) \, ds \, dx \, dy \\
&& =  -\frac{1}{8\pi^2} \left\langle \sum_{j=1}^{3} \left(-2\pi i p_j - \frac{h}{2} \Partial{q_j}\right)^2 f_1 , f_2 \right\rangle \\
&& \hspace{1.5cm} + \int_{\Space{R}{3}} \int_{\Space{R}{6}} e^{-2\pi i qx} f_1 \left( q , p + \frac{h}{2} x \right) \overline{ f_2 (q,p) } \, dq \, dp \frac{1}{\pi \|x\|^2} \, dx \\
&& = \left\langle  -\frac{1}{8\pi^2}  \sum_{j=1}^{3} \left(-2\pi i p_j - \frac{h}{2} \Partial{q_j}\right)^2 f_1 (q,p), f_2 (q,p) \right\rangle \\
&& \hspace{1.5cm} + \left\langle \int_{\Space{R}{3}} e^{-2\pi i qx} \frac{1}{\pi \|x\|^2} f_1 \left( q , p + \frac{h}{2} x \right) \, dx , f_2 (q,p)  \right\rangle. \\
\end{eqnarray*}
From this the result follows in an analogous manner to the last proof.
\end{proof}
Again the operator we obtain is hard to analyse. In this representation we do not get a clear time development. We have now shown that both the Schr\"odinger representation and the $\rho_h$ representation are insufficient when studying the Kepler/Coulomb problem. This leads us to a search for different Hilbert spaces and different representations of the Heisenberg group. We hope to find a space in which time evolution for the Kepler/Coulomb problem is clear.

\section{Spherical Polar Coordinates in \newline $p$-Mechanics} \label{sect:spherpolinpmech}

\indent

It has been shown that the Schr\"odinger equation on $L^2 (\Space{R}{3})$ for the Kepler/Coulomb problem is simplified through the use of spherical polar coordinates \cite[Sect. 12.5]{Merzbacher70} \cite[Chap. 11]{Messiah61}. We now use spherical polar coordinates to develop another form of the unitary irreducible infinite dimensional representation of the Heisenberg group. We first give a summary of spherical polar co-ordinates.

Spherical polar coordinates\index{spherical polar coordinates} have been of great use in solving numerous problems with spherical symmetries from a wide range of disciplines. Spherical polar coordiantes let us map from $\Space{R}{3} \setminus \{ ( \xi_1 , \xi_2 , \xi_3 ) : \xi_1 = \xi_2 = 0 \}$ to the space\index{$\mathbb{SP}^3$} $\mathbb{SP}^3 = \{ (r,\theta ,\phi) : r > 0 , 0 \leq \theta < 2 \pi , 0 < \phi < \pi \}$. The mapping from $\Space{R}{3} \setminus \{ ( \xi_1 , \xi_2 , \xi_3 ) : \xi_1 = \xi_2 = 0 \}$ to $\mathbb{SP}^3$ is defined by \cite[Sect.10.4]{BaxandallLiebeck86}
\begin{equation} \label{eq:spmapr}
r = ( \xi_1^2 + \xi_2^2 + \xi_3^2 )^{\frac{1}{2}} ,
\end{equation}
\begin{equation} \label{eq:spmapth}
\theta =  \left\{ \begin{array}{ll} \tan^{-1} \left( \frac{\xi_2}{\xi_1} \right), & \textrm{if $\xi_2 \geq 0$ and $\xi_1 \neq 0$}, \\
\pi + \tan^{-1} \left( \frac{\xi_2}{\xi_1} \right),  & \textrm{if $\xi_2 < 0$ and $\xi_1 \neq 0$}, \\
\frac{\pi}{2}, & \textrm{$\xi_2 >0$ and $\xi_1 =0$}, \\
\frac{3\pi}{2}, & \textrm{$\xi_2 < 0$ and $\xi_1 =0$},
\end{array} \right.
\end{equation}
\begin{equation} \label{eq:spmapph}
\phi = \left\{ \begin{array}{ll} \sin^{-1} \left( \frac{(\xi_1^2 + \xi_2^2 )^{1/2}}{(\xi_1^2 + \xi_2^2 + \xi_3^2)^{1/2}} \right), & \textrm{if $\xi_3 \geq 0$}, \\
\pi - \sin^{-1} \left( \frac{(\xi_1^2 + \xi_2^2 )^{1/2}}{(\xi_1^2 + \xi_2^2 + \xi_3^2)^{1/2}} \right), & \textrm{if $\xi_3 < 0$}.
\end{array} \right.
\end{equation}
We denote the above map by\index{$\msp$} $\msp$ and take the values of $\tan^{-1}$, $\sin^{-1}$ in $[0,\pi)$, $[0,\frac{\pi}{2})$ respectively. The inverse\index{$\mspi$} mapping, $\mspi$ from $\Space{SP}{3}$ to $\Space{R}{3} \setminus \{ ( \xi_1 , \xi_2 , \xi_3 ) : \xi_1 = \xi_2 = 0 \}$ is defined by
\begin{eqnarray} \label{eq:spmapone}
\xi_1 &=& r \cos ( \theta ) \sin (\phi ) \\ \label{eq:spmaptwo}
\xi_2 &=& r \sin (\theta ) \sin ( \phi ) \\ \label{eq:spmapthree}
\xi_3 &=& r \cos ( \phi ).
\end{eqnarray}

Using these mappings we can transform the space $\ltworthree$ into another Hilbert space\index{$\fsp$}, $\fsp$, by transforming the domain through the map $\msp$.
\begin{Defn}
The space $\fsp$ is defined as
\begin{equation} \nonumber
\fsp = \{ \psi (r,\theta,\phi) = \eta ( \mspi (r,\theta,\phi) ) : \eta \in \ltworthree \}.
\end{equation}
The inner product on $\fsp$ is given by
\begin{equation} \label{eq:iponfsp}
\langle \psi_1 , \psi_2 \rangle_{\fsp} = \int_0^{2\pi} \, \int_0^{\pi} \, \int_{0}^{\infty} \psi_1 (r,\theta,\phi) \overline{\psi_2 (r,\theta,\phi)} r^2 \sin ( \theta ) \, dr \, d\theta \, d\phi.
\end{equation}
\end{Defn}
Note that $\fsp$ is a set of functions with domain $\{ (r,\theta ,\phi) : r > 0 , 0 \leq \theta < 2 \pi , 0 < \phi < \pi \}$.
Now we define a mapping between $\ltworthree$ and $\fsp$.
\begin{Defn}
The mapping\index{$\Phi$} $\Phi : \ltworthree \rightarrow \fsp$ is defined by
\begin{equation} \nonumber
(\Phi \eta) (r, \theta, \phi) = \eta ( \mspi (r,\theta,\phi) ).
\end{equation}
\end{Defn}
The inverse mapping $\Phi^{-1}: \fsp \rightarrow \ltworthree$ is given by $(\Phi^{-1} \psi) (\xi_1, \xi_2 , \xi_3) = \psi ( \msp (\xi_1 , \xi_2 , \xi_3 ) )$. Note that this would give us functions which aren't defined on the set $\{ ( \xi_1 , \xi_2 , \xi_3 ) : \xi_1 = \xi_2 = 0 \}$. For our purposes we can define our functions to be zero at all these points.
\begin{Lemma}
The map $\Phi: \ltworthree \rightarrow \fsp$ is a unitary operator.
\end{Lemma}
\begin{proof}
If $\eta_1$ and $\eta_2$ are two elements of $\ltworthree$ then
\begin{eqnarray} \nonumber
\lefteqn{\langle \Phi \eta_1 , \Phi \eta_2 \rangle} \\ \nonumber
&=& \int_0^{\pi} \, \int_0^{2\pi} \, \int_{0}^{\infty} \eta_1 ( \mspi (r,\theta, \phi )) \overline{ \eta_2 (\mspi(r,\theta, \phi))} r^2 \sin (\phi) \, dr \, d\theta \, d \phi.
\end{eqnarray}
Now if we make the change of variable $\msp:(r,\theta,\phi) \mapsto (\xi_1 , \xi_2 , \xi_3 )$ the Jacobian will cancel out the $r^2 \sin (\theta)$ part of the measure and we will get
\begin{equation} \nonumber
\langle\Phi \eta_1 , \Phi \eta_2 \rangle = \int_{\Space{R}{3}} \eta_1 \overline{\eta_2} \, d\xi.
\end{equation}
This gives us the required result.
\end{proof}
Since $\Phi$ is a unitary operator we have that $\fsp$ is complete with respect to the inner product (\ref{eq:iponfsp}).
We now introduce an infinite dimensional representation of the Heisenberg group on the space $\fsp$.
\begin{Defn}
The spherical polar coordinate infinite dimensional representation of the Heisenberg group, $\rho^P_h$,  on the Hilbert space $\fsp$ is defined by
\begin{eqnarray*}
\lefteqn{(\rho_h^P (s,x,y) \psi) (r,\theta ,\phi)} \\
&& = e^{ -2\pi ihs} e^{\pi ih xy} e^{ 2\pi i \left( x_1 r \cos (\theta) \sin(\phi ) +  x_2 r \sin (\theta) \sin(\phi ) + x_3 r \cos ( \phi ) \right)} \times \\
\lefteqn{\psi \left( \left[ (r \cos (\theta) \sin (\phi ) +hy_1)^2 + (r \sin (\theta) \sin(\phi ) + h y_2 )^2 + ( r \cos ( \phi ) +h y_3 )^2 \right]^{1/2} \right. , } \\
&& \hspace{1cm} \left. \tan^{-1} \left[ \frac{r \sin (\theta) \sin ( \phi) + h y_2}{r \cos (\theta ) \sin (\phi) + h y_1} \right] , \right. \\
&& \left. \sin^{-1} \left[ \frac{((r \cos (\theta) \sin (\phi) + h y_1 )^2  + (r \sin (\theta) \sin (\phi) + h y_2)^2)^{1/2} }{F(r,\theta,\phi,y_1,y_2,y_3) }  \right] \right)
\end{eqnarray*}
$\psi$ is an element of $\fsp$ and
\begin{eqnarray*}
\lefteqn{F(r,\theta,\phi, y_1, y_2 , y_3) }\\
&=& ((r \cos (\theta) \sin (\phi ) +hy_1)^2 + (r \sin (\theta) \sin(\phi ) + h y_2 )^2 + ( r \cos ( \phi ) +h y_3 )^2 )^{1/2}.
\end{eqnarray*}
\end{Defn}
In the above definition we take the value of $\tan^{-1}$ in the range $(0,\pi)$ and the value of $\sin^{-1}$ in the range $(0,2\pi)$. Since
\begin{equation} \nonumber
F(r,\theta,\phi, y_1, y_2 , y_3) \geq ((r \cos (\theta) \sin (\phi) + h y_1 )^2  + (r \sin (\theta) \sin (\phi) + h y_2)^2)^{1/2}
\end{equation}
the domain of $\sin^{-1}$ is satisfied in the above definition.
\begin{Lemma}
The spherical polar coordinate infinite dimensional representation, $\rho_h^P$, is unitarily equivalent to the Schr\"odinger representation $\rho_h^S$. Furthermore the representation $\rho_h^P$ is irreducible and unitary.
\end{Lemma}
\begin{proof}
By a direct calculation it can be shown that $\rho_h^P=\Phi \rho_h^S \Phi^{-1}$. This gives us that $\rho_h^P$ is unitarily equivalent to the unitary irreducible representation $\rho_h^S$. To show $\rho_h^P$ is a unitary operator is trivial since for any $\psi_1 , \psi_2 \in \fsp$ and $g \in \Heisn$
\begin{equation} \nonumber
\langle \rho_h^P (g) \psi_1 , \rho_h^P (g) \psi_2 \rangle = \langle \Phi \rho_h^S (g) \Phi^{-1} \psi_1 , \Phi \rho_h^S (g) \Phi^{-1} \psi_2 \rangle.
\end{equation}
The right hand side of this equation is equal to $\langle \psi_1 , \psi_2 \rangle$ since $\Phi$ , $\rho_h^S$ and $\Phi^{-1}$ are all unitary operators.

To show $\rho_h^P$ is irreducible we use a proof by contradiction. If $\rho_h^P$ is reducible then there exists $\psi_1 , \psi_2 \in \fsp$ such that
\begin{equation} \nonumber
\langle \rho_h^P (s,x,y) \psi_1 , \psi_2 \rangle = 0
\end{equation}
for all $(s,x,y) \in \Heisn$. Since $\Phi^{-1}$ is a unitary map
\begin{equation} \label{eq:oneusedforirred}
\langle \Phi^{-1} \rho_h^P (s,x,y) \psi_1 , \Phi^{-1} \psi_2 \rangle = 0.
\end{equation}
By the definition of $\fsp$ there must exist $\eta_1 , \eta_2 \in \ltworthree$ such that $\psi_1 = \Phi \eta_1$ and $\psi_2 = \Phi \eta_2$. Equation (\ref{eq:oneusedforirred}) now takes the form
\begin{equation} \label{eq:oneusedforirredtwo}
\langle \Phi^{-1} \rho_h^P (s,x,y) \Phi \eta_1 ,  \eta_2 \rangle = 0,
\end{equation}
which is equivalent to
\begin{equation} \nonumber
\langle \rho_h^S (s,x,y) \eta_1 ,  \eta_2 \rangle = 0
\end{equation}
for all $(s,x,y) \in \Heisn$. This implies that the Schr\"odinger representation, $\rho_h^S$, is a reducible representation and hence a contradiction.
\end{proof}
Furthermore since $\rho_h^P$ is unitarily equivalent to $\rho_h^S$ which is unitarily equivalent to $\rho_h$ (see Section \ref{sect:relbetweenfockandotherhilb}) it follows that $\rho_h^P$ is unitarily equivalent to $\rho_h$.

By taking the representation of $p$-mechanical observables under $\rho_h^P$ we will get the corresponding quantum mechanical observables realised as operators on the space $\fsp$. We will go on to show that the angular momentum and Kepler/Coulomb Hamiltonian observables will take a much simpler form under this representation.
\begin{Lemma}
The spherical polar coordinate infinite dimensional representation of the distribution $L_3 = \pzang$ is the operator $\frac{h}{2\pi i} \Partial{\phi}$, that is
\begin{equation} \label{eq:sphrepoflthree}
\rho_h^P \left( L_3 \right) \eta = \frac{h}{2\pi i} \Fracpartial{\eta}{\phi}.
\end{equation}
\end{Lemma}
\begin{proof}
This follows using equation (\ref{eq:repofadistnonaliegp}) and then repeatedly using the multi-dimensional chain rule.
\end{proof}
Interestingly we can get the form of the Laplacian in spherical polar coordinates by taking the spherical polar coordinates representation of the distribution $\zerodelytwo$. Also we could get more complex differential operators by taking different distributions on $\Heisn$.

Furthermore we now show that the one dimensional representations can be used to obtain classical mechanical observables in spherical polar coordinates. Initially we give a Lemma about spherical polar coordinates in Hamiltonian mechanics.
\begin{Lemma}
If we transform our position coordinates $(q_1 , q_2 , q_3) \rightarrow (r, \theta , \phi )$ by the transformation defined in (\ref{eq:spmapr})-(\ref{eq:spmapthree}) then if we take
\begin{eqnarray} \label{eq:sphpolmomr}
p_r &=& \frac{q_1 p_1 + q_2 p_2 + q_3 p_3}{(q_1^2 + q_2^2 + q_3^2 )^{1/2}} \\ \label{eq:sphpolmomth}
p_{\theta} &=& \frac{q_1 q_3 p_1 + q_2 q_3 p_2 - (q_1^2 + q_2^2)p_3}{(q_1^2 + q_2^2)^{1/2}} \\ \label{eq:sphpolmomph}
p_{\phi} &=& -q_2 p_1 + q_1 p_2
\end{eqnarray}
as our new set of momentum coordinates we have a canonical transformation $\Space{R}{6} \setminus \{ (q_1,q_2,q_3,p_1,p_2,p_3) \in \Space{R}{6}: q_1 = q_2 =0 \} \newline \rightarrow \{ (r,\theta,\phi,p_r,p_{\theta},p_{\phi}) \in \Space{R}{6} : r>0, 0 \leq \theta < 2\pi , 0 < \phi < \pi \}$.
\end{Lemma}

\begin{proof}
See \cite[Sect. 5.3]{Jose98} for a proof of this.
\end{proof}

An inversion of equations (\ref{eq:sphpolmomr})-(\ref{eq:sphpolmomph}) gives
\begin{eqnarray}
p_1 &=& p_r \sin ( \theta ) \cos ( \phi ) + \frac{ p_{\theta} \cos (\theta) \cos(\phi)}{r} - \frac{p_{\phi} \sin (\phi)}{r \sin (\theta )} \\
p_2 &=& p_r \sin ( \theta ) \sin ( \phi ) + \frac{ p_{\theta} \cos (\theta ) \sin (\phi)}{r} + \frac{p_{\phi} \cos (\phi)}{r \sin (\theta)} \\
p_3 &=& p_r \cos (\theta) - \frac{ p_{\theta} \sin (\theta)}{r}.
\end{eqnarray}

\begin{Defn}
We define the spherical polar coordinate one dimensional representation of the Heisenberg group on $\Space{C}{}$ by
\begin{eqnarray} \label{eq:onedpherpolrep}
\lefteqn{\rho_{(r,\theta,\phi,p_r,p_{\theta},p_{\phi})} (s,x,y) u } \\ \nonumber
&& = \exp \left( - 2\pi i ( x_1 r \sin ( \theta ) \cos ( \phi ) + x_2 r \sin ( \theta ) \sin (\phi) + x_3 r \cos (\theta) \right) \\ \nonumber
&& \hspace{0.5cm} \times \exp \left( -2\pi i y_1 \left( p_r \sin ( \theta ) \cos ( \phi ) + \frac{ p_{\theta} \cos (\theta) \cos(\phi)}{r} - \frac{p_{\phi} \sin (\phi)}{r \sin (\theta )} \right) \right) \\ \nonumber
&& \hspace{0.5cm} \times \exp \left( -2\pi i y_2 \left( p_r \sin ( \theta ) \sin ( \phi ) + \frac{ p_{\theta} \cos (\theta ) \sin (\phi)}{r} + \frac{p_{\phi} \cos (\phi)}{r \sin (\theta)} \right) \right) \\ \nonumber
&& \hspace{0.5cm} \times \exp \left( -2\pi i y_3 \left( p_r \cos (\theta) - \frac{ p_{\theta} \sin (\theta)}{r} \right) \right) u
\end{eqnarray}
where $u \in \Space{C}{}.$
\end{Defn}
There is a different representation for every element of the set
\begin{eqnarray} \nonumber
\lefteqn{ \left\{ (r,\theta,\phi,p_r,p_{\theta},p_{\phi}) : \right. } \\ \nonumber
&& \hspace{2cm} \left. r > 0 , 0 \leq \theta < 2\pi , 0 < \phi < \pi, p_r \in \Space{R}{}, p_{\theta} \in \Space{R}{}, p_{\phi} \in \Space{R}{}   \right\}.
\end{eqnarray}
\begin{Thm}
The spherical polar coordinate one dimensional representation of the $p$-mechanisation of a classical observable, $f$, is $f$ expressed in spherical polar coordinates $(r,\theta,\phi,p_r,p_{\theta},p_{\phi})$.
\end{Thm}
\begin{proof}
This follows from the definition of $p$-mechanisation and the Fourier inversion formula.
\end{proof}
We now give some examples of observables which take a simpler form under this new one dimensional representation. For example the third component of angular momentum, $L_3$, from equation (\ref{eq:pmechofangmom})  under this representation is just
\begin{equation} \nonumber
\rho_{(r,\theta,\phi,p_r,p_{\theta},p_{\phi})} (L_3) = p_{\phi}.
\end{equation}
Furthermore the Kepler/Coulomb Hamiltonian takes a simple form under this one dimensional representation:
\begin{equation} \nonumber
\rho_{(r,\theta,\phi,p_r,p_{\theta},p_{\phi})} (B_H) = \frac{1}{2} \left( p_r^2 + \frac{p_{\theta}^2}{r^2} + \frac{p_{\phi}^2}{r^2 \sin^2 (\theta)} \right) + \frac{1}{r}.
\end{equation}

We proceed by demonstrating that under the infinite dimensional spherical polar coordinate representation, $\rho_h^P$, some observables take a simpler form. The $\rho_h^P$ representation of the total angular momentum observable is
\begin{equation} \nonumber
\rho_h^P (L^2) = -\frac{h^2}{\sin^2 (\theta)} \left[ \sin (\theta ) \Partial{\theta} \left( \sin ( \theta ) \Partial{\theta} \right) + \Partialtwo{\phi} \right].
\end{equation}
The eigenfunctions\index{eigenfunctions!angular momentum} in $\fsp$ of $\rho_h^P (L^2)$ have been shown to be \cite{Merzbacher70,Messiah61}
\begin{equation} \label{eq:angmomevecinsp}
\psi_{(l,m)} (\theta, \phi) = Y^m_l (\theta, \phi)
\end{equation}
with $l \in \Nat$ and $-l\leq m \leq l$. $Y^m_l$ are the spherical harmonics
\begin{equation} \nonumber
Y^m_l (\theta , \phi ) = \sqrt{\frac{2l+1}{4\pi} \frac{(l-m)!}{(l+m)!}} (-1)^m e^{im\phi} P_l^m ( \cos ( \theta))
\end{equation}
where $P_l^m (x)$ is the associated Legendre function\index{associated Legendre function} \cite[Eq. 11.71]{Merzbacher70}. These eigenfunctions have eigenvalue $l(l+1)\hbar^2$. $\psi_{(l,m)}$ are also eigenfunctions of the operator $\rho_h^P (L_3)$ introduced in equation (\ref{eq:sphrepoflthree}); for this operator they have eigenvalue $m\hbar$.
The spherical polar coordinate infinite dimensional  representation of the Kepler/Coulomb Hamiltonian\index{eigenfunctions!Kepler/Coulomb Hamiltonian} is
\begin{equation} \nonumber
\rho_h^P ( B_H ) \psi (r, \theta , \phi )= \left[ -\frac{h^2}{8\pi^2} \nabla^2 + \frac{1}{r} \right] \psi (r,\theta,\phi).
\end{equation}
$\nabla^2$ denotes the Laplacian in spherical polar coordinates. The bound state (negative energy) eigenfunctions of this operator have been shown to be \cite{Merzbacher70} \cite[Sect. 11.6]{Messiah61}
\begin{eqnarray} \label{eq:kepcouevectorsinsp}
\lefteqn{\psi_{(n,l,m)} (r,\theta,\phi) = \frac{e^{-\kappa r} (2\kappa r )^l }{(2l+1)!} \left[ (2\kappa )^3 \frac{(n+l)!}{2n(n-l-1)!} \right]^{1/2}} \\ \nonumber
&& \hspace{3.5cm} \times F_1 (-n+l+1;2l+2;2\kappa r ) Y_l^m (\theta,\phi)
\end{eqnarray}
where $\kappa =\frac{2\pi}{nh^2}$.
$F_1$ is defined as
\begin{equation} \nonumber
F_1 (a;c;z) = \frac{\Gamma (1-a) \Gamma (c)}{ [ \Gamma (c-a) ]^2} L_{-a}^{c-1} (z)
\end{equation}
$L_r^p$ is the associated Laguerre polynomial\index{associated Laguerre polynomial} defined by
\begin{equation} \nonumber
L^p_{q-p}(z) = (-1)^p \frac{d^p}{dz^p} L_q (z)
\end{equation}
and
\begin{equation} \nonumber
L_q (z) = L_q^0 (z) = e^z \frac{d^q}{dz^q} (e^{-z} z^q ).
\end{equation}

The eigenvalues corresponding to these eigenfunctions are $-\frac{\omega}{n^2}$
\begin{equation} \label{eq:coulombevalue}
\rho_h^P (B_H) \psi_{(n,l,m)} = - \frac{\omega}{n^2} \psi_{(n,l,m)}
\end{equation}
where $\omega=\frac{4\pi^2}{h^2}$.

\begin{Remark}
\emph{ We could transform the $\rho_h$ representation in a similar way to which we have adjusted the Schr\"odinger representation. The equivalent map to $\mathcal{M}_s$ would transform $q$ as before, but it would also change $p$ by the transformation of the momentum space for spherical polar coordinates (see  equations (\ref{eq:sphpolmomr})-(\ref{eq:sphpolmomph})). Unfortunately this representation is of little use for analysing the Kepler/Coulomb problem, but it may be of use for other problems.}
\end{Remark}

\section{Transforming the Position Space} \label{sect:transfposspace}

\indent

In this section we generalise the above treatment of spherical polar coordinates in $p$-mechanics -- we consider a general invertible mapping of position space. We show how starting from a transformation of the position coordinates we can use the Schr\"odinger representation to obtain a corresponding transformation of the momentum coordinates. This transformation of the whole phase space will be a canonical transformation.

Suppose $\mgc: \Space{R}{n} \rightarrow \mathcal{I} \subset \Space{R}{n}$ is an invertible mapping which along with its inverse is differentiable in all its arguments. We assume that $\mathcal{I}$ is an $n$-dimensional subspace of $\Space{R}{n}$. We use the coordinates $(\xi_1, \cdots , \xi_n)$ to denote an element of $\Space{R}{n}$ and $(\zeta_1 , \cdots \zeta_n )$ to denote an element of $\mathcal{I}$. The matrix $D\mgc$ is used to denote the matrix with entries $(D\mgc)_{i,j} = \Fracpartial{\mgc_i}{\xi_j}$. Furthermore if $A$ is a matrix we use $|A|$ to denote its determinant. We assume throughout this section that $|D\mgc| \neq 0$.
\begin{Defn}
$\fgc$ is defined as the image of $\ltworn$ under the mapping\footnote{Throughout this section we use the notation $\eta$ for an element of $\ltworn$ and $\psi$ for an element of $\mathcal{F}$.}
\begin{equation} \label{eq:mappingtocurlyf}
\eta \mapsto \eta \circ \mgci.
\end{equation}
\end{Defn}
We use $\mathcal{N}$ to denote the map from $\ltworn$ to $\fgc$ by (\ref{eq:mappingtocurlyf}). Clearly the inverse of $\mathcal{N}$ is just
\begin{equation} \nonumber
\mathcal{N}^{-1}:\fgc \rightarrow \ltworn \hspace{2cm} \psi \mapsto \psi \circ \mgc
\end{equation}
\begin{Lemma}
If we equip $\fgc$ with the inner product
\begin{equation} \nonumber
\langle \psi_1 , \psi_2 \rangle_{\fgc} = \int_{\mathcal{I}} \psi_1 \overline{\psi_2} \frac{1}{| D\mgc |} \, d\mu
\end{equation}
where $d\mu$ is Lebesgue measure on $\mathcal{I}$ then $\mathcal{N}$ is an isometry, that is
\begin{equation} \nonumber
\langle \eta_1 , \eta_2 \rangle_{\ltworn} = \langle \mathcal{N} \eta_1 , \mathcal{N} \eta_2 \rangle_{\mathcal{F}}
\end{equation}
and $\fgc$ is a Hilbert space.
\end{Lemma}
\begin{proof}
This follows by changing the variable in the integral by $\mgc$. The Jacobian of the transformation will be cancelled out by the $\frac{1}{|D\mathcal{M}|}$. The completeness of $\mathcal{F}$ is a consequence of $\mathcal{N}$ being an isometry.
\end{proof}
Since $\mathcal{N}$ is an isometry the representation $\rho_h^{\mgc} = \mathcal{N} \circ \rho_h^S \circ \mathcal{N}^{-1}$ on $\fgc$ is unitarily equivalent to the Schr\"odinger representation and therefore unitary and irreducible.
\begin{Lemma}
$\rho_h^{\mgc}$ applied to a particular element $\psi \in \fgc$ takes the form
\begin{equation} \nonumber
(\rho_h^{\mgc} \psi) (\zeta) = e^{-2\pi ihs} e^{\pi ih xy} e^{2\pi i x.\mgci (\zeta)} \psi ( \mgc (\mgci (\zeta) +hy)).
\end{equation}
\end{Lemma}
\begin{proof}
If we start with $\psi \in \fgc$ then $\mathcal{N}^{-1} \psi = \eta$ where $\eta \in \ltworn$ such that
\begin{equation} \nonumber
\eta (\xi ) = \psi (\mgc \xi)
\end{equation}
for any $\xi \in \Space{R}{n}$. Now since $(\rho_h^S \eta )(\xi) = e^{-2\pi ihs} e^{\pi ih xy} e^{2\pi i x \xi} \eta (\xi + hy )$ we have
\begin{equation} \nonumber
(\rho_h \mathcal{N}^{-1} \psi ) (\xi) = e^{-2\pi ihs} e^{\pi ih xy} e^{2\pi i x \xi} \psi (\mgc (\xi + hy) ).
\end{equation}
Furthermore
\begin{eqnarray} \nonumber
(\mathcal{N} \rho_h \mathcal{N}^{-1} \psi) (\zeta) &=& [\rho_h (\mathcal{N}^{-1} \psi)] ( \mgci (\zeta)) \\ \nonumber
&=& e^{-2\pi ihs} e^{\pi ih xy} e^{2\pi i x.\mgci (\zeta)} \psi ( \mgc [\mgci (\zeta) +hy]).
\end{eqnarray}
\end{proof}
\begin{Thm} \label{thm:genposttrans}
The $\rhm$ representation of the $p$-mechanical position and momentum observables are\footnote{The notation $\zerodeljx$ was defined in equation (\ref{eq:pmechofangmom})}
\begin{equation} \label{eq:rhmofdelxj}
\left[ \rhm \left(\frac{1}{2\pi i} \zerodeljx \right) \psi \right] (\zeta) = - \mathcal{M}^{-1} (\zeta)_j \psi (\zeta)
\end{equation}
\begin{equation} \label{eq:rhmofdelyj}
\left[ \rhm \left( \frac{1}{2\pi i} \zerodeljy \right) \psi \right] (\zeta) = - \sum_{k=1}^n h \psi_{,k} (\zeta) [(D\mathcal{M})_{k,j} (\mathcal{M}^{-1} (\zeta))] .
\end{equation}
$\psi_{,k}$ is used to denote the differential of the function $\psi$ with respect to its $k$th argument.
\end{Thm}
\begin{proof}
Using equation (\ref{eq:repofadistnonaliegp}) we have for any $\psi_1 , \psi_2 \in \mathcal{F}$ such that \newline $\langle \rhm (s,x,y) \psi_1 , \psi_2 \rangle \in \mathcal{S}(\Heisn)$
\begin{eqnarray} \nonumber
\lefteqn{\langle \rhm (\zerodeljx) \psi_1 , \psi_2 \rangle} \\ \label{eq:thetwolangles}
&=& \langle \langle \rhm (s,x,y) \psi_1 , \psi_2 \rangle , \zerodeljx \rangle \\ \nonumber
&=& - \Partial{x_j} \langle \rhm (s,x,y) \psi_1 , \psi_2 \rangle |_{(s,x,y)=(0,0,0)} \\ \nonumber
&=& -\langle (\pi ihy_j + \mathcal{M}^{-1}(\zeta)_j )\rhm (s,x,y) \psi_1 , \psi_2 \rangle |_{(s,x,y)=(0,0,0)} \\ \nonumber
&=& - \langle \mathcal{M}^{-1}(\zeta)_j \psi_1 , \psi_2 \rangle
\end{eqnarray}
which proves (\ref{eq:rhmofdelxj}). Note that at (\ref{eq:thetwolangles}) the outer brackets $\langle, \rangle$ represent evaluation by a functional while the inner brackets $\langle,\rangle$ represent the inner product on $\fsp$. The proof of (\ref{eq:rhmofdelyj}) is a little more involved.
From equation (\ref{eq:repofadistnonaliegp}) we have
\begin{eqnarray} \nonumber
\lefteqn{\langle \rhm (\zerodeljy) \psi_1 , \psi_2 \rangle} \\ \nonumber
&=& - \Partial{y_j} \langle \rhm (s,x,y) \psi_1 , \psi_2 \rangle|_{(s,x,y)=(0,0,0)} \\ \label{eq:needtousethelemma}
&=& - \langle \pi ihx \rho_h^{\mathcal{M}} (s,x,y) \psi_1 \\ \nonumber
&& \hspace{0.5cm} + e^{-2\pi ihs + \pi ihxy + 2\pi i x. \mathcal{M}^{-1} (\zeta)} \Partial{y_j} \psi_1 (\mathcal{M} [\mathcal{M}^{-1}(\zeta) + hy]) , \psi_2 \rangle|_{(s,x,y)=(0,0,0)}.
\end{eqnarray}
Applying Lemma \ref{lem:lwminaproof} -- which is proved after this theorem --  to equation (\ref{eq:needtousethelemma}) we get
\begin{eqnarray*} \nonumber
\lefteqn{\langle \rhm (\zerodeljy) \psi_1 , \psi_2 \rangle} \\ \nonumber
&=& - \left\langle \sum_{k=1}^n h \psi_{1,k} (\zeta) (D \mathcal{M})_{k,j} (\mathcal{M}^{-1} (\zeta))  , \psi_2 \right\rangle.
\end{eqnarray*}
This proves (\ref{eq:rhmofdelyj}).
\end{proof}
Now we present and prove the Lemma which was used in the proof of Theorem \ref{thm:genposttrans}.
\begin{Lemma} \label{lem:lwminaproof}
We have that
\begin{equation}
\Partial{y_j} \psi (\mathcal{M} [\mathcal{M}^{-1}(\zeta) + hy]) |_{y=0} = \sum_{k=1}^n h \psi_{,k} (\zeta) (D \mathcal{M})_{k,j} (\mathcal{M}^{-1} (\zeta)).
\end{equation}
\end{Lemma}
\begin{proof}
To prove this Lemma we need to use the $n$-dimensional chain rule. The version of the chain rule we use is \cite[Sect. 2.5]{MarsdenTromba03} if $f:\Space{R}{n} \rightarrow \Space{R}{}$ and $A,B: \Space{R}{n} \rightarrow \Space{R}{n}$ then
\begin{eqnarray} \nonumber
\Partial{y_j} f(A \circ B (y) ) = \sum_{k,l=1}^n f_{,k} (A \circ B (y)) \times (DA)_{k,l} (B(y)) \times (DB)_{l,j} (y).
\end{eqnarray}
The $\times$s in the above equation represent the normal multiplication of two scalar values.
If we choose $B$ to be the map $y \rightarrow \mathcal{M}^{-1} (\zeta) + hy$ and $A$ to be the map $z \rightarrow \mathcal{M} (z)$ then $(DB)_{i,j} = \delta_{i,j} h$ and $(DA)_{i,j} = (D\mathcal{M})_{i,j}$ we get
\begin{eqnarray} \nonumber
\lefteqn{\Partial{y_j} \psi (\mathcal{M} [\mathcal{M}^{-1}(\zeta) + hy])} \\
&=& \sum_{k,l=1}^n \psi_{,k } [\mathcal{M} (\mathcal{M}^{-1}(\zeta) +hy)]  (D\mathcal{M}_{k,l}) (\mathcal{M}^{-1} (\zeta) +hy) h \delta_{l,j} \\
&=& \sum_{k=1}^{n} h \psi_{,k} (\mathcal{M}(\mathcal{M}^{-1}(\zeta) +hy) ) (D \mathcal{M})_{k,j} (\mathcal{M}^{-1} (\zeta) + hy)
\end{eqnarray}
which evaluated at $y=0$ is
\begin{equation*}
\sum_{k=1}^n h \psi_{,k} (\zeta) (D \mathcal{M})_{k,j} (\mathcal{M}^{-1} (\zeta)).
\end{equation*}
Hence we have proved the Lemma.
\end{proof}

Theorem \ref{thm:genposttrans} encourages us to choose the one dimensional representation associated to $\mgc$ as
\begin{eqnarray} \nonumber
\lefteqn{\rho_{(\zeta_1 , \cdots , \zeta_n , p_{\zeta_1} , \cdots , p_{\zeta_n} )} }\\ \nonumber
&=& \exp(-2\pi i (x_1 \mgci (\zeta )_1 + \cdots + x_n \mgci (\zeta )_n) ) \\ \nonumber
&& \times \exp \left( -2\pi i \left( y_1 \left( \sum_{j=1}^n (D\mgc)_{(1,j)} (\mgci (\zeta) ) p_{\zeta_j} \right) \right. \right. \\ \nonumber
&& \left. \left. \hspace{2cm} + \cdots + y_n \left( \sum_{j=1}^n (D\mgc)_{(n,j)} (\mgci (\zeta) ) p_{\zeta_j} \right) \right) \right).
\end{eqnarray}

\begin{Proposition}
The mapping
\begin{eqnarray*}
q_j &\mapsto& \mgci (\eta)_j \\
p_j &\mapsto& \left( \sum_{k=1}^n (D\mgc)_{j,k} (\mgci(\eta))) p_{\eta_k} \right)
\end{eqnarray*}
 is a canonical transformation.
\end{Proposition}
\begin{proof}
To prove this we just need to show that the Poisson brackets of $q_i (\eta , p_{\eta} )$ and $p_j (\eta , p_{\eta} )$ is $\delta_{(i,j)}$ in the new coordinates (see equation (\ref{eq:QPcondforct})). We show this by a direct calculation
\begin{eqnarray} \nonumber
\{ q_i, p_j \} &=& \sum_{k=1}^n \left( \Fracpartial{q_i}{\eta_k} \Fracpartial{p_j}{p_{\eta_k}} - \Fracpartial{p_j}{\eta_k} \Fracpartial{q_i}{p_{\eta_k}} \right) \\ \nonumber
&=& \sum_{k,l=1}^n (D \mgci )_{(i,k)} (\eta) (D \mgc)_{(j,l)} (\mgci (\eta)) \delta_{(k,l)} \\ \label{eq:chainruleofzero}
&=& \sum_{k=1}^n (D \mgci )_{(i,k)} (\eta) (D \mgc)_{(j,k)} (\mgci (\eta)).
\end{eqnarray}
The expression at (\ref{eq:chainruleofzero}) by the chain rule is $(D (\mgc \circ \mgci))_{(i,j)}$ which is equal to $DI_{(i,j)}$ where $I$ is the identity operator on $\Space{R}{n}$. So since $DI_{(i,j)}= \delta_{i,j}$ we have
\begin{equation} \nonumber
\{ q_i , p_j \} = \delta_{i,j}.
\end{equation}
\end{proof}

So by the Fourier inversion formula the representation of the $p$-mechanisation of a classical observable $f$ will just be the image of $f$ after the canonical transformation.

\section{The Klauder Coherent States for the Hydrogen atom} \label{sect:klaudcs}

\indent

Ever since Schr\"odinger introduced the harmonic oscillator coherent states, the hunt has been on to find a set of states which have the same properties for the hydrogen atom. Many efforts have been made which possess some of the properties of the harmonic oscillator coherent states, but finding a set of states which possessed all the same properties for the Hydrogen atom was never achieved. One of the best attempts was done by Klauder in his ground breaking paper \cite{Klauder96} ---  a set of coherent states for the hydrogen atom were introduced which had the properties of being: continuous in their label, temporally stable and satisfying a resolution of unity for the bound state portion of the hydrogen atom. Unfortunately they are not minimal uncertainty states, but for our purposes we do not require this property. In this section we give a brief overview of these coherent states. We will exploit these coherent states in Section \ref{sect:klaudrep}.

Before we can define the Kepler/Coulomb\index{coherent states!Kepler/Coulomb} coherent states we need to introduce the angular-momentum coherent states\index{coherent states!angular momentum} adapted to the Kepler/Coulomb problem \cite[Eq. 15]{Klauder96}
\begin{eqnarray} \nonumber
\psi_{(n, \overline{\Omega})} (r,\theta,\phi) &=& \sum_{l=0}^n \sum_{m=-l}^{l} \left[ \frac{ (2l)!}{(l+m)!(l-m)!} \right]^{1/2} \left( \sin \left( \frac{\overline{\theta}}{2} \right) \right)^{l-m}  \left( \cos \left( \frac{\overline{\theta}}{2} \right) \right)^{l+m} \\ \nonumber
&& \hspace{1.5cm} \times e^{-i(m\overline{\phi} + l\overline{\psi} )} \psi_{(n+1,l,m)} (r,\theta,\phi) \, (2l+1)^{1/2}.
\end{eqnarray}
It is important to note that for labeling the coherent states a bar is used over $\overline{\Omega} = (\overline{\theta}, \overline{\phi} ,\overline{\psi})$, to show that they are different from the $\theta$ and $\phi$ in the domain of the function. The functions $\psi_{(n,l,m)} (\rtph)$ are the bound state (negative energy) eigenfunctions (see equation (\ref{eq:coulombevalue})) in $\fsp$ for the Kepler/Coulomb Hamiltonian. We denote by $\bssp$\index{$\bssp$} the subspace of $\fsp$ spanned by the vectors $\psi_{(n,l,m)}$.

We let $\mathcal{AM}_n$\index{$\mathcal{AM}_n$} denote the $n$th angular momentum subspace --- that is the space spanned by the angular momentum eigenfunctions $\psi_{(l,m)}$, from (\ref{eq:angmomevecinsp}) for $0 \leq l \leq n$  and $-l \leq m \leq l$. It is shown in \cite{Klauder96} that these coherent states satisfy a resolution of the identity in the subspace $\mathcal{AM}_n$, that is
\begin{displaymath}
\int \langle \psi , \psi_{(n, \overline{\Omega} )} \rangle \psi_{(n , \overline{\Omega})} \sin ( \overline{\theta} ) \, d\overline{\theta} \, d\overline{\phi} \, d\overline{\psi} =
\left\{ \begin{array}{ll}
\psi, & \textrm{  if $\psi \in \mathcal{AM}_n$; }  \\
0, & \textrm{  otherwise.}
\end{array} \right.
\end{displaymath}
Now we can define the Kepler/Coulomb coherent states\index{coherent states!Kepler/Coulomb} as\footnote{These are by no means the unique choice of Kepler/Coulomb coherent sates. The weights $e^{-\sigma^2}$ and $n!$ may be changed as described in \cite{Klauder96}. In various papers \cite{Fox99,GazeauKlauder99,Crawford00,GazeauMonceau02} various suggestions for other choices of these weights are given.}
\begin{equation} \label{eq:defofkccs}
\psi_{(\sigma, \gamma , \overline{\Omega} )} = e^{-\sigma^2} \sum_{n=0}^{\infty} \left( \frac{\sigma^n \exp \left( - \frac{2\pi  \gamma }{ ih(n+1)^2} \right) }{(n!)^{1/2}}  \right) \psi_{(n, \overline{\Omega})}.
\end{equation}
For later use we define the measure $\nu (\sigma , \gamma , \overline{\Omega} )$ as
\begin{eqnarray} \label{eq:measforkcsp}
\lefteqn{\int f(\sigma , \gamma , \overline{\Omega} ) d\nu (\sigma , \gamma , \overline{\Omega} )} \\ \nonumber
&& = \int_{0}^{\pi} \int_{0}^{2\pi} \int_{0}^{2\pi} \lim_{\Theta \rightarrow \infty} \frac{1}{2\Theta} \int_{-\Theta}^{\Theta}   \int_0^{\infty} f(\sigma , \gamma , \overline{\Omega} ) \, \sin(\overline{\theta}) \, d\sigma \, d\gamma \, d\overline{\theta} \, d\overline{\phi} \, d\overline{\psi}.
\end{eqnarray}
We also define the measure $\mu (r, \theta , \phi )$ by
\begin{equation} \nonumber
\int \psi ( r, \theta, \phi ) d \mu (r, \theta , \phi ) = \int_0^{2\pi} \int_0^{\pi} \int_0^{\infty} \psi ( r, \theta, \phi ) r^2 \sin ( \theta ) \, dr \, d \theta \, d \phi.
\end{equation}

One property of the coherent states defined in equation (\ref{eq:defofkccs}) is that they satisfy a resolution of the identity for the bound states of the Kepler/Coulomb Hamiltonian \cite[Eq. 18]{Klauder96}, that is
\begin{displaymath} \label{eq:resofid}
\int \langle \psi , \psi_{(\sigma, \gamma , \overline{\Omega} )} \rangle_{\fsp} \, \psi_{(\sigma , \gamma , \overline{\Omega})} \, d\nu (\sigma , \gamma , \overline{\Omega} ) =
\left\{ \begin{array}{ll}
\psi , & \textrm{  if $\psi \in \bssp$; }  \\
0 , & \textrm{  otherwise.}
\end{array} \right.
\end{displaymath}

Another property of the coherent states which follows from (\ref{eq:coulombevalue}) is
\begin{equation} \label{eq:effectofkeponklaud}
\rho_h^P (B_H) \psi_{(\sigma, \gamma , \overline{\Omega} )} = e^{-\sigma^2} \sum_{n=0}^{\infty} \left( - \frac{ \omega \sigma^n \exp \left( \frac{2\pi \gamma}{ih(n+1)^2 } \right) }{(n+1)^2 (n!)^{1/2}} \right) \psi_{(n, \overline{\Omega})}.
\end{equation}
This can also be realised as
\begin{equation} \label{eq:alteffectofkeponklaud}
-\frac{2\pi}{ih} \rho_h^P (B_H) \psi_{(\sigma, \gamma , \overline{\Omega} )} = \omega \Partial{\gamma} \psi_{(\sgob)} (\rtph ).
\end{equation}

\section{A Hilbert Space for the Kepler/Coulomb Problem} \label{sect:klaudrep}

\indent

In this section we introduce a new Hilbert space which is suitable for modelling the quantum mechanical Kepler/Coulomb problem. Two models of quantum mechanics are said to be equivalent if all the transition amplitudes are the same \cite{Anderson94}. We show in this section that for a subset of states and a subset of observables
a model using this new Hilbert space will be equivalent to the standard model (that is, the model using the irreducible unitary Schr\"odinger representation on $\ltworthree$). Initially we define a new space.
\begin{Defn}
We define the Kepler/Coulomb space\index{Kepler/Coulomb space}, which we denote $\kcs$\index{$\kcs$}, to be
\begin{equation} \label{eq:defofkcs}
\kcs = \left\{ f(\sigma, \gamma , \overline{\Omega} ) = \int \, \psi(r,\theta, \phi) \overline{\psi_{(\sigma, \gamma , \overline{\Omega})} (r,\theta, \phi)} \, d \mu (r, \theta , \phi ) \, : \, \psi \in \bssp \right\}.
\end{equation}
\end{Defn}
The inner product of $f_1 , f_2 \in \kcs$ is given by
\begin{equation} \nonumber
\langle f_1 , f_2 \rangle = \int \, f_1(\sigma,\gamma,\overline{\Omega}) \overline{f_2 (\sigma , \gamma , \overline{\Omega} )} \, d \nu (\sigma, \gamma , \overline{\Omega} )
\end{equation}
where $\nu$ is the measure defined in equation (\ref{eq:measforkcsp}). We can take the completion of this space with respect to this inner product to obtain a Hilbert space. We have a map $\kcmap: \bssp \rightarrow \kcs$\index{$\kcmap$} given by
\begin{eqnarray} \label{eq:defofkcmap}
\left(\kcmap ( \psi ) \right) (\sigma,\gamma,\overline{\Omega}) &=& \int \, \psi(r,\theta, \phi) \, \overline{\psi_{(\sigma, \gamma , \overline{\Omega})} (r,\theta, \phi)} \, d \mu (r, \theta , \phi )  \\ \nonumber
&=& \langle \psi , \psi_{(\sgob)} \rangle_{\fsp} .
\end{eqnarray}
\begin{Lemma} \label{thm:invofkone}
$\kcmap$ is a unitary operator and has inverse $\kcmap^{-1}: \kcs \rightarrow \bssp$
\begin{equation}
\kcmap^{-1}f = \int f(\sgob) \psi_{(\sgob)} (r,\theta,\phi) \, d \nu (\sgob).
\end{equation}
\end{Lemma}
\begin{proof}
Both of these assertions follow from the fact that the coherent states $\psi_{(\sgob)}$ satisfy a resolution of the identity for the bound states of the Kepler/Coulomb problem.
\end{proof}
\begin{Thm} \label{thm:wheretildecamefrom}
If $A$ is an operator on $\bssp$ and $\psi_1,\psi_2 \in \bssp$, then if we let $\tilde{A} = \kcmap A \kcmap^{-1}$, $f_1 = \kcmap \psi_1$ and $f_2 = \kcmap \psi_2$ we have
\begin{equation} \nonumber
\langle A \psi_1 , \psi_2 \rangle = \langle \tilde{A} f_1 , f_2 \rangle.
\end{equation}
\end{Thm}
\begin{proof}
Using Lemma \ref{thm:invofkone} we have
\begin{eqnarray*}
\langle \tilde{A} f_1 , f_2 \rangle &=& \langle \kcmap A \kcmap^{-1} \kcmap \psi_1 , \kcmap \psi_2 \rangle \\
&=& \langle \kcmap A \psi_1 , \kcmap \psi_2 \rangle \\
&=& \langle A \psi_1 , \psi_2 \rangle.
\end{eqnarray*}
\end{proof}
This theorem means that if we transform the $\fsp$ model of quantum mechanics by the operator $\kcmap$ then our new model is equivalent for operators which preserve $\bssp$ and states which are in $\bssp$. So this new Hilbert space is suitable for modelling quantum mechanics as long as we are only considering operators which preserve $\bssp$ and states which are bound states for the Kepler/Coulomb problem. Unfortunately this model does not extend to all observables and so we can not obtain a representation of the Heisenberg group on this space. We now show that for the Kepler/Coulomb problem the time evolution in our new Hilbert space, $\kcs$, is just a shift in the $\gamma$ variable.
\begin{Thm}
If $\hat{H}$ is the operator on $\bssp$ equal to $\rho_h^P(B_H)$ then\footnote{$\tilde{\hat{H}}$ is continuing the notation which originated in Theorem \ref{thm:wheretildecamefrom}}
\begin{equation}
\tilde{\hat{H}} = \kcmap \hat{H} \kcmap^{-1} = \frac{ih}{2\pi} \omega \Partial{\gamma}.
\end{equation}
\end{Thm}
\begin{proof}
It is clear that $\hat{H}$ will preserve the space $\bssp$ and so is an operator on this space. If we let $f$ be an arbitrary element of $\kcs$ then $f=\kcmap \psi$ for some $\psi \in \bssp$
\begin{eqnarray} \nonumber
\tilde{\hat{H}} f &=& \tilde{\hat{H}} \kcmap \psi \\ \nonumber
&=& \kcmap \hat{H} \psi \\ \nonumber
&=& \langle  \hat{H} \psi , \psi_{(\sgob)} \rangle \\ \label{eq:usedfactofhermi}
&=& \langle \psi , \hat{H} \psi_{(\sgob)} \rangle \\ \label{eq:usedtheequationbefore}
&=& \left\langle \psi , -\frac{ih}{2\pi} \omega \Partial{\gamma} \psi_{(\sgob)} \right\rangle \\ \nonumber
&=& \frac{ih}{2\pi} \omega \Partial{\gamma} \left\langle \psi ,  \psi_{(\sgob)} \right\rangle \\ \nonumber
&=& \frac{ih}{2\pi} \omega \Fracpartial{f}{\gamma}.
\end{eqnarray}
At (\ref{eq:usedfactofhermi}) we have used the fact that $\hat{H}$ is a self adjoint operator on $\fsp$ and at (\ref{eq:usedtheequationbefore}) we have used equation (\ref{eq:alteffectofkeponklaud}).
\end{proof}
The Schr\"odinger equation in $\kcs$ is
\begin{equation} \nonumber
\Fracdiffl{f}{t} = \frac{2\pi}{ih} \tilde{\hat{H}} f=  \omega \Fracpartial{f}{\gamma}.
\end{equation}
So the time evolution of an arbitrary $f(t;\sgob) \in \kcs$ is given by
\begin{equation} \nonumber
f(t;\sgob) = f_0 ( \sigma , \gamma + \omega t , \overline{\Omega} )
\end{equation}
where $f_0 (\sgob )= f (0;\sgob )$, the initial value of the state at time $t=0$. The eigenfunctions of the operator $\tilde{\hat{H}} = \frac{ih}{2\pi} \omega \Partial{\gamma}$  are
\begin{eqnarray} \nonumber
\lefteqn{ f_{(n,l,m)} (\sigma, \gamma, \overline{\Omega})} \\ \nonumber
&=& e^{-\sigma^2} \left( \frac{\sigma^n \exp \left( - \frac{2\pi  \gamma }{ ih(n)^2} \right) }{(n!)^{1/2}}  \right) \left[ \frac{ (2l)!}{(l+m)!(l-m)!} \right]^{1/2} \\ \nonumber
&& \hspace{1cm} \times \left( \sin \left( \frac{\overline{\theta}}{2} \right) \right)^{l-m}  \left( \cos \left( \frac{\overline{\theta}}{2} \right) \right)^{l+m} \\ \nonumber
&& \hspace{1cm} \times e^{-i(m\overline{\phi} + l\overline{\psi} )} \, (2l+1)^{1/2}
\end{eqnarray}
where $n \in \Space{N}{}$, $l\in \Space{N}{}$ such that $0 \leq l \leq n$ and $m \in \Space{Z}{}$ such that $-l \leq m \leq l$. These eigenfunctions will have eigenvalue $-\frac{\omega}{n^2}$ with degeneracy $n^2$. This agrees with the usual quantum mechanical theory. It is important to note that this model is only suitable for calculating probability amplitudes for states which are in $\bssp$ and observables which preserve $\bssp$. However we will describe in Section \ref{sect:genofkepcou} how this can be extended to model a larger set of states.

\section{Generalisations} \label{sect:genofkepcou}

\indent

We now indicate how the above approach for the Kepler/Coulomb problem can be extended to any quantum mechanical system with a discrete spectrum. Furthermore we show that this approach can be extended to include systems with discrete and continuous spectra. This is all done by facilitating the extensions of Klauder's coherent states.

Since Klauder discovered his coherent states for the hydrogen atom there have been many extensions. Majumdar and Sharatchandra have written a paper \cite{MajumdarSharatchandra97} discussing relations between coherent states for the hydrogen atom and the action angle variables\index{action angle variables} for the Kepler problem \cite{Goldstein80}. Fox \cite{Fox99} extended this approach to show how these states could be realised as Gaussians. In \cite{Crawford00} Crawford described an extension which could model general systems with an energy degeneracy. This work used Perelomov's coherent states \cite{Perelomov86} for the degeneracy group. Since these coherent states satisfy both a resolution of the identity for the set of states in question and are temporally stable, the associated Hilbert spaces can be obtained in exactly the same way as in Section \ref{sect:klaudrep}. The proofs will almost follow word for word.

If the set of eigenfunctions for the Hamiltonian in question spans the entire space then $\mathcal{K}_1$ from equation (\ref{eq:defofkcmap}) will be unitary, bijective, invertible and defined on the whole of\footnote{We do not need to necessarily consider the space $\fsp$ here. The map would be defined on the Hilbert space which the eigenfunctions are from. This would usually be $\ltworn$, but as has been shown throughout this thesis another space may be more appropriate for particular systems.} $\fsp$. This means that the Hilbert space we obtain will be able to deal with any observable and any state. Furthermore $\kcmap \rho_h^S \kcmap^{-1}$ will be a unitary irreducible representation of the Heisenberg group which is unitarily equivalent to the Schr\"odinger representation. This representation would be able to model probability amplitudes for any quantum mechanical state and quantum mechanical observable.

We can also extend our approach to systems with both discrete and continuous spectra\index{discrete and continuous spectra}. The extension of the original coherent states to systems with both discrete and continuous spectra is given in
\cite{GazeauKlauder99,GazeauMonceau02}. Since these coherent states satisfy a resolution of the identity and are temporally stable we can obtain another Hilbert space by following the proofs in Section \ref{sect:klaudrep} word for word.

\chapter{Summary and Possible Extensions}

\section{Summary}

\indent

The main focus of this thesis has been demonstrating how the representation theory of the Heisenberg group can be used to model both quantum and classical mechanics. In Chapters \ref{chap:pmechandhn} and \ref{chap:staesandpic} we showed how states and observables from both classical and quantum mechanics could be described using functions/distributions on the Heisenberg group. In doing so we obtained new relations between classical and quantum mechanics. Also using different representations of the Heisenberg group we could simplify calculations which were at the heart of the mathematical formulation of quantum mechanics. In Chapter \ref{chap:staesandpic} we also showed that sometimes it could be more desirable to realise states as integration kernels as opposed to elements of a Hilbert space. By taking these new approaches we managed to simplify the proof of the classical limit of coherent states.

In Chapter \ref{chap:forcedosc} we showed how $p$-mechanics could be used to model some actual physical systems. In doing this we showed that the dynamics of the forced and harmonic oscillators could be modelled using $p$-mechanics. The classical and quantum dynamics would come from the same source separated by the one and infinite dimensional representations respectively. Again in this chapter we had more evidence that by using $p$-mechanics we could simplify some quantum mechanical calculations.

In Chapter \ref{chap:cantransf} we used $p$-mechanics to examine the relation between classical and quantum canonical transformations. One of the main features of this chapter was demonstrating how using a Hilbert space such as\footnote{A space such as $\fock$ or the Fock--Segal--Bargmann space would be just as useful.}  $\hilbh$ can be advantageous when modelling quantum phenomena. In \cite{MelloMoshinsky75,MoshinskySeligman78} Moshinsky and his collaborators used the eigenfunctions of the position and momentum operators on $\ltworn$ -- from the rigged Hilbert space formulation -- to generate a system of differential equations. Instead we used coherent states which were in the actual Hilbert space -- as opposed to the associated triple of rigged Hilbert spaces -- to derive a system of integral equations. The existence of reproducing kernels in $\fock$ and $\hilbh$ replaced the need for delta functions. Also our equations for non-linear transformations did not rely on the property that quantum mechanical observables are elements of the algebra generated by the position and momentum operators. In \cite{MelloMoshinsky75,MoshinskySeligman78} all the quantum mechanical operators are derived using this algebra condition --- in Chapter \ref{chap:cantransf} we used an integral transform instead. This integral transform at first made our equations look less desirable but it was shown that for some examples they take a simple form.

In Chapter \ref{chap:keplercoulomb} we showed that in certain cases it is advantageous to consider representations of $\Heisn$ other than the standard Schr\"odinger representation. The spherical polar coordinate representations showed that the spherical polar coordinate realisation of both classical and quantum mechanics can be derived from the same source. Also it was shown that for certain observables these new representations will give a simpler form than using the standard representations from \cite[Thm. 2.2]{Kisil02.1}. We also showed in this chapter that choosing Hilbert spaces other than $\ltworn$ can be advantageous when analysing the Kepler/Coulomb problem. The Hilbert space we derived in Section \ref{sect:klaudrep} was able to clearly represent the dynamics for the Kepler/Coulomb problem. However this space was limited since it could not model all the observables and states which are involved in the $\ltworn$ model of quantum mechanics -- this also meant there did not exist a unitary irreducible representation of the Heisenberg group on this space.

\section{Possible Extensions}

\indent

In Chapters \ref{chap:pmechandhn} and \ref{chap:staesandpic} we only looked at classical observables which were defined on the phase space $\Space{R}{2n}$. In the general formulation of classical mechanics a general symplectic manifold is used for the phase space. One interesting and very important extension of this work would be to extend the framework of $p$-mechanics to observables which are functions defined on manifolds.

The most immediate extension of the work in Chapter \ref{chap:cantransf} would be to look at more complex canonical transformations especially some more non-linear transformations.  Another interesting extension would be to look at the role of Egorov's Theorem \cite{Kisil02.2} in infinitesimal $p$-mechanical transformations. Egorov's Theorem \cite{Egorov86} has always been posed in the language of pseudodifferential operators on $\Space{R}{n}$; this idea could be extended to our space $\hilbh$ with pseudodifferential operators being replaced by Toeplitz operators as in \cite{Howe80.1}. Another possible extension would be to use more general coherent states to derive different systems of equations. In \cite{Perelomov86} different sets of coherent states for different Lie groups are presented; it would be interesting to see how equations (\ref{eq:thefinaleqnforkernsinctsof}) would change for different Lie groups. Also these new systems of equations may be more suitable for particular problems. Furthermore in Section \ref{sect:genofkepcou} we described how you can choose a system of coherent states which is suitable for a particular system. It may be of interest to see how this can be used to generate systems of canonical transformation equations which are particularly suitable for different systems.

One extension of the work in Chapter \ref{chap:keplercoulomb} would be to choose different weights for the coherent states. The choice of weights for the coherent states in equation (\ref{eq:defofkccs}) would have an effect on the Hilbert space derived in Section \ref{sect:klaudrep}. The choice of particular weights to coincide with physical requirements is a subject currently being heavily researched \cite{Crawford00,GazeauKlauder99,GazeauMonceau02}. It would be interesting to see if these states would generate a Hilbert space which satisfied certain physical requirements.  On the classical side of things an interesting extension of Chapter \ref{chap:keplercoulomb} would be to try and adapt the aforementioned important work of Moser\cite{Moser70} and Souriau \cite{Souriau74} into the $p$-mechanical construction. This would require the extension of $p$-mechanics to deal with classical observables defined on manifolds.

\appendix

\chapter{}

\section{Some Useful Formulae and Results} \label{app:useful}

\indent

In this section we present some results and formulae which are used to underpin the work in this thesis.
An equation used throughout this thesis is
\begin{equation}
\label{eq:waveletwithx}
\int_{\Space{R}{}} \exp ( -a x^2 + 2b x ) \, dx = \left( \frac{\pi}{a} \right)^{\frac{1}{2}} \exp \left( \frac{b^2}{a} \right).
\end{equation}
where $a>0$. A similar equation \cite[p337]{GradshteynRyzhik80}
which we repeatedly use is
\begin{equation} \label{eq:waveletwithxforalln}
\int_{\Space{R}{}} x^n \exp(-ax^2+2bx) \, dx = \frac{1}{2^{n-1} a} \left( \frac{\pi}{a}\right)^{\frac{1}{2}} \frac{d^{n-1}}{db^{n-1}} \left( b \exp \left( \frac{b^2}{a} \right) \right)
\end{equation}
providing $a>0$.  This equation for the particular value of $n=1$ is well known:
\begin{equation} \label{eq:waveletwith}
\int_{\Space{R}{}} x \exp ( -a x^2 + 2b x ) \, dx = \left( \frac{\pi}{a} \right)^{\frac{1}{2}} \left( \frac{b}{a} \right) \exp \left( \frac{b^2}{a} \right).
\end{equation}

One theorem that is used throughout this thesis is Fubini's Theorem on changing\index{Fubini's Theorem} the order of integration.
\begin{Thm}[Fubini's Theorem]\label{thm:fubini}\cite[Sect. 15]{Zwillinger92}
If $f(x,y)$ is an integrable function on $\Space{R}{n} \times \Space{R}{m}$ then
\begin{equation}
\int_{\Space{R}{n} \times \Space{R}{m} } f(x,y) \, dx \, dy = \int_{\Space{R}{m}} \left( \int_{\Space{R}{n}} f(x,y) \, dx \right) \, dy = \int_{\Space{R}{n}} \left( \int_{\Space{R}{m}} f(x,y) \, dy \right) \, dx.
\end{equation}
\end{Thm}
A proof of this can be found in \cite[Sect. 35.3]{KolmogorovFomin75}. In Appendix \ref{app:distributions} we present a similar result using distributions instead of functions.

The next Lemma\index{Schur's Lemma} is a version of one of the most important results in representation theory.
\begin{Lemma}[Schur's Lemma]\cite[Chap. 0, Prop. 4.1]{Taylor86} \label{lem:schurslemma}
A representation, $\rho$, of a group $G$ on a Hilbert space is irreducible if and only if for any bounded linear operator $U$
\begin{equation}
U \rho (g) = \rho (g) U \hspace{1cm} \forall g \in G \hspace{1cm} \implies \hspace{1cm} U = c I
\end{equation}
where $c$ is a constant and $I$ is the identity operator.
\end{Lemma}

\section{Vector Fields and Differential Forms on $\Space{R}{2n}$} \label{app:vfsanddfs}

\indent

In this section we give a brief overview of vector fields and differential forms on $\Space{R}{2n}$. Vector fields and differential forms are usually described in the language of manifolds. For this thesis we only discuss objects on $\Space{R}{2n}$ so we do not describe these concepts in complete generality. A good description of manifolds and their relation to classical mechanics is given in \cite{MarsdenRatiu99}.

The set of tangent vectors\index{tangent vector} at a point $(q',p') \in \Space{R}{2n}$ can be realised as the set of functionals on $\cinf (\Space{R}{2n})$ of the form
\begin{equation} \label{eq:tangentvector}
f(q,p) \mapsto \sum_{i=1}^{n} a_i \Fracpartial{f}{q_i} |_{x=(q',p')} + b_i \Fracpartial{f}{p_i} |_{x=(q',p')}.
\end{equation}
The space of all tangent vectors at a point $(q',p')$ is denoted as $T_{(q',p')} \Space{R}{2n}$. A vector field\index{vector field} on $\Space{R}{2n}$ associates to each point of $\Space{R}{2n}$ a tangent vector at that point. A vector field can be realised by a differential operator of the form
\begin{equation} \label{eq:vectorfield}
\sum_{i=1}^n \left( a_i (q,p) \Partial{q_i} + b_i (q,p) \Partial{p_i} \right)
\end{equation}
where now $a_i$ and $b_i$ are $C^{\infty}$ functions on $\Space{R}{2n}$.  If we multiply a vector field by a $C^{\infty}$ function it will be another vector field.

A differential one-form\index{differential one-form} on $\Space{R}{2n}$ is a map from the set of vector fields to the space of $C^{\infty}$ functions on $\Space{R}{2n}$. The differential one-forms $dq_1, \cdots , dq_n, \newline dp_1,\cdots,dp_n$ are defined by
\begin{equation} \label{eq:eachdifferentialform}
dq_i \left(a(q,p) \Partial{q_j} \right) = a(q,p) \, \delta_{ij} \hspace{3cm} dp_i \left(b(q,p) \Partial{p_j} \right) = b(q,p) \delta_{ij}.
\end{equation}
Any differential one-form on $\Space{R}{2n}$ can be written in the form
\begin{equation} \label{eq:differentialform}
\sum_{i=1}^n (a_i (q,p) dq_i + b_i (q,p) dp_i)
\end{equation}
where again $a_i$ and $b_i$ are functions on $\Space{R}{2n}$. If we multiply a differential one-form by a function it will be another differential one-form. For any function $f \in \cinf (\Space{R}{2n})$ the associated differential one-form $df$ is
\begin{equation} \label{eq:differentialformassoctoafctn}
df = \sum_{i=1}^n \Fracpartial{f}{q_i} dq_i + \Fracpartial{f}{p_i} dp_i .
\end{equation}

A two-form\index{two-form} is a map which sends two vector fields to a $C^{\infty}$ function on $\Space{R}{2n}$. If we have two one-forms $\alpha$ and $\beta$ the wedge product\index{wedge product} of $\alpha$ and $\beta$ will be the two-form
\begin{equation}
(\alpha \wedge \beta ) (X,Y) = \alpha (X) \beta (Y) - \alpha (Y) \beta (X)
\end{equation}
where $X,Y$ are vector fields on $\Space{R}{2n}$.
If we have a one-form $\alpha = f dg$ where $f,g \in \cinf (\Space{R}{2n})$ then the two-form $d\alpha$ (called the exterior derivative of $\alpha$) is
\begin{equation} \nonumber
d\alpha = df \wedge dg.
\end{equation}
This is a very specific form of the exterior derivative. For a general overview of the exterior derivative see for example \cite{MarsdenRatiu99}.

All the above equations can be modified in the natural way to replace $\Space{R}{2n}$ with any open subset of $\Space{R}{n}$. For the rest of this section we use the space $\Space{R}{n}$ as opposed to $\Space{R}{2n}$. Suppose $f:\Space{R}{n} \rightarrow \Space{R}{n}$ is a differentiable map then the derivative of this map at point $x$ (denoted $T_x f$) is the map from $T_x (\Space{R}{n})$ to $T_{f(x)} (\Space{R}{n})$ defined as
\begin{equation} \label{eq:diffofmapdefn}
((T_x f) X) (c) = X (c \circ f).
\end{equation}
Here $X$ is a tangent vector realised as a functional on $\cinf (\Space{R}{n})$ and $c \in \cinf (\Space{R}{n})$. In terms of differential operators this is
\begin{equation} \label{eq:diffofmapintermsofdiff}
T_x f : \sum_{i=1}^n a_i (x') \Partial{x_i} |_{x'=x} \mapsto \sum_{i=1}^n \sum_{j=1}^n (Df)_{i,j} a_j (x') \Partial{x_i} |_{x'=f(x)}.
\end{equation}

\section{Lie Groups and their Representations} \label{app:liegps}

\indent

The purpose of this appendix is to introduce Lie groups, Lie algebras and all the surrounding machinery. Throughout this section we only consider finite dimensional Lie groups. Lie groups\index{Lie groups} in general are defined using manifolds. All the Lie groups needed in this thesis can be defined without manifolds so we just refer the reader to \cite[Chap. 9]{MarsdenRatiu99} for this general definition and related theory. For the purposes of this thesis we define Lie groups in a simpler manner.
\begin{Defn}
A Lie group, $G$, is a group which is homeomorphic to a subset of $\Space{R}{n}$ such that the group multiplication map and the inversion map are both analytic.
\end{Defn}
We now define a nilpotent Lie group. First we need to consider a sequence of subgroups. For any Lie group $G$ there is a sequence
\begin{equation} \label{eq:nilpotentseq}
G_0 = G \supset G_1 \supset \cdots \supset G_k \supset \cdots
\end{equation}
where $G_k$ is the closed subgroup of $G_{k-1}$ generated by elements of the type $g_1 g_2 g_1^{-1} g_2^{-1}$, $g_1 \in G$, $g_2 \in G_{k-1}$.
\begin{Defn}
A nilpotent Lie group is a Lie\index{nilpotent Lie group} group for which sequence (\ref{eq:nilpotentseq}) terminates, that is $G_l = \{ e \}$ for all $l$ larger than some $k$.
\end{Defn}
The Heisenberg group is a nilpotent Lie group; sequence (\ref{eq:nilpotentseq}) for this example is
\begin{equation} \nonumber
G \supset Z \supset \{ e \},
\end{equation}
where $Z= \{ (s,0,0) : s \in \Space{R}{} \}$, the centre of $\Heisn$.

In order to define the Lie algebra\index{Lie algebra} of a Lie group we need to introduce left--invariant vector fields\index{left--invariant vector fields}.
$\lambda_l (g)$ is used to denote the left shift on the set of functions defined on $G$, that is
\begin{equation} \label{eq:defofleftshift}
(\lambda_l(g) f)(h) = f(g^{-1} h)
\end{equation}
A vector field (see equation (\ref{eq:vectorfield})) on the Lie group $G$ is an object which at every point, $g \in G$, of the Lie group will give a tangent vector (see equation (\ref{eq:tangentvector})) $X(g)$ at $g$. A vector field, $X$, is left invariant\index{left invariant vector field} if
\begin{equation} \nonumber
([T_h (\lambda_l (g))] X)(h) = X(g^{-1}h)
\end{equation}
where $T_h(\lambda_l (g))$ is the differential (see equation (\ref{eq:diffofmapdefn})) of the left shift map. One realisation of the Lie algebra, $\LieA$, associated to a Lie group is the set of vectors spanned by the left-invariant vector fields. Equivalent realisations of the Lie algebra associated to a Lie group are given in \cite[Chap. 6]{Kirillov76}.

Now if $X_{\xi}$ is the left invariant vector field corresponding to $\xi \in \LieA$ then there exists a unique integral curve \cite[Sect. 9.1]{MarsdenRatiu99} $\gamma_{\xi}: \Space{R}{} \rightarrow G$ such that $\gamma_{\xi} (0) = e$ and $\gamma'_{\xi} (t) = X_{\xi} (\gamma_{\xi} (t))$. This gives us an exponential map from $\LieA$ to $G$ by
\begin{equation} \nonumber
\exp (\xi) = \gamma_{\xi} (1).
\end{equation}

We can consider functions which map from a Lie group to $\Space{C}{}$. Differentiating these functions is done in the natural way, and we now show how to integrate these functions. Left invariant Haar\index{Haar measure} measure, $dg$, on a Lie group, $G$, is a measure such that for any integrable function on $G$
\begin{equation} \nonumber
\int_G f(h g) \, dg = \int_G f(g) \, dg.
\end{equation}
Right-invariant Haar measure is defined analogously. Left and right invariant Haar measures for a Lie group may or may not coincide. If they do coincide then the measure is called unimodular. Now we know how to integrate these functions we can define spaces such as $L^1 (G)$ and the Hilbert space $L^2 (G)$ in the usual way.

The representation\index{representation!group} of a group $G$ on a Hilbert space $H$ is a family of operators
\begin{equation} \label{eq:repofaliegp}
\rho (g) : H \rightarrow H, \hspace{1.5cm} g \in G,
\end{equation}
which satisfy the algebra homomorphism property
\begin{equation}
\rho ( g_1 g_2 ) = \rho (g_1 ) \rho (g_2)
\end{equation}
and the identity
\begin{equation}
\rho (e) = I.
\end{equation}
Furthermore $\rho$ is a unitary representation if
\begin{equation} \nonumber
\rho (g)' = \rho (g)^{-1} = \rho (g^{-1}).
\end{equation}
The representation\index{representation!function} of a function\footnote{The space $\czeroinfg$ is defined in Appendix \ref{app:distributions}.} $f \in \czeroinfg$ is defined as
\begin{equation} \label{eq:repofafctnonaliegp}
\rho (f) v = \int_{G} f(g) \rho (g) v \, dg
\end{equation}
for any $v \in H$. The representation of a distribution is defined in Appendix \ref{app:distributions}.

The convolution\index{convolution!functions} of two functions $f_1 , f_2 \in L^1 (G)$  is given by
\begin{equation} \label{eq:defofconv}
(f_1 * f_2) (g) = \int_G f_1 (h) f_2(h^{-1} g) \, dh = \int_G f_1 (gh^{-1})f_2 (h) \, dh .
\end{equation}
The two definitions of convolution given above are equivalent due to the invariance of Haar measure. Other forms of convolution are given by
\begin{eqnarray} \label{eq:defofequivformofconv}
(f_1*f_2)(g) &=& \int_G (f_1 (h)) \lambda_l (g)\tilde{f}_2 (h) \, dh \\ \nonumber
&=& \int_G \lambda_r (g^{-1}) \tilde{f_1} (h) f_2 (h) \, dh \\ \nonumber
&=& \int f_1 (h) \lambda_l (h) f_2 (g) \, dh
\end{eqnarray}
where $\tilde{f}(g) = f(g^{-1})$ and $\lambda_l, \lambda_r$ are the left and right regular representations respectively. Furthermore if we assume our space has an $L^2$ inner product then the convolution of two functions $f_1 , f_2$ can also be realised as
\begin{equation} \nonumber
(f_1 * f_2 )(g) = \langle f_1 , \lambda_l (g) \tilde{f_2} \rangle = \langle \lambda_r (g^{-1}) f_1 , f_2 \rangle
\end{equation}
where $\tilde{f} (g)$ is now $\overline{f(g^{-1})}$. Convolutions involving distributions are defined in Appendix \ref{app:distributions}.

If $X$ is an element of the Lie algebra, $\mathfrak{g}$ associated with the Lie group $G$, then the representation of $X$ is defined as
\begin{equation} \label{eq:repofliealg}
\rho (X)u = \lim_{h \rightarrow 0} \frac{\rho(e^{hX})u - u}{h}.
\end{equation}

\section{Induced Representations} \label{app:indrep}

\indent

Induced representations \cite[Chap. 13]{Kirillov76} , \cite[Sect. 4.2]{AliAntGaz00} are a large part of representation theory. Here we give a brief overview of the parts of the subject relevant to this thesis. The theory is based around starting with the representation of a subgroup, then extending this to a representation of the whole group.

Let $H$ be a closed subgroup of a nilpotent Lie group $G$ and let $\rho$ be a representation of the subgroup $H$ onto some Hilbert space $V$.  The space $L(G,H,\rho)$ is defined as the set of measurable functions from $G$ to $V$ such that $F(gh) = \rho (h) F(g)$ for all $h \in H$. The representation, $\eta$, of $G$ on $L(G,H,\rho)$ defined by
\begin{equation} \label{eq:firstinducedrep}
\eta (g) F(g_1) = F( g^{-1} g_1 )
\end{equation}
is called the representation induced in the sense of Mackey by $\rho$. This map preserves the space $L(G,H,\rho)$ since for any $h \in H$
\begin{equation} \nonumber
(\eta(g)F)(g_1 h) = F(g^{-1} g_1 h) = \rho(h) F( g^{-1} g_1) = \rho (h) (\eta(g)F)(g_1).
\end{equation}
 An inner product on $L(G,H,\rho)$ is given by
\begin{equation} \label{eq:ipforindrep}
\langle F_1 , F_2  \rangle_{L(G,H,\rho)} = \int_G \langle F_1 (g) , F_2 (g) \rangle_V d\mu (g).
\end{equation}
The measure $d\mu (g)$ is chosen so that $\eta$ becomes a unitary representation. A general construction of the measure $\mu$ is given in \cite[Sect 13.2]{Kirillov76}; for our purposes the choice of $\mu$ is very simple (see Section \ref{sect:indreponhn}). $L^2 (G,H,\rho)$ is the Hilbert space associated with the inner product (\ref{eq:ipforindrep}).

There is another realisation of induced representations which are also relevant to this thesis. We use $X$ to denote the homogeneous space\footnote{We use the notation $g^{-1} H$ for an element of X rather than $gH$ since we are considering left shifts.} $G / H$. If $\sigma$ is a measurable mapping which extracts from each coset a particular element, that is $\sigma(g_i^{-1}H) \in g_i^{-1} H$ then we have the following lemma.
\begin{Lemma} \label{eq:decompforindrep}
If $g \in G$ then there exists a unique $x \in X$ and a unique $h \in H$ such that $g= \sigma (x) h$.
\end{Lemma}
\begin{proof}
Since the cosets $g_i^{-1}H$ partition $G$, $g$ must be an element of one and only one coset $x=g_i^{-1} H \in X$ and so $g=g_i^{-1} a$ for a particular $a\in H$. Furthermore $\sigma (g_i^{-1} H) = g_i^{-1} h$ for some $h \in H$. Since both $a$ and $h$ are in $H$, $v=h^{-1}a$ must also be in $H$. Finally we have
\begin{equation} \nonumber
g= g_i^{-1} a = g_i^{-1} h h^{-1} a = \sigma(x) v,
\end{equation}
where $x=g_i^{-1}H$ and $v$ are uniquely defined.
\end{proof}
Note that the $x \in X$ here is the coset in which $g$ lies.
$L^2(X)$ is the set of measurable functions on $X$ which are square integrable with respect to the invariant measure on $X$ which is derived from the Haar measure on $G$ in the usual way \cite[Chap. 9]{Kirillov76}. There is an isometry between $L^2 (G,H,\rho)$ and $L^2 (X)$ by associating $f\in L^2 (X)$ to $F\in L^2 (G,H,\rho)$ by
\begin{equation} \label{eq:isombetweenltwogandltwox}
f(x)=F(\sigma(x)),
\end{equation}
see \cite[Sect. 13.2]{Kirillov76} for more details of this.
\begin{Thm} \label{thm:secondindrep}
Under this isometry, representation $\eta$ from (\ref{eq:firstinducedrep}) goes into the representation $\phi$ on $L^2(X)$
\begin{equation} \nonumber
[\phi(g)F](x) = \rho (v) f(g^{-1} x)
\end{equation}
where $v=\sigma (g^{-1}x)^{-1} g^{-1} \sigma(x)$.
\end{Thm}
\begin{proof}
By a direct calculation
\begin{equation} \nonumber
[\phi(g)f](x) = \eta(g) F(\sigma(x)) = F(g^{-1} \sigma(x)).
\end{equation}
Since $g^{-1} \sigma(x) \in G$ is in the coset $g^{-1} x$ we can use Lemma \ref{eq:decompforindrep} to obtain a unique $v \in H$ such that $g^{-1} \sigma (x) =  \sigma(g^{-1}x)v$. This implies that
\begin{equation}
[\phi(g) f] (x) = F(\sigma (g^{-1}x) v ) = \rho(v)F(\sigma (g^{-1} x)) = \rho (v) f (g^{-1} x).
\end{equation}
\end{proof}

\section{Distributions} \label{app:distributions}

\indent

Throughout the development of quantum mechanics the classical notion of a function has been insufficient. This led to much development in the theory\index{distributions} of distributions \cite{GelfandShilov77} \cite[Chap. V]{ReedSimon80} \cite[Sect. 3.3]{KirillovGvishiani82} \cite{Treves67}. It should be realised that the theory of distributions is a mathematical field in its own right and also has applications in many other areas of applied mathematics. The basic idea of distributions is to choose a test space of functions -- which will be a set of functions with certain required properties\footnote{Informally these are sometimes referred to as spaces of sufficiently "nice" functions.} --  then to consider all operations on the dual space to this.

Before we can develop the theory of distributions we need to introduce a test\index{test space} space\index{$\czeroinf$}, $\mathcal{D} (\Space{R}{n})$,
\begin{equation} \nonumber
\mathcal{D} (\Space{R}{n}) = \{ \phi \in \cinf ( \Space{R}{n} ) : \phi \textrm{  has bounded support } \}.
\end{equation}
$\cinf ( \Space{R}{n} )$ is the space of functions on $\Space{R}{n}$ with continuous derivatives of all orders. The support of a continuous function is the closure of the set on which $\phi (x)$ is non-zero. The space $\mathcal{D} (\Space{R}{n})$  is often denoted\index{$\mathcal{D} (\Space{R}{n})$} by $\czeroinf$.
\begin{Defn}
The space of all distributions on $\Space{R}{n}$, denoted $\mathcal{D}'(\Space{R}{n})$, is\index{$\mathcal{D}'(\Space{R}{n})$} the set of continuous linear functionals (that is the dual space) on $\mathcal{D} (\Space{R}{n})$.
\end{Defn}

The action of a distribution $f$ on a test function $\phi$ is represented by $\langle f, \phi \rangle$. Constructing topologies on $\mathcal{D}(\Space{R}{n})$ and $\mathcal{D}'(\Space{R}{n})$ is a delicate operation which we do not go into here; we refer the reader to \cite{ReedSimon80}.

We often need to consider different test spaces, the smaller the test space we choose the larger the space of distributions we obtain. Another commonly used test space is the Schwartz space (or functions of rapid decrease) $\schw$. The Schwartz\index{Schwartz space} space, $\schw$, is the space\index{$\schw$} of all $\cinf$ functions, $\phi$, for which
\begin{equation} \nonumber
\| \phi \|_{k,q} = \sup_{x \in \Space{R}{n}} | x^k \phi^{(q)} (x) | < \infty
\end{equation}
for any multi-indices $k=k_1 , \hdots , k_n$, $q=q_1 , \hdots , q_n $. Here $x^k= x_1^{k_1}. \hdots . x_n^{k_n}$ and $\phi^{(q)} (x) = \frac { \partial^{q_1 + \hdots q_n} \phi  }{\partial x_1^{q_1} \hdots \partial x_n^{q_n} } (x)$.
An example of an element in $\schw$ is $e^{-x^2}$.
 $\schw$ is given a Fr\'echet space topology \cite{ReedSimon80} \cite[Chap.10, Example 4]{Treves67} by the semi-norms $\| \cdot \|_{k,q}$.
\begin{Defn}
The dual space to $\schw$ is called the space of tempered distributions\index{tempered distributions} and is denoted\index{$\schwp$} by $\schwp$.
\end{Defn}
The topology on $\schwp$ is derived in the natural way from the Fr\'echet space topology on $\schw$. The Fourier transform
\begin{equation} \label{eq:ftonschw}
\left( \mathcal{F} \phi \right) (y) = \int_{\Space{R}{n}} \phi (x) e^{-2\pi i xy} \, dx
\end{equation}
is an isomorphism from $\schw$ to $\schw$ \cite[Thm. 25.1]{Treves67}. If $f \in \schwp$ then the Fourier transform of $f$ denoted by $\mathcal{F} f$ is defined by
\begin{equation} \label{eq:ftonschwp}
\langle f, \mathcal{F} \phi \rangle = \langle \mathcal{F} f , \phi \rangle.
\end{equation}
The Fourier transform is an isomorphism from $\schwp$ to $\schwp$ \cite[Thm. 25.6]{Treves67}.
We now define one more space of distributions.
\begin{Defn}
The space\index{distributions with compact support} of distributions with compact support, $\mathcal{E}'(\Space{R}{n})$, is the\index{$\mathcal{E}'(\Space{R}{n})$} dual space to $\cinf (\rn)$.
\end{Defn}

The following inclusions clearly hold
\begin{equation} \nonumber
\czeroinf \subset \schw \subset \cinf (\rn)
\end{equation}
which implies that
\begin{equation} \nonumber
\mathcal{E}'(\rn) \subset \schwp  \subset \mathcal{D}' (\rn).
\end{equation}
We can add two distributions $f,g$ together by
\begin{equation} \nonumber
\langle f+g,\phi \rangle = \langle f , \phi \rangle + \langle g, \phi \rangle.
\end{equation}
The differentiation of a distribution is defined as follows
\begin{equation} \nonumber
\left\langle \Fracpartial{f}{x_i} , \phi \right\rangle = - \left\langle f, \Fracpartial{\phi}{x_i} \right\rangle.
\end{equation}

We can clearly replace $\Space{R}{n}$ by a Lie group $G$ to obtain the spaces $\mathcal{D}'(G)$, $\mathcal{E}'(G)$ and $\mathcal{S}'(G)$. The case of taking $G$ to be the Heisenberg group is used throughout this thesis.

We now show how representations of distributions\index{representation!distribution} are defined \cite{Taylor86}. If $\rho$ is a representation on a Hilbert space of a Lie group $G$ and $k$ is a distribution on $G$ then the representation of $k$ is defined by
\begin{equation} \label{eq:repofadistnonaliegp}
\langle \rho (k) v_1 , v_2 \rangle_H = \langle \langle \rho (g) v_1 , v_2 \rangle_H , k \rangle .
\end{equation}
where $v_1 , v_2 \in H$ such that $\langle \rho (g) v_1 , v_2 \rangle$ is in the test space. The brackets $\langle , \rangle_H$ represent the inner product on the Hilbert space $H$, whereas the brackets without the $H$ subscript are the action of a functional acting on an element of the test space.

The convolution\index{convolution!distributions} of two functions on a non-commutative nilpotent Lie group was defined in Appendix \ref{app:liegps}; we now extend this notion to the convolution of distributions on a non-commutative nilpotent Lie group. We first define the convolution of a distribution and an element of the test space. If $f \in \schwpg$ and $\phi \in \schwg$ then their convolution is defined in a similar way to (\ref{eq:defofequivformofconv})
\begin{equation} \nonumber
(f*\phi)(g) = \langle f , \lambda_l (g) \tilde{\phi} \rangle
\end{equation}
where $\tilde{\phi} (h) = \overline{\phi(h^{-1})}$.
We move on to define the convolution of two distributions which itself is a distribution. If $f_1 , f_2 \in \schwpg$ and $\phi \in \schwg$ then
\begin{equation} \label{eq:convoftwodistsforgeng}
\langle f_1 * f_2 , \phi \rangle = \int f_1 (g)  \left( \tilde{f_2} * \phi \right) (g) \, dg .
\end{equation}
For more discussion about these notions see \cite{Taylor86}. We complete this appendix with a result known as Fubini's theorem for distributions -- it is an analogy of Theorem \ref{thm:fubini}.
\begin{Thm} \label{thm:fubinifordist}
If $F_1$ is a distribution on $\Space{R}{m}$ and $F_2$ is a distribution on $\Space{R}{n}$ then for every test function, $\phi$, on $\Space{R}{m} \times \Space{R}{n}$
\begin{equation}
\langle F_1 , \langle F_2 , \phi \rangle \rangle = \langle F_2 , \langle F_1 , \phi \rangle \rangle.
\end{equation}
\end{Thm}
For a proof of this theorem see \cite[Thm. 40.4]{Treves67}.

\bibliography{everything}

\def\cprime{$'$} \def\cprime{$'$}
\begin{thebibliography}{10}

\bibitem{AliAntGazMueller95}
S.~Twareque Ali, J.-P. Antoine, J.-P. Gazeau, and U.~A. Mueller.
\newblock Coherent states and their generalizations: a mathematical overview.
\newblock {\em Rev. Math. Phys.}, 7(7):1013--1104, 1995.

\bibitem{AliAntGaz00}
Syed~Twareque Ali, Jean-Pierre Antoine, and Jean-Pierre Gazeau.
\newblock {\em Coherent states, wavelets and their generalizations}.
\newblock Graduate Texts in Contemporary Physics. Springer-Verlag, New York,
  2000.

\bibitem{Anderson94}
Arlen Anderson.
\newblock Canonical transformations in quantum mechanics.
\newblock {\em Ann. Physics}, 232(2):292--331, 1994.

\bibitem{AntoineVause81}
J.-P. Antoine and M.~Vause.
\newblock Partial inner product spaces of entire functions.
\newblock {\em Ann. Inst. H. Poincar\'e Sect. A (N.S.)}, 35(3):195--224, 1981.

\bibitem{Antoine99}
Jean-Pierre Antoine.
\newblock Partial inner product spaces of analytic functions.
\newblock In {\em Generalized functions, operator theory, and dynamical systems
  (Brussels, 1997)}, volume 399 of {\em Chapman \& Hall/CRC Res. Notes Math.},
  pages 26--47. Chapman \& Hall/CRC, Boca Raton, FL, 1999.

\bibitem{Arnold90}
V.~I. Arnold.
\newblock {\em Mathematical methods of classical mechanics}.
\newblock Springer-Verlag, New York, 1990.
\newblock Translated from the 1974 Russian original by K. Vogtmann and A.
  Weinstein, Corrected reprint of the second (1989) edition.

\bibitem{Bargmann61}
V.~Bargmann.
\newblock On a {H}ilbert space of analytic functions and an associated integral
  transform.
\newblock {\em Comm. Pure Appl. Math.}, 14:187--214, 1961.

\bibitem{BaxandallLiebeck86}
Peter Baxandall and Hans Liebeck.
\newblock {\em Vector calculus}.
\newblock Oxford Applied Mathematics and Computing Science Series. The
  Clarendon Press Oxford University Press, New York, 1986.

\bibitem{Berezin72}
F.~A. Berezin.
\newblock Covariant and contravariant symbols of operators.
\newblock {\em Izv. Akad. Nauk SSSR Ser. Mat.}, 36:1134--1167, 1972.

\bibitem{Berezin75}
F.~A. Berezin.
\newblock General concept of quantization.
\newblock {\em Comm. Math. Phys.}, 40:153--174, 1975.

\bibitem{Bohm01}
Arno Bohm.
\newblock {\em Quantum mechanics: foundations and applications}.
\newblock Springer-Verlag, New York, third edition, 2001.
\newblock Prepared with Mark Loewe.

\bibitem{BransdenJoachin00}
B.H. Bransden and Joachin C.J.
\newblock {\em Quantum Mechanics}.
\newblock Prentice Hall, Harlow, second edition, 2000.

\bibitem{Brodlie02}
Alastair Brodlie.
\newblock Classical and quantum coherent states.
\newblock {\em Internat. J. Theoret. Phys.}, 42(8):1707--1731, 2003.

\bibitem{Brodlie04}
Alastair Brodlie.
\newblock Nonlinear canonical transformations in classical and quantum
  mechanics.
\newblock {\em J. Math. Phys.}, 45(8):3413--3431, 2004.

\bibitem{Brodlie04.1}
Alastair Brodlie.
\newblock The representation theory of the {H}eisenberg group and beyond.
\newblock In {\em XIth International Conference on Symmetry Methods in Physics:
  Conference Proceedings}, 2004.
\newblock To appear.

\bibitem{KisilBrodlie02}
Alastair Brodlie and V.V. Kisil.
\newblock States and observables in $p$-mechanics.
\newblock In {\em Advances in Mathematics Research,V}, pages 101--136. Nova
  Science, 2003.
\newblock E-print:arXiv:quant-ph/0304023.

\bibitem{CocolicchioViggiano00}
D.~Cocolicchio and M.~Viggiano.
\newblock The squeeze expansion and the dissipative effects in coupled
  oscillations.
\newblock In {\em Advanced special functions and applications (Melfi, 1999)},
  volume~1 of {\em Proc. Melfi Sch. Adv. Top. Math. Phys.}, pages 277--290.
  Aracne, Rome, 2000.

\bibitem{CooperPellegrini99}
Richard~K Cooper and Claudio Pellegrini.
\newblock {\em Modern analytic mechanics}.
\newblock Kluwer academic/Plenum publishers, New York, 1999.

\bibitem{Crawford00}
MGA Crawford.
\newblock Temporally stable coherent states in energy-degenerate systems: The
  hydrogen atom.
\newblock {\em Phys. Rev. A}, 62(1):012104, 2000.

\bibitem{ZachosCurtrightFairlie98}
Thomas Curtright, David Fairlie, and Cosmas~K. Zachos.
\newblock Features of time-independent {W}igner functions.
\newblock {\em Phys. Rev. D (3)}, 58(2):025002, 14, 1998.

\bibitem{Dirac47}
P.A.M. Dirac.
\newblock {\em The Principles of Quantum Mechanics}.
\newblock Oxford University Press, Clarendon, 1947.
\newblock Third Edition.

\bibitem{DirlKasperkovitzMoshinsky88}
R.~Dirl, P.~Kasperkovitz, and M.~Moshinsky.
\newblock Wigner distribution functions and the representation of a
  nonbijective canonical transformation in quantum mechanics.
\newblock {\em J. Phys. A}, 21(8):1835--1846, 1988.

\bibitem{Egorov86}
Yu.~V. Egorov.
\newblock {\em Linear differential equations of principal type}.
\newblock Contemporary Soviet Mathematics. Consultants Bureau, New York, 1986.
\newblock Translated from the Russian by Dang Prem Kumar.

\bibitem{Fedosov02}
Boris Fedosov.
\newblock Deformation quantization: pro and contra.
\newblock In {\em Quantization, Poisson brackets and beyond (Manchester,
  2001)}, volume 315 of {\em Contemp. Math.}, pages 1--7. Amer. Math. Soc.,
  Providence, RI, 2002.

\bibitem{Folland89}
Gerald~B. Folland.
\newblock {\em Harmonic analysis in phase space}.
\newblock Princeton University Press, Princeton, NJ, 1989.

\bibitem{Fox99}
Ronald~F. Fox.
\newblock Generalized coherent states.
\newblock {\em Phys. Rev. A (3)}, 59(5):3241--3255, 1999.

\bibitem{GarciaMoshinsky80}
G.~Garc{\'{\i}}a-Calder{\'o}n and M.~Moshinsky.
\newblock Wigner distribution functions and the representation of canonical
  transformations in quantum mechanics.
\newblock {\em J. Phys. A}, 13(6):L185--L188, 1980.

\bibitem{GazeauKlauder99}
Jean~Pierre Gazeau and John~R. Klauder.
\newblock Coherent states for systems with discrete and continuous spectrum.
\newblock {\em J. Phys. A}, 32(1):123--132, 1999.

\bibitem{GazeauMonceau02}
Jean-Pierre Gazeau and Pascal Monceau.
\newblock Generalized coherent states for arbitrary quantum systems.
\newblock In {\em Conf\'erence Mosh\'e Flato 1999, Vol. II (Dijon)}, volume~22
  of {\em Math. Phys. Stud.}, pages 131--144. Kluwer Acad. Publ., Dordrecht,
  2000.

\bibitem{GelfandShilov77}
I.~M. Gel{\cprime}fand and G.~E. Shilov.
\newblock {\em Generalized functions. {V}ol. 1}.
\newblock Academic Press [Harcourt Brace Jovanovich Publishers], New York, 1964
  [1977].
\newblock Properties and operations, Translated from the Russian by Eugene
  Saletan.

\bibitem{GelfandVilenkin77}
I.~M. Gel{\cprime}fand and N.~Ya. Vilenkin.
\newblock {\em Generalized functions. {V}ol. 4}.
\newblock Academic Press [Harcourt Brace Jovanovich Publishers], New York, 1964
  [1977].
\newblock Applications of harmonic analysis, Translated from the Russian by
  Amiel Feinstein.

\bibitem{Goldstein80}
Herbert Goldstein.
\newblock {\em Classical mechanics}.
\newblock Addison-Wesley Publishing Co., Reading, Mass., second edition, 1980.
\newblock Addison-Wesley Series in Physics.

\bibitem{Gotay80}
Mark~J. Gotay.
\newblock Functorial geometric quantization and {V}an {H}ove's theorem.
\newblock {\em Internat. J. Theoret. Phys.}, 19(2):139--161, 1980.

\bibitem{GradshteynRyzhik80}
I.~S. Gradshteyn and I.~M. Ryzhik.
\newblock {\em Table of integrals, series, and products}.
\newblock Academic Press [Harcourt Brace Jovanovich Publishers], New York,
  1980.
\newblock Corrected and enlarged edition edited by Alan Jeffrey, Incorporating
  the fourth edition edited by Yu. V. Geronimus [Yu. V. Geronimus]\ and M. Yu.
  Tseytlin [M. Yu. Tse\u\i tlin], Translated from the Russian.

\bibitem{GuilleminSternberg90}
Victor Guillemin and Shlomo Sternberg.
\newblock {\em Variations on a theme by {K}epler}.
\newblock American Mathematical Society, Providence, RI, 1990.

\bibitem{HanKimNoz95}
D.~Han, Y.~S. Kim, and Marilyn~E. Noz.
\newblock {${\rm O}(3,3)$}-like symmetries of coupled harmonic oscillators.
\newblock {\em J. Math. Phys.}, 36(8):3940--3954, 1995.

\bibitem{HanKimNozYeh93}
D.~Han, Y.~S. Kim, Marilyn~E. Noz, and Leehwa Yeh.
\newblock Symmetries of two-mode squeezed states.
\newblock {\em J. Math. Phys.}, 34(12):5493--5508, 1993.

\bibitem{Hepp74}
Klaus Hepp.
\newblock The classical limit for quantum mechanical correlation functions.
\newblock {\em Comm. Math. Phys.}, 35:265--277, 1974.

\bibitem{Honerkamp98}
Josef Honerkamp.
\newblock {\em Statistical physics}.
\newblock Springer-Verlag, Berlin, 1998.
\newblock An advanced approach with applications, Translated from the German
  manuscript by Thomas Filk.

\bibitem{Howe80.1}
Roger Howe.
\newblock Quantum mechanics and partial differential equations.
\newblock {\em J. Funct. Anal.}, 38(2):188--254, 1980.

\bibitem{IguriCastagnino99}
S.~Iguri and M.~Castagnino.
\newblock The formulation of quantum mechanics in terms of nuclear algebras.
\newblock {\em Internat. J. Theoret. Phys.}, 38(1):143--164, 1999.
\newblock Irreversibility and cosmology. Fundamental aspects of quantum
  mechanics (Peyresq, 1997).

\bibitem{Jose98}
Jorge~V. Jos{\'e} and Eugene~J. Saletan.
\newblock {\em Classical dynamics}.
\newblock Cambridge University Press, Cambridge, 1998.
\newblock A contemporary approach.

\bibitem{Kirillov76}
A.~A. Kirillov.
\newblock {\em Elements of the theory of representations}.
\newblock Springer-Verlag, Berlin, 1976.
\newblock Translated from the Russian by Edwin Hewitt, Grundlehren der
  Mathematischen Wissenschaften, Band 220.

\bibitem{Kirillov99}
A.~A. Kirillov.
\newblock Merits and demerits of the orbit method.
\newblock {\em Bull. Amer. Math. Soc. (N.S.)}, 36(4):433--488, 1999.

\bibitem{KirillovGvishiani82}
A.~A. Kirillov and A.~D. Gvishiani.
\newblock {\em Theorems and problems in functional analysis}.
\newblock Problem Books in Mathematics. Springer-Verlag, New York, 1982.
\newblock Translated from the Russian by Harold H. McFaden.

\bibitem{Kirillov94}
A.A. Kirillov.
\newblock {\em Representation theory and noncommutative harmonic analysis.
  {I}}, volume~22 of {\em Encyclopaedia of Mathematical Sciences}.
\newblock Springer-Verlag, Berlin, 1994.
\newblock Fundamental concepts. Representations of Virasoro and affine
  algebras, A translation of {\it Current problems in mathematics. Fundamental
  directions. Vol.\ 22} (Russian), Akad.\ Nauk SSSR, Vsesoyuz.\ Inst.\ Nauchn.\
  i Tekhn.\ Inform., Moscow, 1988 [MR 88k:22001], Translation by V. Sou\v cek,
  Translation edited by A. A. Kirillov.

\bibitem{Kisil96.1}
Vladimir~V. Kisil.
\newblock Plain mechanics: classical and quantum.
\newblock {\em J. Natur. Geom.}, 9(1):1--14, 1996.

\bibitem{Kisil99}
Vladimir~V. Kisil.
\newblock Wavelets in {B}anach spaces.
\newblock {\em Acta Appl. Math.}, 59(1):79--109, 1999.

\bibitem{Kisil02.2}
Vladimir~V. Kisil.
\newblock Meeting {Descartes} and {Klein} somewhere in a noncommutative space.
\newblock In A.~Fokas, J.~Halliwell, T.~Kibble, and B.~Zegarlinski, editors,
  {\em Highlights of Mathematical Physics}, pages 165--189. AMS, 2002.
\newblock E-print:arXiv{math-ph/0112059}.

\bibitem{Kisil02}
Vladimir~V. Kisil.
\newblock Quantum and classical brackets.
\newblock {\em Internat. J. Theoret. Phys.}, 41(1):63--77, 2002.
\newblock E-print:arXiv:math-ph/0007030.

\bibitem{Kisil02.1}
Vladimir~V. Kisil.
\newblock $p$-{Mechanics} as a physical theory: an introduction.
\newblock {\em J. Phys. A}, 37:183--204, 2004.
\newblock E-print:arXiv{quant-ph/0212101}.

\bibitem{Klauder96}
John~R. Klauder.
\newblock Coherent states for the hydrogen atom.
\newblock {\em J. Phys. A}, 29(12):L293--L298, 1996.

\bibitem{KolmogorovFomin75}
A.~N. Kolmogorov and S.~V. Fom{\={\i}}n.
\newblock {\em Introductory real analysis}.
\newblock Dover Publications Inc., New York, 1975.
\newblock Translated from the second Russian edition and edited by Richard A.
  Silverman, Corrected reprinting.

\bibitem{Kurunoglu62}
Behram Kur{\c{s}}uno{\u{g}}lu.
\newblock {\em Modern quantum theory}.
\newblock W. H. Freeman and Co., San Francisco, Calif., 1962.

\bibitem{Liboff80}
Richard Liboff.
\newblock {\em Introductory quantum mechanics}.
\newblock Holden-Day Inc., San Francisco, 1980.

\bibitem{MajumdarSharatchandra97}
Pushan Majumdar and H.~S. Sharatchandra.
\newblock Coherent states for the hydrogen atom.
\newblock {\em Phys. Rev. A (3)}, 56(5):R3322--R3325, 1997.

\bibitem{MarsdenRatiu99}
Jerrold~E. Marsden and Tudor~S. Ratiu.
\newblock {\em Introduction to mechanics and symmetry}.
\newblock Springer-Verlag, New York, second edition, 1999.
\newblock A basic exposition of classical mechanical systems.

\bibitem{MarsdenTromba03}
Jerrold~E. Marsden and Anthony Tromba.
\newblock {\em Vector Calculus}.
\newblock W.H. Freeman and co., fifth edition, 2003.

\bibitem{Martinez83}
Jos{\'e} Martinez.
\newblock Diagrammatic solution of the forced oscillator.
\newblock {\em European J. Phys.}, 4(4):221--227 (1984), 1983.

\bibitem{MelloMoshinsky75}
P.~A. Mello and M.~Moshinsky.
\newblock Nonlinear canonical transformations and their representations in
  quantum mechanics.
\newblock {\em J. Mathematical Phys.}, 16(10):2017--2028, 1975.

\bibitem{Merzbacher70}
Eugen Merzbacher.
\newblock {\em Quantum mechanics}.
\newblock John Wiley \& Sons Inc., New York, 1998.

\bibitem{Messiah61}
Albert Messiah.
\newblock {\em Quantum mechanics. {V}ol. {I}}.
\newblock Translated from the French by G. M. Temmer. North-Holland Publishing
  Co., Amsterdam, 1961.

\bibitem{Moser70}
J.~Moser.
\newblock Regularization of {K}epler's problem and the averaging method on a
  manifold.
\newblock {\em Comm. Pure Appl. Math.}, 23:609--636, 1970.

\bibitem{MoshinskySeligman78}
M.~Moshinsky and T.~H. Seligman.
\newblock Canonical transformations to action and angle variables and their
  representations in quantum mechanics.
\newblock {\em Ann. Physics}, 114(1-2):243--272, 1978.

\bibitem{MoshinskySeligman79}
M.~Moshinsky and T.~H. Seligman.
\newblock Canonical transformations to action and angle variables and their
  representation in quantum mechanics. {II}. {T}he {C}oulomb problem.
\newblock {\em Ann. Physics}, 120(2):402--422, 1979.

\bibitem{Perelomov86}
A.~Perelomov.
\newblock {\em Generalized coherent states and their applications}.
\newblock Texts and Monographs in Physics. Springer-Verlag, Berlin, 1986.

\bibitem{Kisil97}
Oleg~V. Prezhdo and Vladimir~V. Kisil.
\newblock Mixing quantum and classical mechanics.
\newblock {\em Phys. Rev. A (3)}, 56(1):162--175, 1997.

\bibitem{ReedSimon80}
Michael Reed and Barry Simon.
\newblock {\em Methods of modern mathematical physics. {I}}.
\newblock Academic Press Inc. [Harcourt Brace Jovanovich Publishers], New York,
  second edition, 1980.
\newblock Functional analysis.

\bibitem{Roberts66}
J.~E. Roberts.
\newblock Rigged {H}ilbert spaces in quantum mechanics.
\newblock {\em Comm. Math. Phys.}, 3:98--119, 1966.

\bibitem{Ruelle66}
David Ruelle.
\newblock States of physical systems.
\newblock {\em Comm. Math. Phys.}, 3:133--150, 1966.

\bibitem{Simms73}
D.~J. Simms.
\newblock Bohr-{S}ommerfeld orbits and quantizable symplectic manifolds.
\newblock {\em Proc. Cambridge Philos. Soc.}, 73:489--491, 1973.

\bibitem{Simms74}
D.~J. Simms.
\newblock Geometric quantization of energy levels in the {K}epler problem.
\newblock In {\em Symposia Mathematica, Vol. XIV (Convegno di Geometria
  Simplettica e Fisica Matematica, INDAM, Rome, 1973)}, pages 125--137.
  Academic Press, London, 1974.

\bibitem{Sniatycki80}
Jedrzej {\'S}niatycki.
\newblock {\em Geometric quantization and quantum mechanics}.
\newblock Springer-Verlag, New York, 1980.

\bibitem{Souriau74}
Jean-Marie Souriau.
\newblock Sur la vari\'et\'e de {K}\'epler.
\newblock In {\em Symposia Mathematica, Vol. XIV (Convegno di Geometria
  Simplettica e Fisica Matematica, INDAM, Rome, 1973)}, pages 343--360.
  Academic Press, London, 1974.

\bibitem{Taylor86}
Michael~E. Taylor.
\newblock {\em Noncommutative harmonic analysis}.
\newblock American Mathematical Society, Providence, RI, 1986.

\bibitem{Treves67}
Fran{\c{c}}ois Tr{\`e}ves.
\newblock {\em Topological vector spaces, distributions and kernels}.
\newblock Academic Press, New York, 1967.

\bibitem{Vilenkin68}
N.~Ja. Vilenkin.
\newblock {\em Special functions and the theory of group representations}.
\newblock American Mathematical Society, Providence, R. I., 1968.

\bibitem{Bohm02}
S.~Wickramasekara and A.~Bohm.
\newblock Symmetry representations in the rigged {H}ilbert space formulation of
  quantum mechanics.
\newblock {\em J. Phys. A}, 35(3):807--829, 2002.

\bibitem{Woodhouse92}
N.~M.~J. Woodhouse.
\newblock {\em Geometric quantization}.
\newblock The Clarendon Press Oxford University Press, New York, second
  edition, 1992.
\newblock Oxford Science Publications.

\bibitem{Zachos02a}
Cosmas Zachos.
\newblock Deformation quantization: quantum mechanics lives and works in
  phase-space.
\newblock {\em Internat. J. Modern Phys. A}, 17(3):297--316, 2002.
\newblock E:print:arXiv:hep-th/0110114.

\bibitem{Zwillinger92}
Daniel Zwillinger.
\newblock {\em Handbook of integration}.
\newblock Jones and Bartlett Publishers, Boston, MA, 1992.

\end{thebibliography}

\bibliographystyle{plain}

\printindex

\end{document}